\documentclass[a4paper,fleqn,usenatbib]{mnras}

\usepackage{newtxtext,newtxmath}

\usepackage[T1]{fontenc}
\usepackage{ae,aecompl}

\usepackage{graphicx}	
\usepackage{hyperref}   
\usepackage{amsmath}	
\usepackage{amssymb}	
\usepackage{longtable}  

\usepackage{eso-pic}

\AddToShipoutPictureBG*{%
  \AtPageUpperLeft{%
    \hspace{0.75\paperwidth}%
    \raisebox{-3.5\baselineskip}{%
      \makebox[0pt][l]{\textnormal{DES-2017-0231}}
}}}%

\AddToShipoutPictureBG*{%
  \AtPageUpperLeft{%
    \hspace{0.75\paperwidth}%
    \raisebox{-4.5\baselineskip}{%
      \makebox[0pt][l]{\textnormal{FERMILAB-PUB-18-256-AE}}
}}}%

\title[High-z Galaxies in DES]{Candidate Massive Galaxies at $z \sim 4$ in the Dark Energy Survey}

\author[P. Guarnieri et al.]{
Pierandrea Guarnieri,$^{1}$\thanks{E-mail: pierandrea.guarnieri@port.ac.uk}
Claudia Maraston,$^{1}$ 
Daniel Thomas,$^{1}$
Janine Pforr,$^{2}$ \newauthor
Violeta Gonzalez-Perez,$^{1}$
James Etherington,$^{1}$
Joakim Carlsen, \newauthor
Xan Morice-Atkinson,$^{1}$
Christopher J. Conselice,$^{3}$
Julia Gschwend,$^{4,5}$ \newauthor
Matias Carrasco Kind,$^{6,7}$
Tim Abbott,$^{8}$
Sahar Allam,$^{9}$
David Brooks,$^{10}$
David Burke,$^{11,12}$ \newauthor
Aurelio Carnero Rosell,$^{4,5}$
Jorge Carretero,$^{13}$
Carlos Cunha,$^{9}$
Chris D'Andrea,$^{14}$ \newauthor
Luiz da Costa,$^{4,5}$
Juan De Vincente,$^{15}$
Darren DePoy,$^{16}$
H. Thomas Diehl,$^{9}$
Peter Doel,$^{10}$ \newauthor
Josh Frieman,$^{9,11}$
Juan Garcia-Bellido,$^{17}$
Daniel Gruen,$^{11,12}$
Gaston Gutierrez,$^{9}$ \newauthor
Dominic Hanley,$^{1}$
Devon Hollowood,$^{18}$
Klaus Honscheid,$^{19,20}$
David James,$^{21}$ \newauthor
Tesla Jeltema,$^{18}$ 
Kyler Kuehn,$^{22}$
Marcos Lima,$^{4,23}$
Marcio A.~G.~Maia,$^{4,5}$ \newauthor
Jennifer Marshall,$^{16}$ 
Paul Martini,$^{19,20}$
Peter Melchior,$^{24}$
Felipe Menanteau,$^{25,26}$ \newauthor
Ramon Miquel,$^{13,27}$
Andres Plazas Malagon,$^{28}$
Samuel Richardson,$^{1}$
Kathy Romer,$^{29}$  \newauthor
Eusebio Sanchez,$^{15}$
Vic Scarpine,$^{9}$ 
Rafe Schindler,$^{12}$
Ignacio Sevilla,$^{15}$  \newauthor
Mathew Smith,$^{30}$
Marcelle Soares-Santos,$^{31}$
Flavia Sobreira,$^{4,32}$
Eric Suchyta,$^{33}$ \newauthor
Gregory Tarle,$^{34}$
Alistair Walker,$^{6}$
William Wester$^{7}$
\\
$^{1}$Institute of Cosmology and Gravitation, University of Portsmouth, Dennis Sciama Building, Burnaby Road, Portsmouth PO1 3FX, UK \\
$^{2}$ESTEC: European Space Research and Technology Centre, Keplerlaan 1, 2201 AZ Noordwijk, Netherlands \\
$^{3}$University of Nottingham, School of Physics \& Astronomy, Nottingham, NG7 2RD, UK \\
$^{4}$Laboratorio Interinstitucional de e-Astronomia - LIneA, Rua Gal. Jose Cristino 77, Rio de Janeiro, RJ - 20921-400, Brazil \\
$^{5}$Observatorio Nacional, Rua Gal. Jose Cristino 77, Rio de Janeiro, RJ - 20921-400, Brazil \\ 
$^{6}$Department of Astronomy, University of Illinois at Urbana-Champaign, 1002 W. Green Street, Urbana, IL 61801, USA \\
$^{7}$National Center for Supercomputing Applications, 1205 West Clark St., Urbana, IL 61801, USA \\
$^{8}$Cerro Tololo Inter-American Observatory, National Optical Astronomy Observatory, Casilla 603, La Serena, Chile \\ 
$^{9}$Fermi National Accelerator Laboratory, P. O. Box 500, Batavia, IL 60510, USA \\
$^{10}$Department of Physics \& Astronomy, University College London, Gower Street, London, WC1E 6BT, UK \\ 
$^{11}$Kavli Institute for Particle Astrophysics \& Cosmology, P. O. Box 2450, Stanford University, Stanford, CA 94305, USA \\
$^{12}$SLAC National Accelerator Laboratory, Menlo Park, CA 94025, USA \\
$^{13}$Institut de Fisica d'Altes Energies (IFAE), The Barcelona Institute of Science and Technology, Campus UAB, 08193 Bellaterra \\ (Barcelona) Spain \\
$^{14}$Department of Physics and Astronomy, University of Pennsylvania, Philadelphia, PA 19104, USA \\
$^{15}$Centro de Investigaciones Energeticas, Medioambientales y Tecnologicas (CIEMAT), Madrid, Spain \\
$^{16}$George P. and Cynthia Woods Mitchell Institute for Fundamental Physics and Astronomy, and Department of Physics and Astronomy, \\ Texas A\&M University, College Station, TX 77843,  USA \\
$^{17}$Instituto de Fisica Teorica UAM/CSIC, Universidad Autonoma de Madrid, 28049 Madrid, Spain \\
$^{18}$Santa Cruz Institute for Particle Physics, Santa Cruz, CA 95064, USA \\
$^{19}$Center for Cosmology and Astro-Particle Physics, The Ohio State University, Columbus, OH 43210, USA \\
$^{20}$Department of Physics, The Ohio State University, Columbus, OH 43210, USA \\
$^{21}$Harvard-Smithsonian Center for Astrophysics, Cambridge, MA 02138, USA \\
$^{22}$Australian Astronomical Observatory, North Ryde, NSW 2113, Australia \\
$^{23}$Departamento de Fisica Matematica, Instituto de Fisica, Universidade de Sao Paulo, CP 66318, Sao Paulo, SP, 05314-970, Brazil \\
$^{24}$Department of Astrophysical Sciences, Princeton University, Peyton Hall, Princeton, NJ 08544, USA \\
$^{25}$Department of Astronomy, University of Illinois at Urbana-Champaign, 1002 W. Green Street, Urbana, IL 61801, USA \\
$^{26}$National Center for Supercomputing Applications, 1205 West Clark St., Urbana, IL 61801, USA \\
$^{27}$Institucio Catalana de Recerca i Estudis Avancats, E-08010 Barcelona, Spain
$^{28}$Jet Propulsion Laboratory, \\ California Institute of Technology, 4800 Oak Grove Dr., Pasadena, CA 91109, USA \\
$^{29}$Department of Physics and Astronomy, Pevensey Building, University of Sussex, Brighton, BN1 9QH, UK \\
$^{30}$School of Physics and Astronomy, University of Southampton,  Southampton, SO17 1BJ, UK \\
$^{31}$Brandeis University, Physics Department, 415 South Street, Waltham MA 02453 \\
$^{32}$Instituto de Fisica Gleb Wataghin, Universidade Estadual de Campinas, 13083-859, Campinas, SP, Brazil \\
$^{33}$Computer Science and Mathematics Division, Oak Ridge National Laboratory, Oak Ridge, TN 37831 \\
$^{34}$Department of Physics, University of Michigan, Ann Arbor, MI 48109, USA
}

\date{Accepted XXX. Received YYY; in original form ZZZ}

\pubyear{2018}

\begin{document}	
\label{firstpage}
\pagerange{\pageref{firstpage}--\pageref{lastpage}}
\maketitle

\begin{abstract}
Using stellar population models, we predicted that the Dark Energy Survey (DES) - due to its special combination of area (5000 deg.sq.) and depth ($i = 24.3$) - would be in the position to detect massive ($\gtrsim 10^{11}$ M$_{\odot}$) galaxies at $z \sim 4$. We confront those theoretical calculations with the first $\sim 150$~deg.~sq.~of DES data reaching nominal depth. From a catalogue containing $\sim 5$~million sources, $\sim26000$~were found to have observed-frame $g-r$~vs~$r-i$~colours within the locus predicted for $z \sim 4$~massive galaxies. We further removed contamination by stars and artefacts, obtaining 606 galaxies lining up by the model selection box. We obtained their photometric redshifts and physical properties by fitting model templates spanning a wide range of star formation histories, reddening and redshift. Key to constrain the models is the addition, to the optical DES bands $g$, $r$, $i$, $z$, and $Y$, of near-IR $J$, $H$, $K_{s}$~data from the Vista Hemisphere Survey. We further applied several quality cuts to the fitting results, including goodness of fit and a unimodal redshift probability distribution. We finally select 233 candidates whose photometric redshift probability distribution function peaks around $z\sim4$, have high stellar masses ($\log($M$^{*}$/M$_{\odot})\sim 11.7$ for a Salpeter IMF) and ages around 0.1 Gyr, i.e.~formation redshift around 5. These properties match those of the progenitors of the most massive galaxies in the local universe. This is an ideal sample for spectroscopic follow-up to select the fraction of galaxies which is truly at high redshift. These initial results and those at the survey completion, which we shall push to higher redshifts, will set unprecedented constraints on galaxy formation, evolution, and the re-ionisation epoch.
\end{abstract}

\begin{keywords}
galaxies: evolution -- galaxies: high-redshift 
\end{keywords}



\section{Introduction}
\label{sec:intro}

The formation and evolution of the most massive galaxies in the universe remains an open problem in cosmology and astrophysics. 

The fossil stellar population in the local universe shows that the most massive galaxies host the oldest stellar populations and that they should have formed around $z \sim 5$ \citep*{CoSoBa_1999, Thomas_etal2005, Thomas_etal2010, CoBa_2008}. Radial gradients in stellar populations which are flat in age and element abundance ratios \citep{Mehlert_etal2003, Pipino_etal2007, Goddard_etal2017, CALIFA} suggest that the early formation is a global property of the galaxy rather than of just its inner core. 

Within the hierarchical galaxy formation paradigm \citep{WhiteRees1978}, the most massive objects assemble last, at relatively low redshift ($z\sim0.5$) even if their building blocks may contain ancient stellar populations that later merge \citep{DeLucia_etal2006, RicciardelliANDFranceschini_2010}. This model implies a scarcity of massive galaxies at high redshift and their gradual build-up towards our epoch~\citep{Violeta_etal2009}. 
Hence, one key approach for constraining galaxy formation on a cosmological scale is to search for the progenitors of the most massive galaxies at an increasingly larger look-back time.

Massive galaxies are indeed being found spectroscopically at increasingly higher redshifts, in the range $z \sim 1.5-3.0$~\citep{Lonoce_etal2015, Yan_etal2004, Onodera_etal2012, Cimatti_etal2004, Kriek_etal2016, Straatman_etal2014, Conselice_etal2007, Whitaker_etal2013} and even $z \sim 3-4$ \citep{Mancini_etal2009, Santini_etal2009, Caputi_etal2012, Caputi_etal2015, Guo_2013, Ilbert_etal2013, Muzzin_etal2013, Stefanon_etal2013, Marsan_etal2017}. At such high redshifts, usually the word `massive' refers to stellar masses up to $\sim~10^{11}~M_{\odot}$. The highest value reported so far is a $1.7 \times 10^{11}~M_{\odot}$~galaxy at a spectroscopic redshift of 3.717 \citep{Glazebrook_etal2017}. The detection of such an impressively massive galaxy at such a high redshift is a challenge to galaxy formation models. We shall return to these works in relation to our project.
 
In order to bridge the fossil record with the formation event and trace galaxy evolution over cosmic time, many works have attacked the problem in a statistical sense, by probing number density evolution as a function of galaxy mass. Studies of the galaxy mass function over the past decade reached the uniform conclusion that the abundance of the most massive galaxies ($M/M_{\odot}>10^{11.5}$) hardly evolves since $z\sim1$ \citep*{CimattiDaddiRenzini2006, Wake_etal2006, Pozzetti_etal2010, Marchesini_etal2010, Muzzin_etal2013, Violeta_etal2009, Mortlock_etal2015}. One caveat to these studies has been that the observational database was drawn from small area, deep surveys, which carry the problem of cosmic variance. This is particularly severe at the highest mass where the galaxy mass function is steep and errors on photometric data are large. However, this has been recently solved by using the cosmological SDSS-III/BOSS survey (10,000 deg.~sq.). This survey has allowed the calculation of the galaxy mass function around $M^{*}\sim 10^{12}~M_{\odot}$~with unprecedented statistics \citep{Maraston_etal2013} thanks to the large area and the target selection centred on massive galaxies. The conclusion of this work is that the abundance of the most massive galaxies is constant in the redshift range 0.4-0.6, and larger than what is predicted by galaxy formation models. Bundy et al. (2017) using deeper photometry from the so-called Stripe82 region reached the same conclusion and showed that it is robust against the way stellar masses are calculated. Furthermore, \citet{Etherington_etal2017}, using data from the Dark Energy Survey (DES) survey (see below), showed that the evolution of the high-mass end of the galaxy mass function does not seem to depend on the environment. \cite{Thomas_etal2010} reached the same conclusion performing a different analysis, namely using the chemical information in the fossil stellar population properties of the most massive galaxies. They showed that ages, metallicities and chemical abundance ratios of the most massive galaxies do not depend on the environment, leading to the inference that their formation and evolution are mainly driven by internal processes, reinforcing downsizing as the evolution paradigm for these galaxies (Peng et al. 2010).

As just mentioned, probing the massive end of the galaxy population requires a wide survey area. The Dark Energy Survey (DES) is a galaxy survey aimed at probing cosmic acceleration. The survey is collecting galaxy photometric data in the southern sky at magnitude depths of $\sim $25.5, 25.0, 24.4, 23.9 and 22.0 in the $g, r, i, z$ and $Y$ bands respectively, for a very large portion of sky (5000 deg. sq.)  \citep{Rossetto_etl2011, DEScollaboration}. At completion ($\sim2019$), DES will have observed on the order of 300 million galaxies.
 
\citet{Davies_etal_2013} (hereafter D13) forecasted that - by virtue of its suitable combination of area and depth - DES is currently the best survey to detect the rare, massive ($\sim 10^{12}M_{\odot}$) galaxies at high redshift ($z \gtrsim 4$), should these exist (see Figure 2 in D13).
D13 used stellar population models spanning a wide range of properties (e.g.~age, metallicity, star formation history, stellar mass and dust reddening) to model galaxies as a function of DES magnitude, colours and redshift, identifying colour-colour selection maps for redshifts $z \sim 4, 5$, and $6$ (see Figures 7,8,9 in D13). 

The scope of the present paper is to apply the D13 theoretical selection maps to a sample of real DES data for the first time, specifically the latest available set of Year 3 (Y3) data, in order to find candidate massive high-$z$~galaxies. The data we use consist of observations completed on a limited sky region ($\sim 150$ sq.~deg.) probed since the Science Verification (SV) programme (thus at the DES nominal depth) in order to test the observational process and general workflow.

We first proceed by plotting the new DES data on the D13 colour-colour plots, and then calculate the photometric redshifts and physical properties of those sources falling into the predicted boxes for $z\sim4$ galaxies, after carefully removing artefacts of various kinds.
We focus on $z\sim4$, rather than on $z\sim5$ or $z\sim6$, as if these rare massive galaxies exist they are more likely to be observed at lower redshifts. We need to maximise our chances given the smaller area covered by the SV footprint compared to the one that will be available at DES completion.

Calculations of template-based photometric redshift are the common procedure at high redshift, but we shall also discuss the effect of using alternative redshifts from the DES neural network pipeline \citep{NeuralNet}. Instrumental to the robustness of our fitting is the availability of near-IR bands from the Vista Hemisphere Survey (VHS)~\citep{VHS, Banerji_etal2015}, which could extend the baseline photometry to a total of 8 filters. It is interesting to test whether the final results are consistent with the D13 predictions, which were based on the sole $g$, $r$, $i$, $z$, $Y$~DES magnitudes.

We then analyse in detail the fitting results for all candidates and conservatively retain only those obeying several quality criteria, including a unimodal probability distribution function in redshift, a good $\chi_{r}^{2}$ and other model parameters. 

At the end of the procedure we select 233 individual galaxies, of which some are selected with both reddening options. We find 109 using the SMC reddening law and 203 using the Calzetti law. For these, we examine their properties (including mass, age, SFR, SFH) and draw initial conclusions on galaxy evolution, using also galaxy formation simulations as a comparison. 

The paper is structured as follows. Section~\ref{sec:method} describes our method to find high-$z$ candidates, which includes an initial colour selection (Section~\ref{sec:colourselection}) and photometric template fittings to calculate photometric redshifts and physical properties (Section~\ref{sec:photometricfitting}). In Section~\ref{sec:Results} we describe in detail our best candidate selection, including their physical properties. In the same section we detail further tests we completed to check the reliability of our sample. In Section~\ref{sec:FittingStraatman14} we perform a comparison with the literature. Lastly, in Section~\ref{sec:Conclusions} we summarise our work and discuss our key findings as well as their relevance for future work and in the context of current research.

\section{Method}
\label{sec:method}

\subsection{Overview}

Our aim is to identify the most likely high redshift ($z \sim 4$) massive galaxy candidates within a dataset of $\sim$~4.9 million objects. Starting from the simulations performed by D13 we proceed using real DES data in this context for the first time. In this paper we focused on the $z \sim 4$ case in order to maximise the chance to find objects in the small area covered by the SV data.\footnote{An initial look at the $z \sim 5$ and $6$ selection maps (from D13) did not result in any high-$z$ candidates, but this does not exclude the possibility that they exist and it may be due simply to the fact that the SV data cover only $\sim 150$ deg.~sq. We shall pursue the higher redshift bins when we look at the entire database.} The process we followed in order to identify the best candidates is comprised of different steps. We summarise them here and discuss each of them in separate sections below.

\begin{figure*}
\centering
\includegraphics[width=0.49\textwidth]{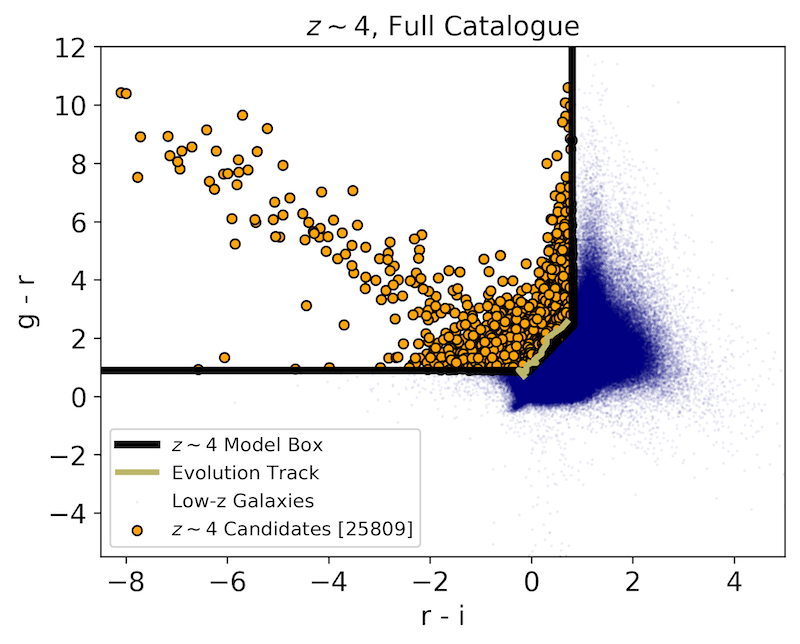}
\includegraphics[width=0.49\textwidth]{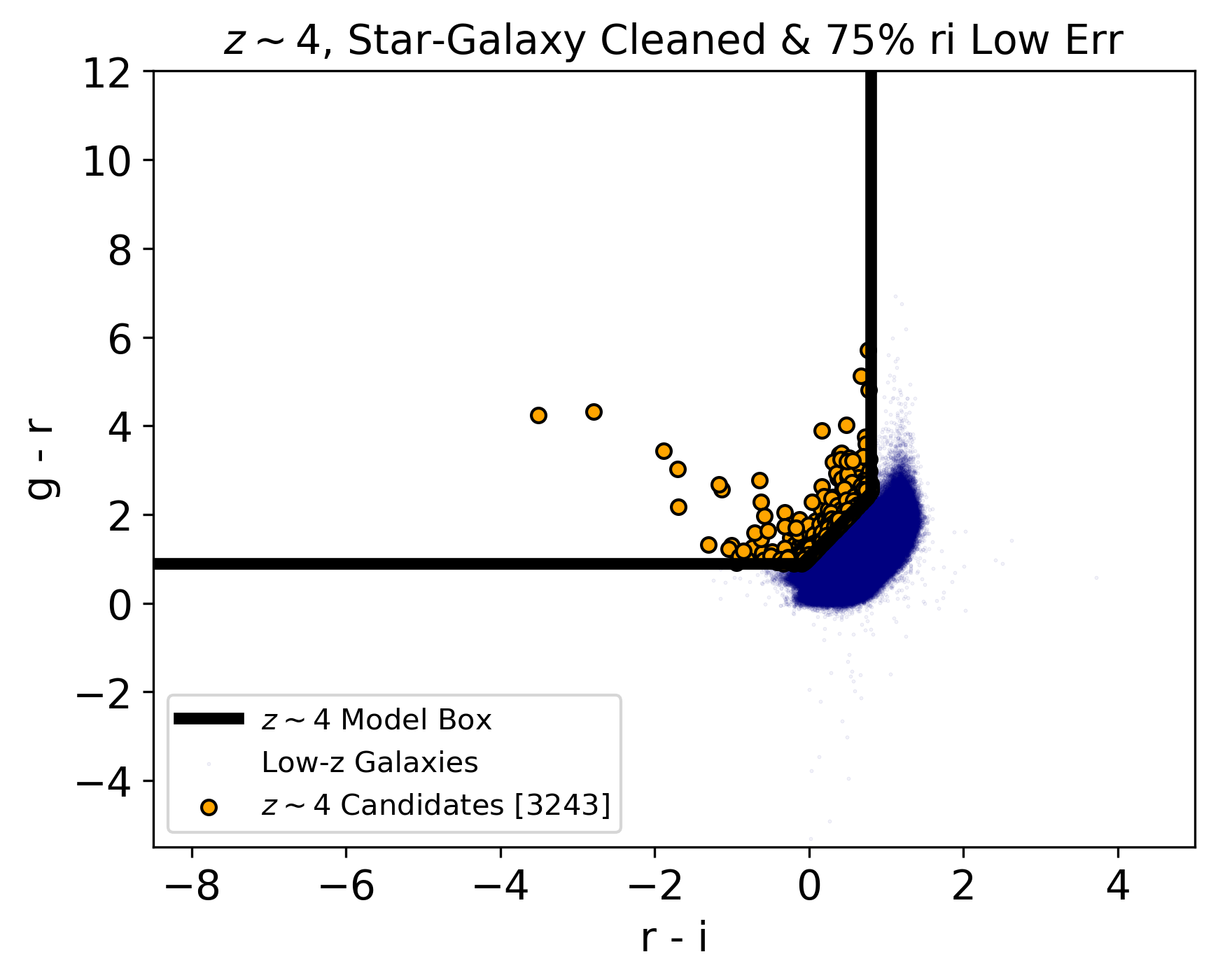}
\includegraphics[width=0.49\textwidth]{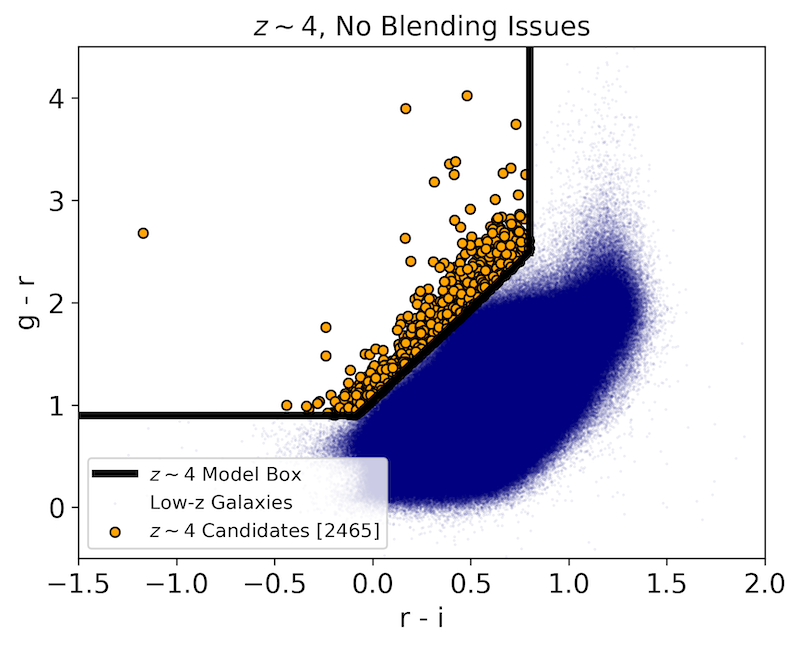}
\includegraphics[width=0.50\textwidth]{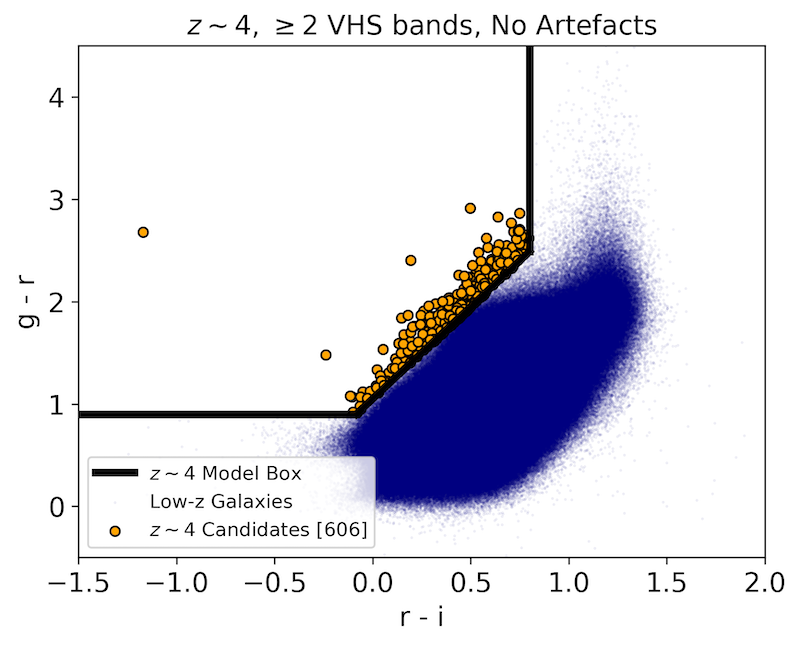}
\caption{Colour-colour selection maps for selecting massive $z\sim4$ candidates using DES photometry, as from D13. Black solid lines define the predicted selection boxes for $z\sim4$ (red g-r, blue r-i relative to the black line), with the khaki line in the top left panel showing the redshift evolution track for a model galaxy as illustration. Orange dots correspond to sources identified as high-$z$~candidate galaxies lying in the model selection area. Objects that fall outside the selection area are plotted as a density map in blue (darker is denser). The top-left panel shows the full Commodore-selected Y3 catalogue without any cut applied. The top-right panel shows the map after the cuts for star-galaxy separation and against large errors in $r$~and $i$. The bottom-left panel shows the final selection map after removal of objects potentially affected by blending and the bottom-right one contains only those sources with at least 2 VHS bands and no visual artefacts near the galaxies in the image cutouts. We use MAG\_DETMODEL AB photometry.}
\label{fig:CCmapsonly}
\end{figure*}

At first all Y3 data were placed on the D13 colour-colour plots (Figure~\ref{fig:CCmapsonly}, top left-hand panel). Then, the catalogue was scanned to remove sources that could potentially be stars and/or those affected by the largest errors. The sources passing these two criteria (non-stars and small errors) are plotted in Figure~\ref{fig:CCmapsonly}, top right-hand panel. From this point, we focus on the objects lying within the D13 selection box for massive, $z\sim4$~galaxies (solid black line in Figure~\ref{fig:CCmapsonly}). These candidates were further pruned of those potentially affected by blending (using the DES pipeline flags). The result of this further selection is shown in Figure~\ref{fig:CCmapsonly}, lower left-hand panel. Last, VHS archive data were matched to the selected sources in order to extend their DES photometry with $J$, $H$, and $K_{s}$ bands. Some sources lacked one or more VHS bands and, as explained later, this has been taken into account when estimating the reliability of our fits. The matching was done automatically by inserting the RA and Dec coordinates of each DES object in the VHS data access pages\footnote{\href{url}{http://horus.roe.ac.uk/vsa/index.html}} to find the closest VHS source. The maximum matching radius was kept to the web utility default value of 5 arcsec. However, we have visually checked every single image for each source to confirm that the DES and VHS photometry was matched correctly.

Those sources passing the condition of having data in at least two VHS bands populate the selection box in the lower right-hand panel of Figure~\ref{fig:CCmapsonly}. For these we calculated the photometric redshift and stellar population properties using a template fitting procedure. Photometric redshifts and their probability distribution functions were compared to those calculated by the DES pipeline working groups (more details in Section~\ref{sec:datacatandcuts}).

Lastly, a full-fledged analysis of the results for each candidate was performed in order to identify only those that are convincing $z\sim 4$ galaxies.

\subsection{Initial Selection}
\label{sec:colourselection}

\subsubsection{Catalogue of DES Data \& Colour Selection Cuts}
\label{sec:datacatandcuts}
We used photometric data in the $g$, $r$, $i$, $z$, and $Y$ bands from the DES Y3 Gold 2.0 release, which contains the latest, highest quality photometry for DES. Among the magnitude options, we use MAG\_DETMODEL photometry (in the AB system), as it refers to the same physical aperture hence it is optimal for template fitting. As described in~\citet{Melchior+2015}, magnitudes are measured by SExtractor in each filter using a model fit to the surface brightness of the source in each image. The detection image for each object was created by the DES pipeline by linearly combining the $r$, $i$, and $z$ images~\citep{Abbott+2018}.

From the Y3 catalogue we wanted to choose those objects that had been observed since the SV stage, meaning that their photometry matches the full nominal depth of the survey. In order to select them, we used the {\it COMMODORE} catalogue \citep[for details see][and Capozzi et al., {\it sub.}]{Etherington_etal2017} which refers to the SV data and provides, among other entries\footnote{ The COMMODORE catalogue also contains the DES SV photometry (which was not used for this work as the Y3 one is of higher quality) and galaxy physical properties obtained with the same models and procedure as here, but using the neural network redshift as fixed redshift for the given source.} the sky position (RA and Dec), the neural network redshift and a flag for performing star-galaxy separation.

The crossmatch between Y3 and the COMMODORE SV data resulted in $\sim 4.9$ million sources. The rest of the sources in Y3 (the vast majority, amounting to the impressive figure of 394 million objects) will need further observations for reaching the same depth levels.
 
\begin{figure*}
\centering
\includegraphics[width=0.49\textwidth]{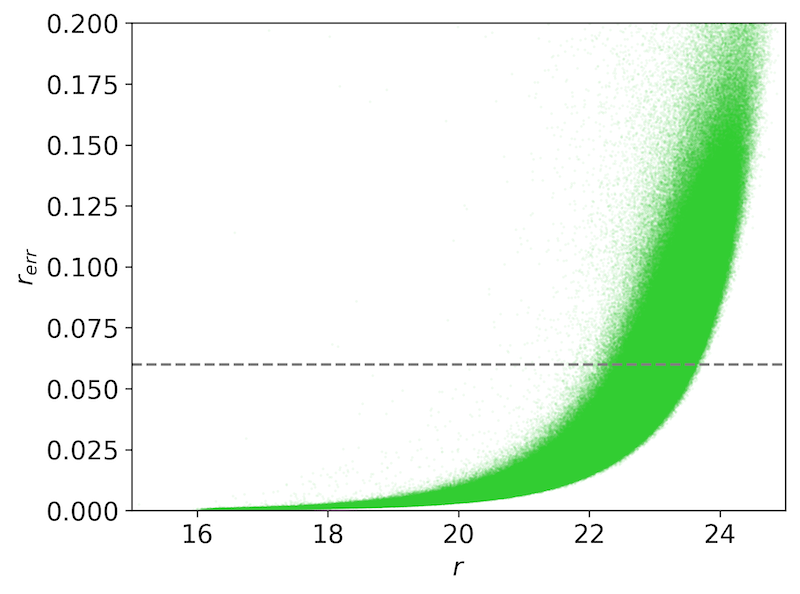}
\includegraphics[width=0.50\textwidth]{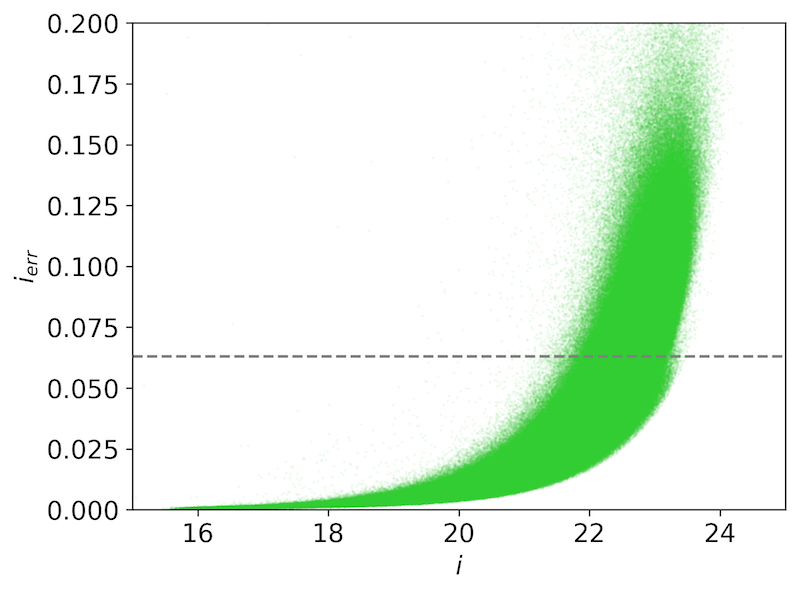}
\caption{Error against magnitude for $r$ and $i$ bands. The horizontal lines mark the error cuts we applied (i.e.~$r_{err} < 0.060$ and $i_{err} < 0.063$). These identify the effective magnitude limits of our analysis as $\sim23.7$ and $\sim23.5$ for the $r$ and $i$ bands, respectively.}
\label{fig:errVSmag}
\end{figure*}

First, we considered the star-galaxy separation parameter included in the catalogue and validated by the {\it COMMODORE} team \citep[and detailed in][]{Kim_etal2015}. This method uses a supervised machine learning technique to provide the probability of a given source of being either a galaxy or a star. Note that this procedure may lead to the exclusion of compact galaxies, and many high-z massive galaxies are compact~\citep{Straatman_etal2014}, but it makes it much more likely that we avoid star or pure AGN contamination (this last point is discussed in more detail in Section~\ref{sec:AGNcontamination}). Quantitatively, we kept those sources with $\geq 99.977\%$ probability of being a galaxy\footnote{For DES users: the entry we used is called TPZ$_{SG\_CLASS}$ and we excluded sources with TPZ$_{SG\_CLASS}$ > 0.00023}. This first cleaning left us $73\%$ of the original sample (i.e.~$\sim 3.7$ million sources). We further performed a cut in photometric errors. We examined the error distributions in $r$ and $i$ separately and we conservatively decided to remove the tail of largest errors (i.e.~$r_{err} < 0.060$ mag and $i_{err} < 0.063$ mag), which means removing $\sim 25\%$~of objects in each band. 
These cuts effectively identify the magnitude limits of our work in the $r$ and $i$ bands, which are $\sim23.7$ and $\sim23.5$, respectively. This can be appreciated by looking at Figure~\ref{fig:errVSmag}, where we plot error against magnitude for the $r$ and $i$ bands (left-hand and right-hand panel, respectively) of our full catalogue after star-galaxy separation. The horizontal lines indicate our error cuts.

We then matched objects satisfying the error cuts in both bands.
Note that we considered these bands as they were the ones used for the theoretical colour selection, see Section~\ref{sec:CCmap}.  We left the $g$~band free of constraints as it is the drop-out band at high redshifts (see Figure~\ref{fig:9bestcandidates} and visit this \href{http://icg.port.ac.uk/~guarniep/}{link}\footnote{Un-hidden hyperlink: \url{http://icg.port.ac.uk/~guarniep/}} for several examples).

After the star-galaxy separation and the lowest error cut, we were left with $\sim 2.7$ million candidate galaxies. 
Furthermore, we removed all galaxies whose photometry was marked by the DES pipeline as being potentially affected by blending.
Lastly, we extended the photometric baseline of the DES data with near-IR VHS data, and we decided to consider only those sources for which we could add at least two additional photometric data to the DES data (see Section~\ref{sec:VHSdataExtension} for more details).

It should be noted that the Y3 Gold 2.0 catalogue also contains a photometric redshift estimate for each source as calculated by the DES pipeline. This redshift is calculated using the so-called Bayesian Photometric Redshift (BPZ) code~\citep{Hoyle_etal2017}. The prior used in BPZ strongly disfavours high-z solutions for bright galaxies. This does not mean, however, that higher redshift objects do not exist and this is exactly what we aimed to find out. Later in the paper we shall also perform a comparison of the physical properties we would have obtained for galaxies had we used the $z_{BPZ}$.

\subsubsection{Theoretical Colour - Colour Selection Maps}
\label{sec:CCmap}

As recalled in the Introduction, D13 investigated colour combinations of DES bands for stellar population models \citep{Maraston_2005, Maraston_etal2006} with various parameters such as age, star formation history, dust, and stellar mass, as placed at various redshifts, in order to select those corresponding to the mass-redshift combinations of interest. The resulting selection boxes display regions within the colour-colour diagrams where high-$z$ objects should be found. 

Hence, our first step was to plot the DES galaxy candidates on the D13 $g - r$ vs $r - i$ plots in order to single out those entering the $z \gtrsim 4$ box. Note that by using this mapping first it is possible to substantially reduce the number of candidates for fitting, which is helpful because running a template fitting code for photometric redshift and physical properties for millions of sources is very time consuming. Even more importantly, using such a colour selection box is crucial to maximise the likelihood of the candidates to truly be at high redshift. 

The colour-colour $g-r$ vs $r-i$~diagram using Y3 photometry for the sources in the {\it COMMODORE} catalogue is shown in the top-left panel of Figure~\ref{fig:CCmapsonly}, where the selection box for $z\sim4$~objects is highlighted with a black solid line. In the same panel, the khaki line depicts the redshift evolution track for a model galaxy as taken from D13. The vast majority of the sources are consistent with being low-redshift objects (plotted as small, blue points; the darker, the denser). The high-$z$ candidates falling in the selection area are plotted in orange. They total 25809 sources (top-left panel). After application of the cuts described in Section~\ref{sec:datacatandcuts} - namely star-galaxy separation and removal of sources with highest error in bands $r$ and $i$ (for a total of  25\% in each of the two bands) - we are left with 3243 objects as potential high-$z$ candidates (top-right panel). We then removed candidates with blending issues, which left us with 2465 galaxies (bottom-left panel).
Further visual inspection of the cutout images allowed us to remove any source that showed artefact traits, such as satellite trails and black-out areas, near the selected source. This does not mean that their photometry is necessarily compromised (as these were not flagged by the pipeline), but we decided to exclude them as we could not verify their photometric quality. This is done along with further considerations on the number of near-IR bands we could match to our galaxies, as described in the next section.

\subsubsection{Extending the Photometry to the Near-IR}
\label{sec:VHSdataExtension}
As is well known, the accuracy of spectro-photometric model fitting depends on the number of available data points and especially on the baseline in wavelength they cover \citep[e.g.][]{Pforr_etal2012, Pforr_etal2013, Banerji_etal2008, Banerji_etal2015}. In order to strengthen the reliability of our photometric fitting procedure, as mentioned in the previous sections, we looked for additional bands for the sources within the colour-colour map. We were able to cross-match the DES optical data with the VHS survey \citep{VHS}, thereby extending our photometric catalogue to the near-infrared bands $J$, $H$, and $K_{s}$\footnote{The VHS photometry, in the Vega magnitude system, was converted to the AB system to match the DES one according to the following relations: $J_{AB} = J_{Vega} + 0.916$, $H_{AB} = H_{Vega} + 1.366$, and $K_{sAB} = K_{sVega} + 1.827$. Source: \href{url}{http://casu.ast.cam.ac.uk/surveys-projects/vista/technical/filter-set}}. We used the {\it Petrosian magnitudes} from the VHS Data Release 5, which rely on the Petrosian radii of galaxies to determine the photometric aperture. This allows to recover the flux also for extended sources, making them ideal to work in combination with the DES DET\_MODEL magnitudes discussed earlier. We do not expect the atmospheric seeing to have an effect on our template fitting results since we use extended magnitude types \citep[DES DET\_MODEL and VHS {\it Petrosian};][]{Cross+2012} and the median seeing among the eight DES+VHS bands was shown to be similar \citep{Banerji_etal2015}. Additionally, the point spread function (PSF) full width at half maximum (FWHM) of the DES and VHS camera and telescope configurations are 0.49 arcsec\footnote{\href{url}{https://www.noao.edu/meetings/decam/media/ \\ DECam\_Technical\_specifications.pdf}} and 0.51 arcsec\footnote{\href{url}{https://www.eso.org/sci/facilities/paranal/telescopes/\\vista.html}}, respectively. Therefore we consider DES and VHS photometry to be compatible without further manipulation.

We noted that the majority of DES sources does not have all three VHS bands available. There could be several reasons for this. First of all, VHS has not imaged in the $H$ band all regions overlapping with DES \citep{Reed_etal2017}. Additionally, VHS is shallower than DES and the optimal depth depends on the nature of the candidates. For example, young star forming objects could be faint in the rest-frame optical (sampled by the VHS at high-$z$) simply because they are dominated by massive hot stars. 

In order to retain only those candidates for which the model fitting would be better constrained (see Section~\ref{sec:visualisationAndAnalysis}), we decided to focus on objects having observation in at least 2 VHS bands, which means to fit a minimum of 7 photometric bands.

This further selection criterium, along with the removal of artefacts as described in the previous section, led us to identify 606 galaxies (bottom-right panel of Figure~\ref{fig:CCmapsonly}). These correspond to $\sim 0.01$\% of the whole 5-million sources DES catalogue. Template fittings and analysis have been performed on these 606 objects only, as described in the next section.

\subsection{Determining Redshift and Physical Properties}
\label{sec:photometricfitting}

\subsubsection{Template Fitting Procedure}

In order to confirm candidate $z \sim 4$ massive galaxies, the redshift and physical properties of the sources selected up to this stage were calculated. This was done using the photometric redshift code HyperZ \citep{HyperZ} combined with ancillary scripts for the calculation of the stellar mass, as in our previous works \citep{Daddi_etal2005, Maraston_etal2006}. HyperZ compares model spectral energy distributions of stellar populations (which are referred to as templates) to observed photometric data, and selects the best models using a $\chi^2$ minimisation method.
HyperZ outputs the photometric redshift, the best-fitting template, and a reduced $\chi^{2}$ value ($\chi_{r}^{2}$) for the best-fitting template, calculated as $\chi^{2}/(N-1)$, where $N$ is the number of filters.

Additionally, a series of input parameters can be modified in order to more finely control the way HyperZ operates. These are: age limits, magnitude limits, redshift range and binning, reddening law, and template setup. We discuss them later.
 We have explored different combinations for fitting and model setup, which we now describe. We should say in advance that the final results are robust against these parameter variations.

Each galaxy spectral energy distribution model (the template) is calculated assuming a star formation history (SFH, detailing the mode of star formation, e.g. single burst, exponentially-declining star formation, etc.), an age (t parameter, which runs from the start of star formation at $t=0$ through the galaxy evolution, at logarithmic time steps from 1 Myr to the age of the Universe at the given redshift and assumed cosmology), a chemical composition, and a reddening by dust. Each model is redshifted at various values of redshift and a $\chi^2$ is calculated for each redshifted model. 

The redshift range we explored here varied from 0 to 6 in steps of 0.05, consistently for each fitting run. A variegated selection of 32 sets of model spectral energy distributions based on the \citet{Maraston_2005} evolutionary population synthesis models (M05) was used at each run, spanning a wide variety of star formation histories~\citep[SFHs; as in][]{Maraston_etal2006}. These include single-bursts simple stellar populations, $\tau$ (exponentially declining), truncated (constant until an instantaneous decline to zero, to simulate rapid quenching), and constant SFHs, each of them calculated for a grid of 221 ages and four metallicities ranging between 1/5 to twice solar. All the runs were repeated for two reddening laws: the so-called `SMC' law \citep{SMC_prevot, SMC_bouchet} and the well-known `Calzetti' law \citep{CalzettiLaw}. For each the extinction parameter $A_{V}$~was allowed to vary between 0 and 3, in steps of 0.5. These two reddening laws were selected because they are maximally different among the options offered by HyperZ and they are appropriate for different classes of high-$z$~galaxies. \citet{Maraston_etal2006}, by exploring all options in HyperZ, concluded that these two are those identifying the best fits in most cases of $z\sim2$~galaxies, and that while Calzetti's law is calibrated with starbursts, the SMC seems to be more appropriate for passive galaxies \citep[as also concluded by][]{KriekConroy2013}. We assumed a Salpeter (1955) IMF for all model options. Furthermore, we used an age cut to retain only solutions older than 0.1 Gyr, which is commonly used in order to avoid age-dust degeneracy pushing the fits towards low ages; we have tested that our results, as far our best candidates are concerned, do not depend on this choice (see Section \ref{sec:AgeLimPhotFits}). 
Lastly, we ran the code with loose absolute magnitude limits (i.e.~between -12 and -30) for all sources as we found that thanks to using at least 7 photometric bands (DES + VHS) we would instead obtain stable results and also avoid the risk of over-fitting.

Relevant to this work, we have expanded the public version of HyperZ by adding the calculation of the redshift probability distribution function (PDF; see also Pforr et al., 2018 {\it in press}). 
HyperZ provides the probabilities associated with the $\chi^{2}$ of all the fitted models. We calculate the PDF for the photometric redshift by summing up all the probabilities for each redshift step (0 to 6, in this case) and then normalising by the number of models (32). The sum of the discrete probabilities of each model over the redshift range equals to one. As we use the same redshift bin (0.05) throughout the explored redshift range (0 to 6) for all fitting runs and the model templates contain the same number of ages (221) we should be minimising the risk of artificially favouring particular solutions. \footnote{We note, however, that as galaxy colours may vary non-linearly, the colour-space may not be evenly sampled in spite of an homogeneous age sampling in all templates.}
The photometric redshift PDF is critical in order to distinguish high-probability high-redshift from lower-redshift distributions. 
We shall use PDFs, among other indicators, in order to determine the robustness of our final candidates. 

\subsubsection{Visualisation and Analysis of the Results}
\label{sec:visualisationAndAnalysis}
\begin{figure*}
\centering
\includegraphics[width=0.49\textwidth]{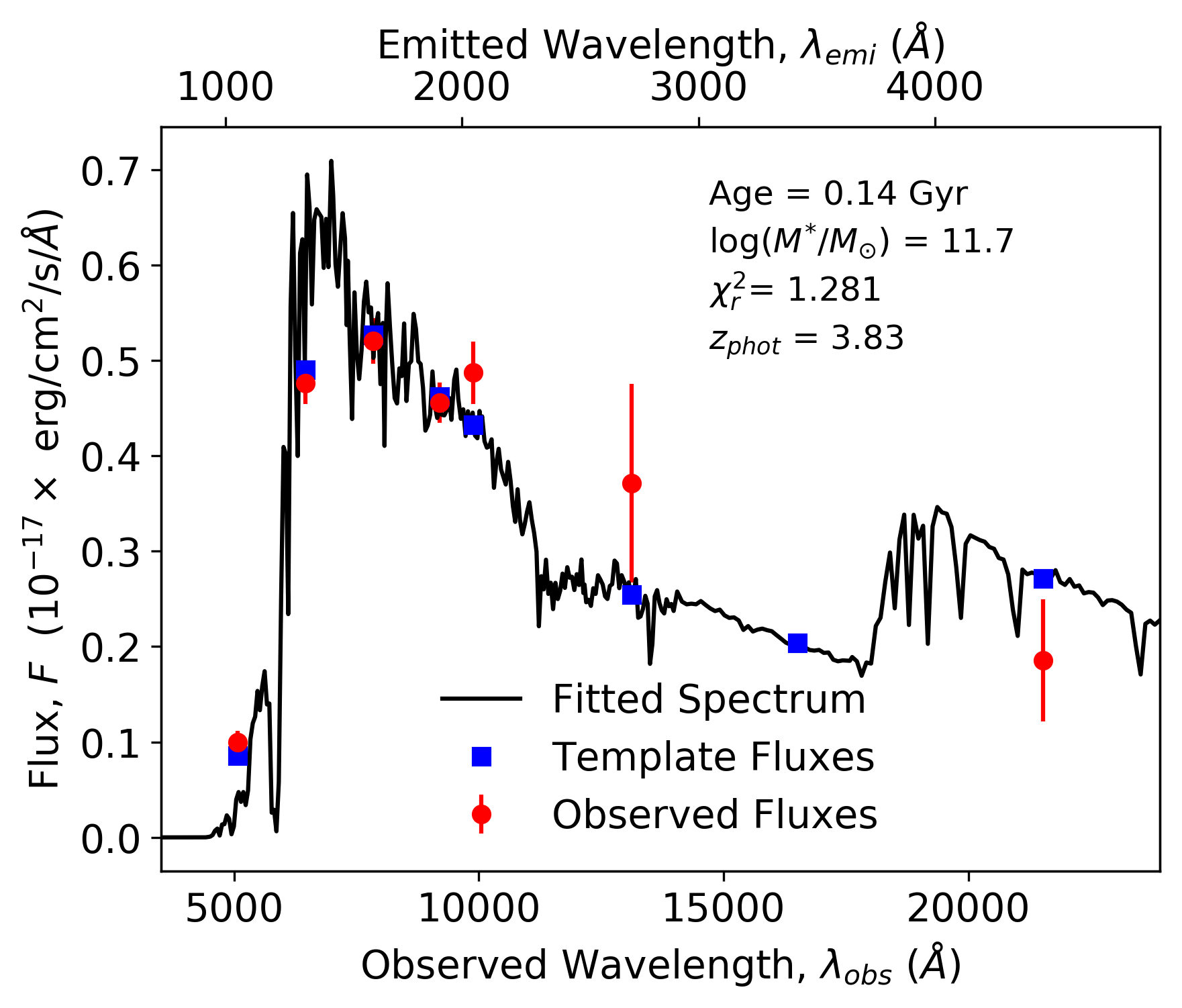}
\includegraphics[width=0.50\textwidth]{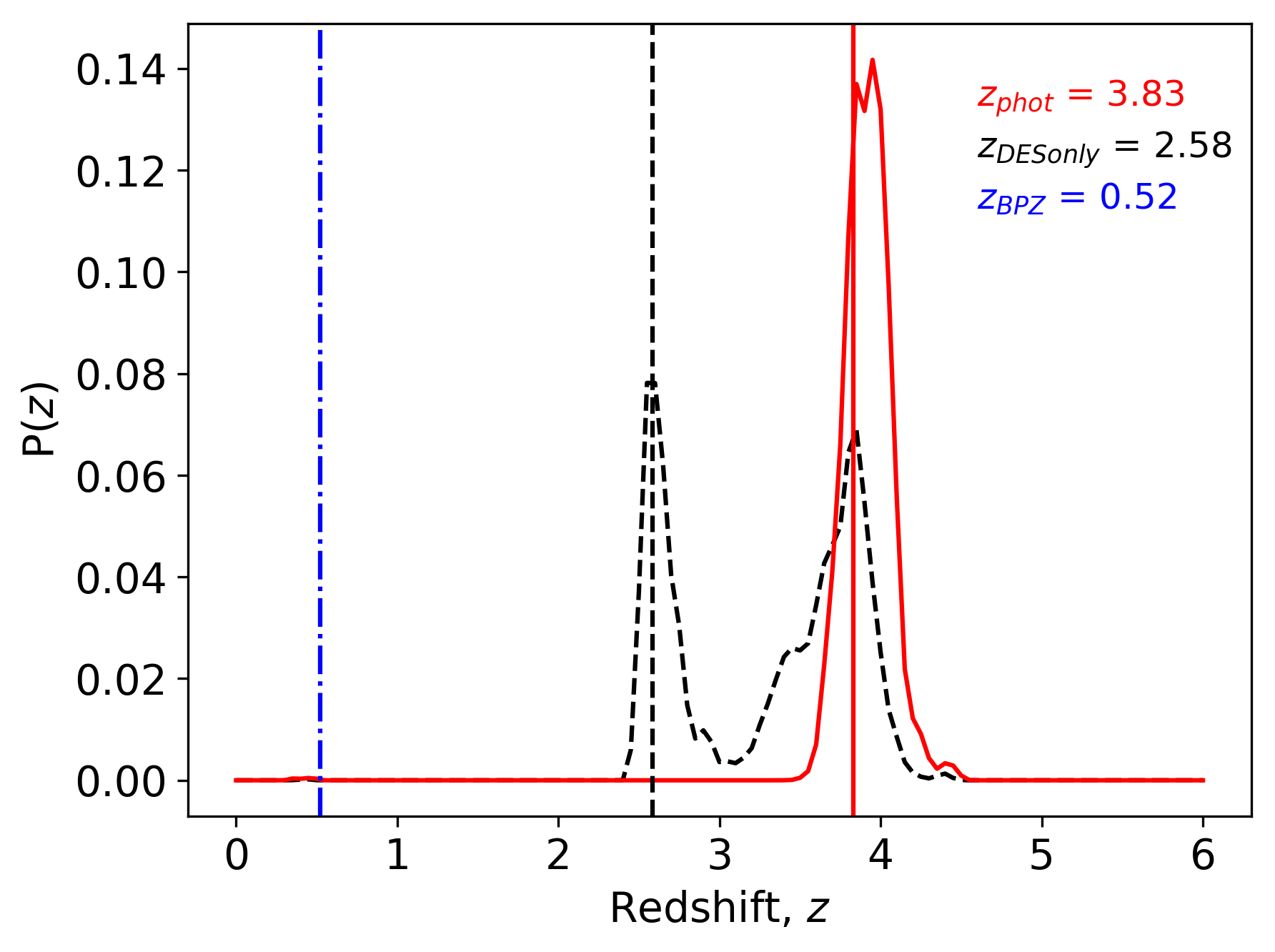}
\includegraphics[width=0.49\textwidth]{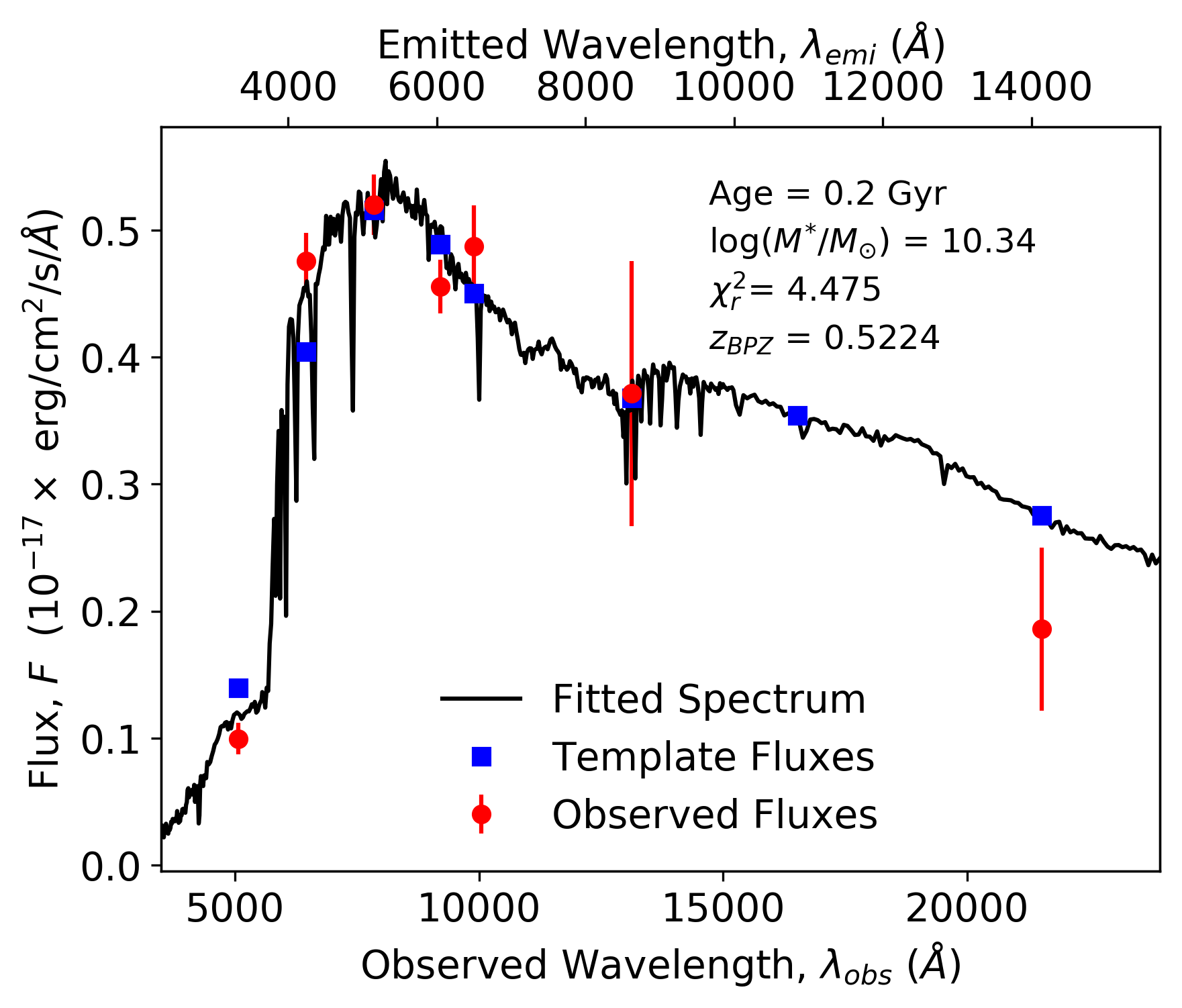}
\includegraphics[width=0.49\textwidth]{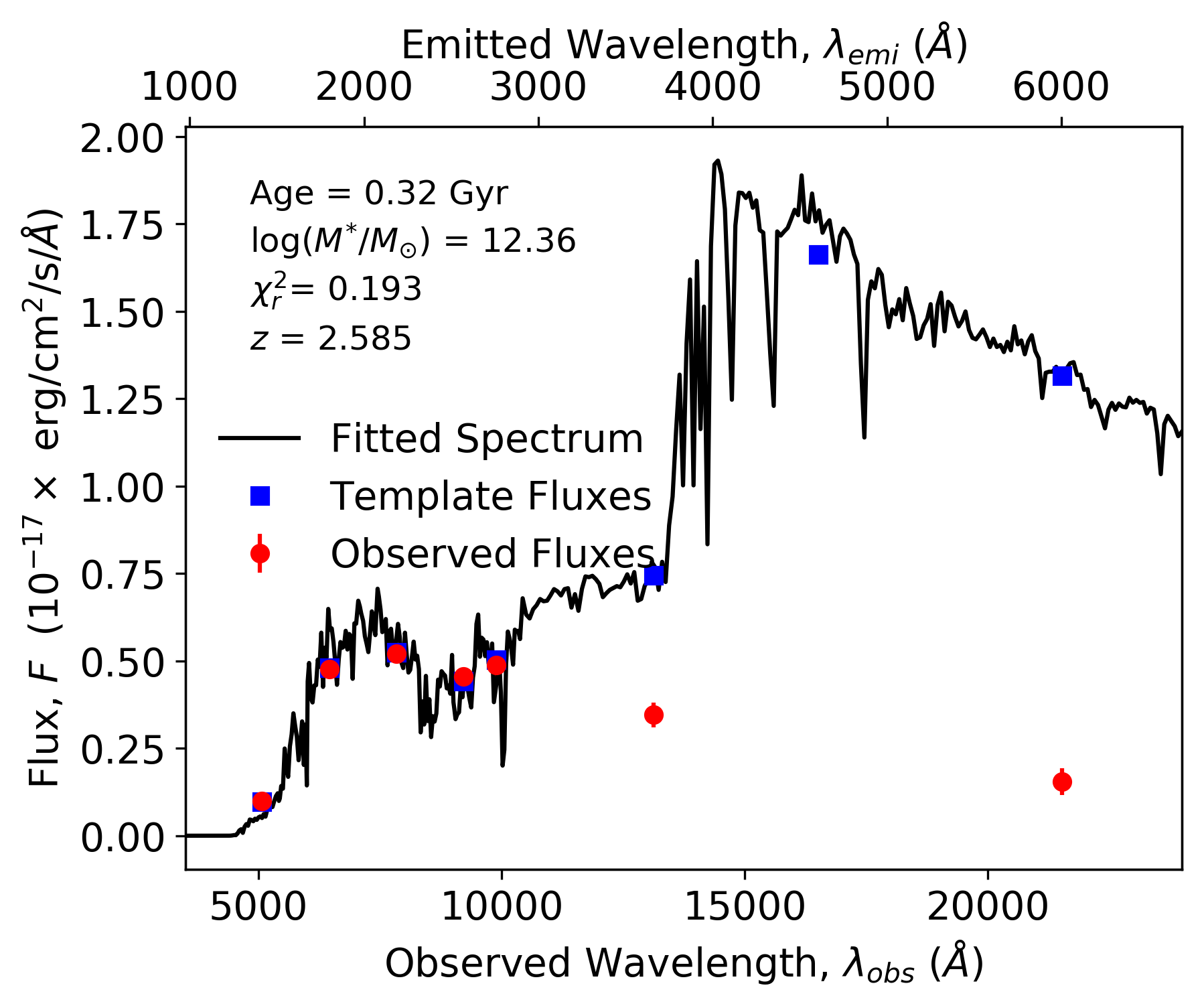}
\caption{A comprehensive display of template fitting results (for object with ID 494790027 as an example). \textbf{Top-left:} Template fitting showing the best fitting model spectrum (solid black line) with overlaid data (red circles) with error bars, and the template fluxes (blue squares) to which the data have been fit. The derived properties age (Gyr) and stellar mass (in log and solar units) are labelled, along with the reduced $\chi^{2}$~and redshift of the best fitting model. \textbf{Top-right:} The PDF of the template-fitting redshift is shown as a solid red line (the best result is marked by a vertical line), the same for the case of only using DES bands for the fitting is shown in dashed black, and the DES BPZ value in dotted-dashed blue. \textbf{Bottom-left:} Template fitting having fixed the redshift to the value of $z$ fixed according to the DES BPZ. This results in a considerably worse fit than when the redshift is allowed to vary (top-left), suggesting that the most likely solution is the high-redshift one. \textbf{Bottom-right:} Fitting as in top-left, but only for the DES bands while the VHS bands are plotted for reference; the redshift result confirms a high redshift  (i.e.~3.28 instead of 3.80), but it can be seen that, since the overlaid VHS band fluxes are far from the model template fluxes, the result does not represent the correct template.}
\label{fig:Fitting_PDF_Examples}
\end{figure*}

For each candidate, we then analysed simultaneously both the best-fit result (the photometric data along with the best-fitting model) as well as the photometric redshift PDF. Examples are shown in the top two panels of Figure~\ref{fig:Fitting_PDF_Examples}.  

In the model fitting plot (left-hand side) the photometric data (red) are matched to the template fluxes (blue), corresponding to the best fitted model (solid black line) when adjusted for the response function of the telescope camera in each band. The physical parameters of the best fitting model, the age (in Gyr), the stellar mass (in $M_{\odot}$) and the $\chi_{r}^{2}$ are labelled. Axis labels include observed and emitted wavelengths. 

The probability distribution of photometric redshifts is shown in the upper right-hand panel (in red for the curve relative to the fit in the top-left panel), where a vertical line is drawn at the value corresponding to the best-fitting model.

In some cases, the most probable solution (i.e. the peak of the curve) may not correspond to the redshift of the best fitted model. This happens when several less probable solutions (model fits with worse $\chi_{r}^{2}$) have similar redshift value and therefore sum up to show a higher peak in the PDF plots. It can happen that the less likely, but numerous solutions sum up such as to create another maximum which competes with the one of the best-fitted solution. As it will be described later, we only accept as massive high-z galaxies those for which the best fitted model's redshift matches the most probable solution (i.e.~peak of the curve).

Additionally, we also indicate the value of the $z_{BPZ}$ as calculated by the DES pipeline (dotted-dashed vertical blue line). It is important to stress again that the DES $z_{BPZ}$ was trained with sources up to $z \sim 1.3$ therefore this procedure could not produce any solution at higher redshift.

For each candidate, we further tested the effect of assuming the DES BPZ redshift as its true redshift, and explored the resulting quality of fit and the derived physical parameters. To this aim we run a different version of HyperZ (named HyperZ-spec), which fixes the redshift to a known value, and repeated the fitting runs keeping the stellar population model setup as before. The result for the same high-$z$~candidate can be seen in Figure~\ref{fig:Fitting_PDF_Examples} (bottom-left). 
In this case we must compare absolute and not reduced $\chi^{2}$ values as the $\chi_{r}^{2}$ figures are calculated in the same way in both codes (as $\chi^{2}/(N-1)$), but in HyperZ-spec the redshift is fixed.

We find that the absolute $\chi^2$~of the fit at the fixed DES BPZ redshift is considerably larger than the one obtained leaving the redshift free (26.850 vs 7.686).
The $\Delta \chi^2$ in this example ($\sim 19$) implies that the low-redshift solution is inconsistent with the best-fit high-redshift solution at the $>2\sigma$ confidence level. The best fit corresponding to the DES BPZ redshift is also visually less convincing. The physical properties (lower mass and older age) are consistent with a low-redshift solution.

Lastly, we tested the effect of fitting only DES data vs fitting DES+VHS data.  An example is shown in Figure~\ref{fig:Fitting_PDF_Examples} (bottom-right).
The fit obtained without VHS bands (hence with the 5 DES bands only) is different in terms of physical parameters (compared to the top-left panel fit, in which VHS bands were fit as well) and the very low $\chi_{r}^{2}$ value (0.045) indicates that DES-only fits are prone to over-fitting due to the low number of bands. 
When the same galaxy is fitted with the additional VHS bands (top-left panel) we verify, as we saw earlier, that the `only-DES' fitting did not correspond to the real solution, showing the importance of extending the photometric data to, in this case, the near-IR (see Section~\ref{sec:DESonlyVSDESVHS} for a quantitative evaluation on the effects of using the VHS bands in model fitting). The PDF of this type of fit is plotted in black on the top-right panel plot.

\begin{figure*}
\centering
\includegraphics[width=0.49\textwidth]{Fitting_5_RL4.png}
\includegraphics[width=0.50\textwidth]{PDF_5_RL4.png}
\includegraphics[width=0.49\textwidth]{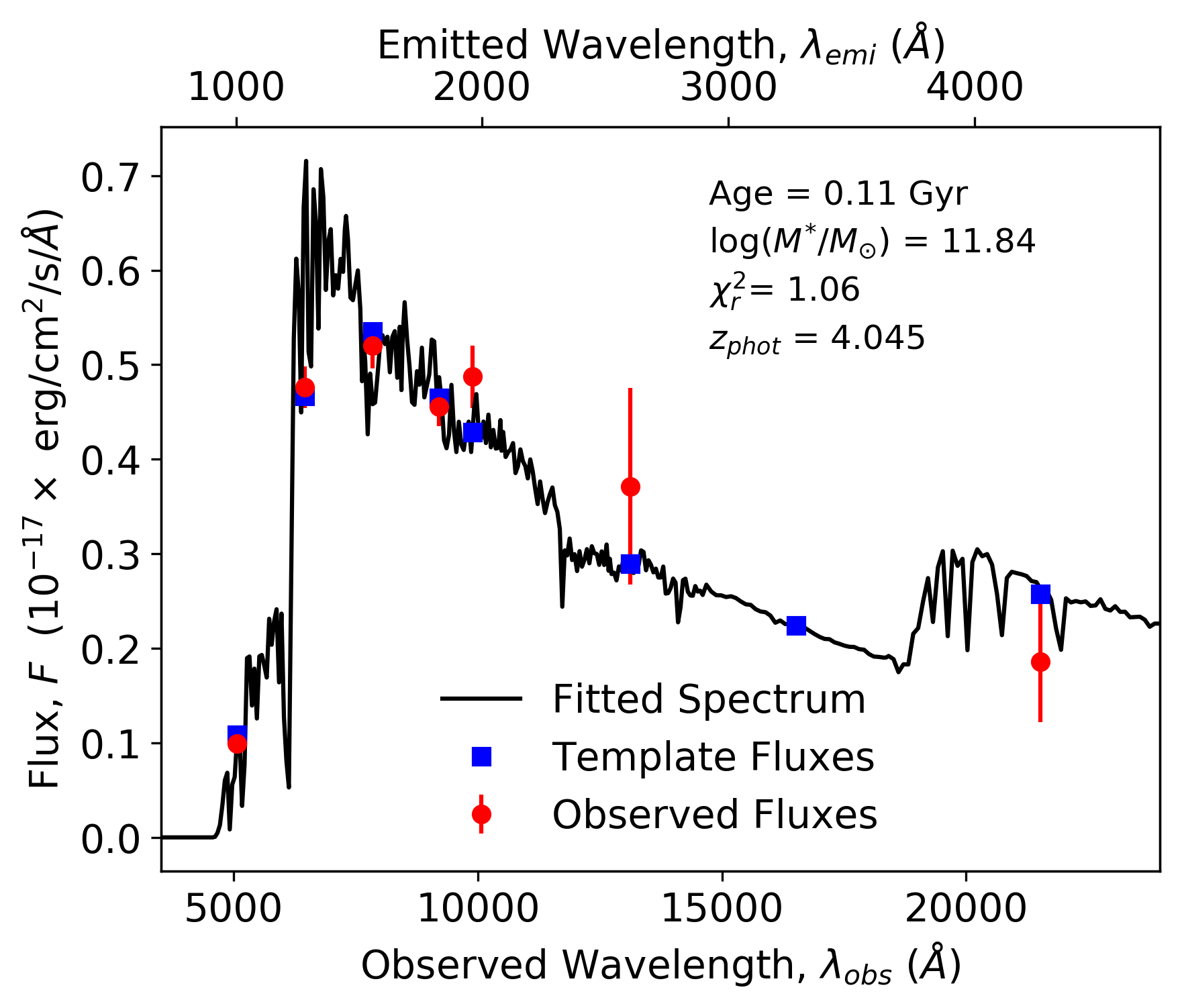}
\includegraphics[width=0.50\textwidth]{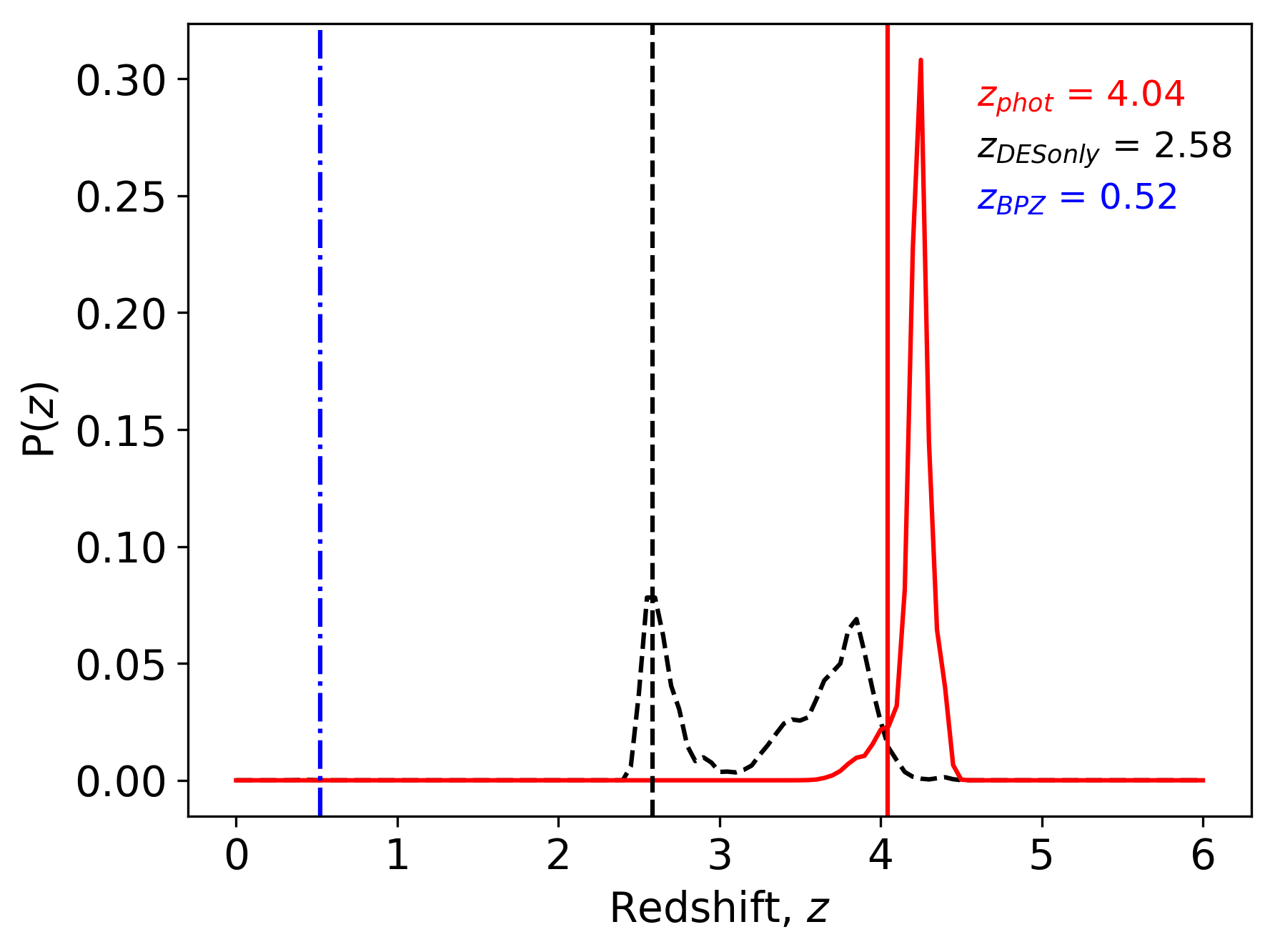}
\caption{Legend as Figure~\ref{fig:Fitting_PDF_Examples}. Fittings and PDFs for a high-$z$ candidate for the various reddening laws (one in each row respectively), from top to bottom: SMC law, and Calzetti law. Compared to the SMC result for this object, the solution obtained with the Calzetti law has got a negligibly higher redshift and lower $\chi_{r}^{2}$, and displays minor changes in mass and age. ID of object shown here: 494790027.}
\label{fig:3ReddeningLaws}
\end{figure*}

In Figure~\ref{fig:3ReddeningLaws}, as an illustration, we show the effect of the assumed reddening prescription on the fitting results. 
For this specific galaxy, the fit performed with the SMC-law releases a slightly lower redshift and $\chi_{r}^{2}$ value, along with minor changes in terms of mass and age, compared to the results obtained using the Calzetti-law. A similar trend is seen in the correspondent PDFs. The PDFs referred to the fits with only DES bands are somewhat broader, but still peak at high redshift.

\section{Results}
\label{sec:Results}

Starting from $\sim5$ million sources, the exclusion of stars, blended sources, galaxies with uncertain data (large errors) and objects lacking VHS photometry was followed by a filtering with the theoretical colour-colour selection maps and a final visual inspection, which left us with 606 candidate massive high-$z$~galaxies. These candidates were submitted to stellar population model fitting.

Since we focus on the high-redshift population in this paper, we proceed with a series of selection cuts (described in the next section) in order to identify the best high-$z$, massive galaxy candidates. In future studies, we shall explore the wider zoology of all sources falling into the selection box.

\subsection{Selection of the Best Candidates}
\label{sec:SelBestCandidates}
From now on, we consider the results from HyperZ runs for all 606 sources as obtained with the two reddening laws mentioned earlier. This results in two fitting results for each object. We then proceed to select the most secure sub-sample of galaxies at $z \geq 3$. To this aim, we performed three additional cuts, namely: i) we excluded those galaxies with unphysical stellar mass ($\log_{10}(M^{*}/M_{\odot}) > 12.5$); ii) we excluded objects whose fits have $\chi_{r}^{2} >3$ \citep[this value was chosen considering the number of fitted bands and values typical of high-$z$~galaxies, e.g.][]{Maraston_etal2006}; iii) we excluded objects for which the probability of a $z \geq 3$ solution is less than 95\% (corresponding to a $2\sigma$ confidence level), determined by looking at the PDFs of each object.

The number of individual galaxies passing all selection cuts is 233. With respect to the adopted reddening, 109 fitting results pass our selection criteria when using the SMC law and 203 with the Calzetti law case. This means that some galaxies satisfy the selection criteria for both reddening laws and therefore have been selected twice.
The sample of 233 galaxies constitutes our {\it best candidate} pool. Their fitting properties are given in full in Appendix~\ref{app:FullGoldenSampleResults}.

\begin{figure*}
	\centering
	\includegraphics[width=0.49\textwidth]{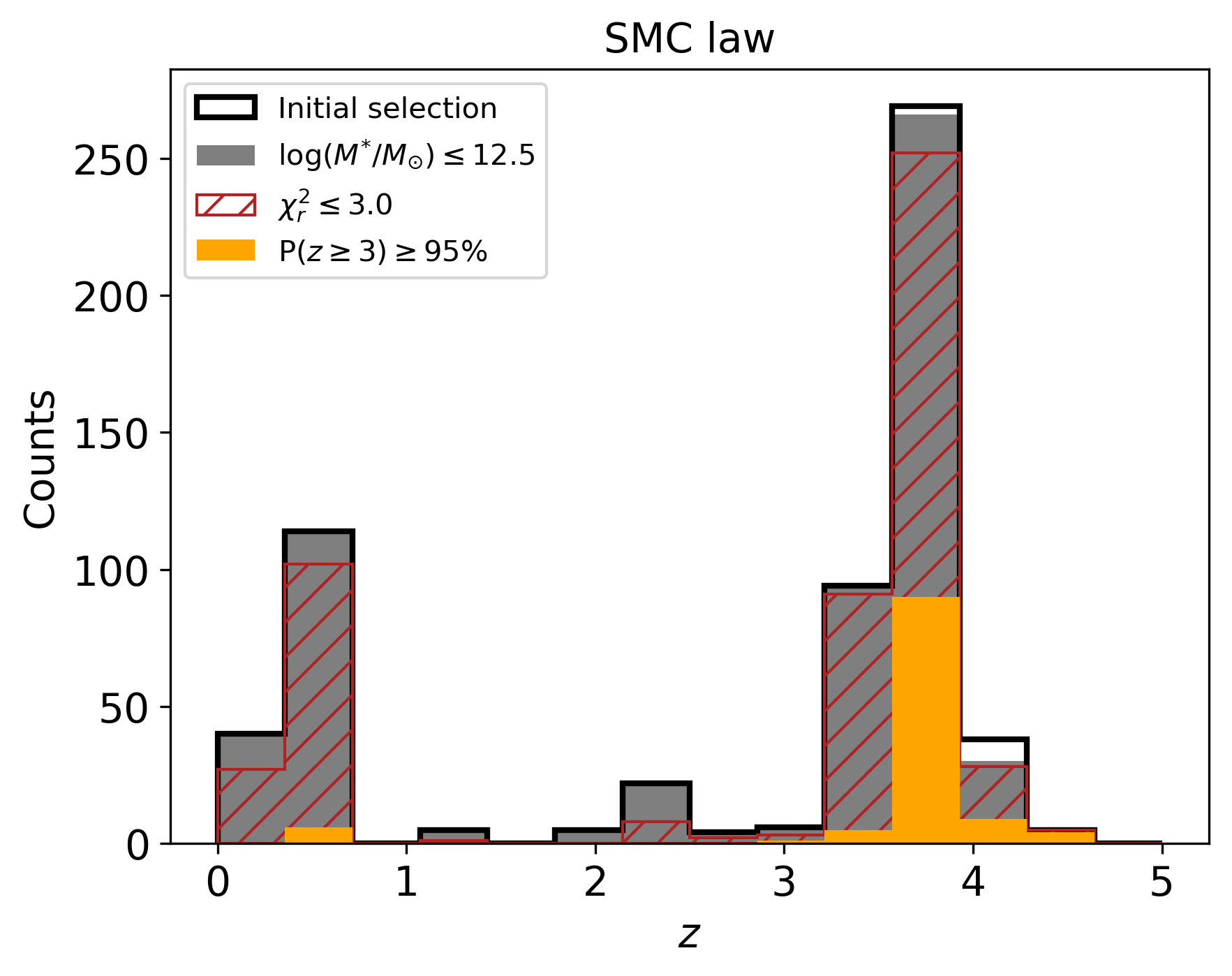}
	\includegraphics[width=0.49\textwidth]{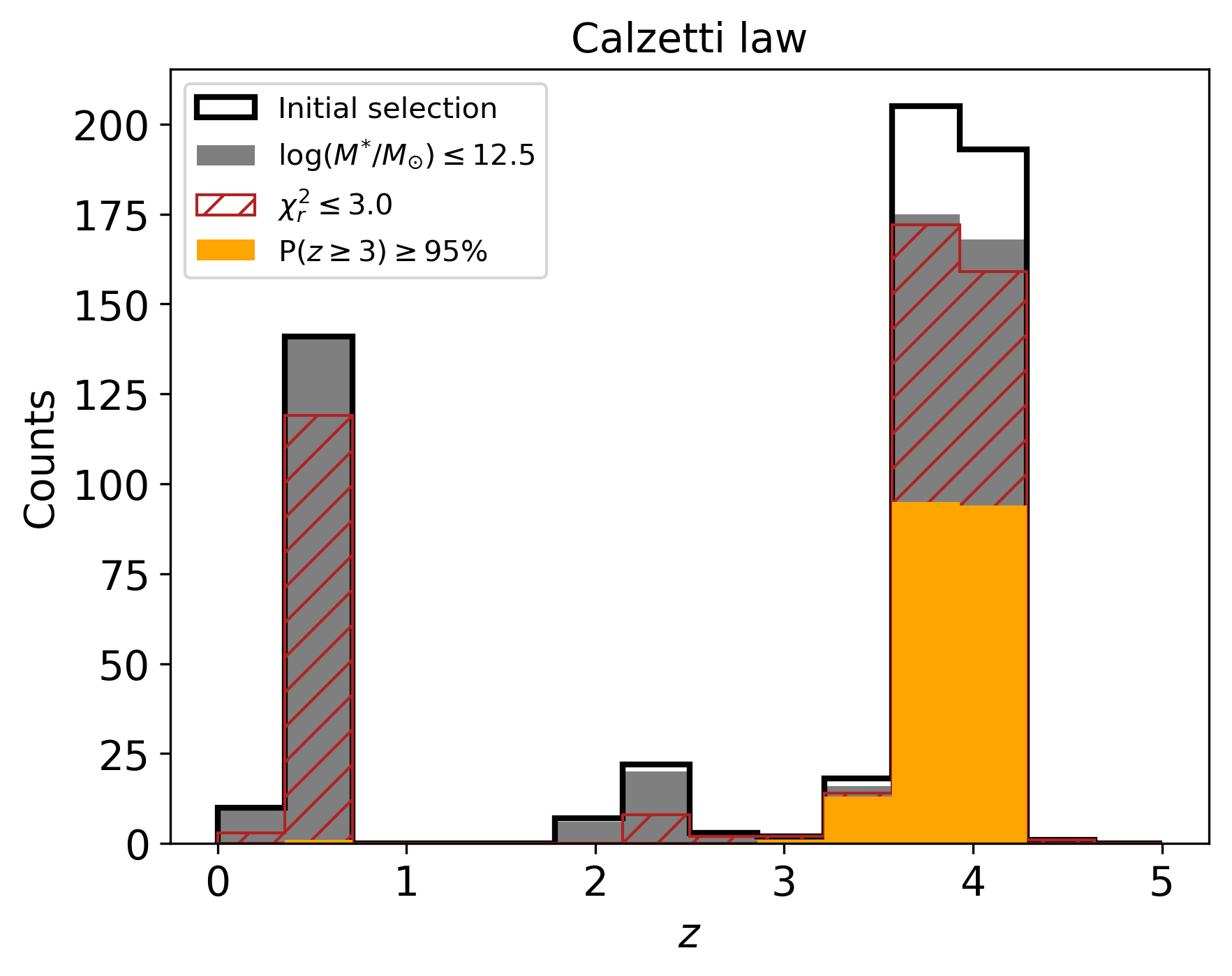}
    \caption{Redshift distribution for the 606 galaxies as found in the selection box of Figure~\ref{fig:CCmapsonly} (bottom-right panel). The different cuts, applied in series, shown on the histograms illustrate our selection procedure, namely: all fitted galaxies (solid black line); those with a physical mass (filled grey bars); those fitted with a $\chi_{r}^{2} \leq 3.0$ (hatched red); those passing our photometric redshift PDF criteria (orange bars). The two panels display the cases of SMC law on the left and Calzetti law on the right.}
    \label{fig:zDist78}
\end{figure*}

The redshift distribution before and after all the cuts described above is shown in Figure~\ref{fig:zDist78} (here we also show template-fitting photometric redshift results for galaxies at $z < 3$) for the SMC-type of runs and the Calzetti-type in the left-hand and right-hand panels, respectively. The 606 sources from the $z \sim 4$ selection box of Figure~\ref{fig:CCmapsonly} (bottom-right panel) are shown by the solid black line; the objects with $\log_{10}(M^{*}/M_{\odot}) > 12.5$ are identified by a grey shaded area; the red hatching indicates galaxies with $\chi_{r}^{2} >3$; the shaded orange area finally highlights the sources with high-$z$ probability of at least 95\% (note again that the best redshift solution may not correspond to the most probable one). The cuts are applied in series, as shown.

If we consider an average mass value of $\sim 10^{11.7}~M_{\odot}$, the expected counts according to \citet{DeLucia_etal2006} galaxy formation models, as plotted in Figure 1 of D13, are $\sim 1000$ at completion of DES. For our case ($\sim 2.7$ million sources, when including star-galaxy separation and error cuts, instead of the expected value of 300 million upon completion, i.e.~1\%) this would mean $\sim 10$ objects. We find 233 candidates, a value which lies an order of magnitude above the prediction of those models. Before drawing conclusions, however, we intend to spectroscopically confirm our best candidates.

We should also stress again that we have been very conservative for this initial paper. For example, galaxies with non-detections in VHS bands could still be high-$z$~objects perhaps dominated by very young stars and faint in rest-frame optical, and objects with potential blending issues are not necessarily low-redshift interlopers. It will be the subject of future work to study the excluded objects.

\subsubsection{On AGN contamination}
\label{sec:AGNcontamination}

A further source of uncertainty is the random presence of AGNs in the best candidates, whose effect could be to make some magnitudes brighter thereby possibly affecting the derived stellar mass. On the other hand, the AGN can be obscured and buried in the centre of the galaxy host which would not affect our template-fitting results.
High-$z$ galaxies dominated by AGNs are expected to look like point-like sources (even though not all point-like sources are expected to host an AGN and not all galaxies hosting an AGN would have their flux dominated by it). A way to quantify the point-likeness of a source is to compare the magnitudes of the object over an extended aperture with its PSF magnitude, as for a pure AGN these two quantities are the same.  For all our best candidates we have evaluated a parameter, dubbed $\sigma_{AGN}$, which is meant to quantify this magnitude difference and is defined as:
\begin{equation}
	\sigma_{AGN} = | \frac{i - i_{PSF}}{\sqrt{i_{err}^{2} + i_{PSF err}^{2}}} |,
\end{equation}
where $i$ and $i_{PSF}$ are an object extended and PSF $i$-band magnitudes, respectively, and $err$~are their respective errors.

\begin{figure}
	\centering
	\includegraphics[width=\columnwidth]{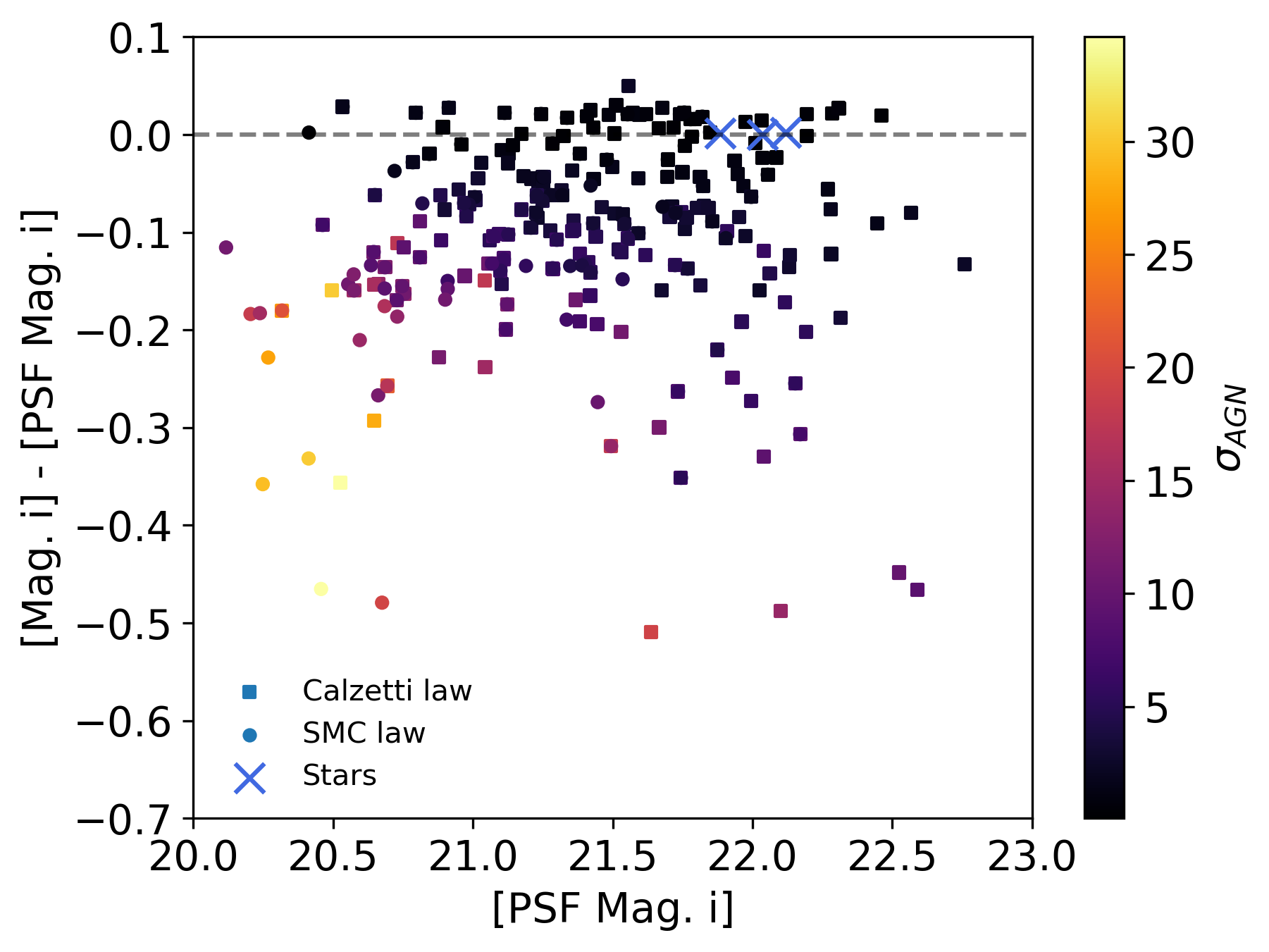}
    \caption{Difference between measured magnitude (MAG\_DETMODEL) and PSF magnitude in the $i$ band for the best candidates (distinguishing between those found with the SMC or the Calzetti law) plotted against the PSF $i$ magnitude. Symbols  are coloured by their $\sigma_{AGN}$ values (see text). The further away a galaxy is from the zero-level (dashed line), the brighter it is compared to the PSF magnitude, suggesting that it is resolved and less likely to be an AGN. Three stars are also plotted as blue crosses and they lie on the zero-level line, as expected for clear point sources.}
    \label{fig:psfvsmag}
\end{figure}

Figure~\ref{fig:psfvsmag} show the results, by plotting the magnitude difference to the PSF vs the PSF magnitude, with each object being coloured according to the $\sigma_{AGN}$~parameter, a 0 value of which means a complete point-like case. We also plot as blue crosses three, randomly selected objects from the COMMODORE catalogue with $\sim 100\%$~probability of being stars. They do lie precisely at the PSF zero-level (dashed line), as expected.

We see that our candidates span a range in `extension', but several lie on the point-source line although they have been classified as galaxies rather than stars by the DES pipeline. This may imply that they lie at very high redshift or that they host an AGN or a combination of both. As using this information in any quantitative way would be completely arbitrary, we just provide the $\sigma_{AGN}$ values in the data tables along with all other template-fitting results (Appendix~\ref{app:FullGoldenSampleResults}), but we do not include it in our selection process.

\subsection{Properties of the Best Candidates}
\label{sec:PropGoldenSample}
The fitting results for the best candidates are given in Tables~\ref{tab:app-SMC} and \ref{tab:app-Calzetti} for the two reddening laws, respectively. The results include, along with the photometric redshift, mass, and age, the star formation history (SFH) and its corresponding metallicity, Z/H (Z$_{\odot}$), of the best-fit template, and the absolute magnitude in the $i$ band as calculated by HyperZ.
All the outlined results are obtained using the photometry, with relevant errors, reported in Table~\ref{tab:app:MagGoldenSample}.

\begin{figure*}
\includegraphics[width=0.70\textwidth]{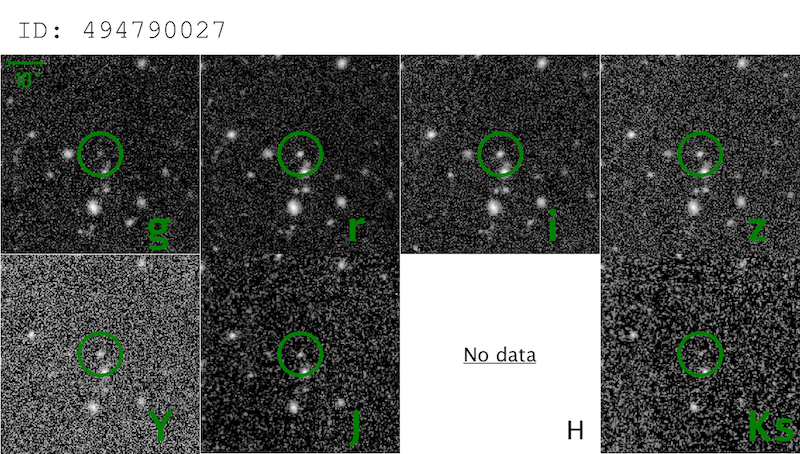}
\includegraphics[width=0.49\textwidth]{Fitting_5_RL4.png}
\includegraphics[width=0.50\textwidth]{PDF_5_RL4.png}
\includegraphics[width=0.49\textwidth]{Fitting-spec_5.png}
\includegraphics[width=0.49\textwidth]{Fitting-DESonly_5.png}
\caption{The observation images for all available bands (DES and VHS), the best fit, the PDF, the fit with fixed redshift according to $z_{BPZ}$, and the fit only using DES bands for one of our best candidates. For the galaxy images the green circles indicate the sky position of the galaxy; if the circle is red it means that the photometry is not available for that given filter. If VHS observations are not available, for one or more filters, the image is replaced by text saying so. For the plots below the images (bottom four panels) the legend is the same as in Figure~\ref{fig:Fitting_PDF_Examples}. 
The rest of the best candidate plots and images can be found following this \href{http://icg.port.ac.uk/~guarniep/}{link}.}
\label{fig:9bestcandidates}
\end{figure*}

Figure~\ref{fig:9bestcandidates} shows all relevant plots for one candidate as an example. The same plots for the other candidates can be downloaded at this \href{http://icg.port.ac.uk/~guarniep/}{link}. 

First we show image cutouts in the various photometric bands available, with a green circle indicating the galaxy position. Notice how, as expected, the galaxy is barely visible in the bluest band ($g$~band), which is the drop-out band at these redshifts. We also point out when bands are not available.
The panels below the tiles display: the spectrophotometric model fitting (upper-left), the redshift PDFs (upper-right),  the spectrophotometric model fitting obtained when the redshift is fixed to the $z_{BPZ}$ value (bottom-left), and when only the DES bands are used (bottom-right).

In the spectrophotometric model fitting plots (similar to what was shown before in Figure~\ref{fig:Fitting_PDF_Examples}), the black line shows the best-fitting model spectrum, the blue squares the fluxes of that model in the DES and VHS filters and the red points the observed data in the same filters. Some key parameters of the model\footnote{The other model parameters are given in Table~\ref{tab:app-SMC}} are printed on the plot, the age (in Gyr) and the stellar mass (in $\log M_{\odot}$), along with the $\chi_{r}^{2}$. 

The upper right plot shows the photometric redshift PDF using all available bands (red), DES only bands (black), and the best redshift from the DES pipeline run of BPZ (blue). Not surprisingly, the DES BPZ redshift favours lower solutions, typically around 0.3. In order to probe the type of fitting we would obtain had we assumed the low-$z$~solution as the correct one, we use the DES BPZ value as the true redshift and run HyperZ-spec, keeping all other parameters identical.

\begin{figure}
\centering
\includegraphics[width=\columnwidth]{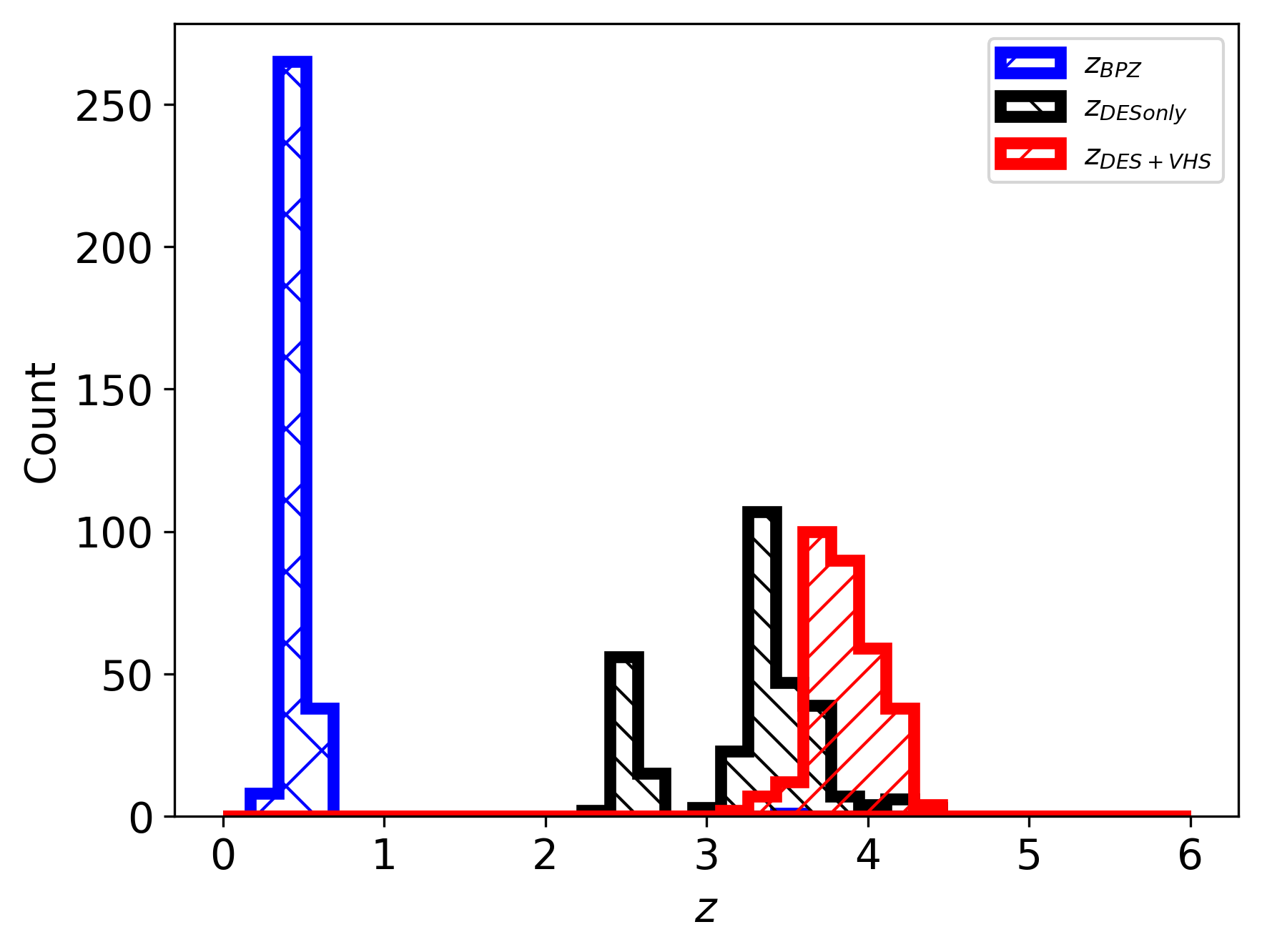}
\caption{The distributions of $z_{DES+VHS}$, $z_{DESonly}$, and $z_{BPZ}$ for the best candidate galaxies are plotted as red, black, and blue histograms, respectively. It can be seen that even when using only DES bands to fit these galaxies, the results still tend to be at high-$z$ and relatively close to the $z_{DES+VHS}$ ones; on the other hand, as explained earlier, $z_{BPZ}$ values stay below $z=1$.}
\label{fig:GoldenSamplez_comparison}
\end{figure}

The result for this type of run is shown, for instance, in the bottom-left panel of Figure~\ref{fig:9bestcandidates} (and at this \href{http://icg.port.ac.uk/~guarniep/}{URL} for other objects). As expected due to the small redshift, the model fitting results in a lower mass galaxy ($\log M\sim10.3~M_{\odot}$), though with a similar age, and generally a worse fit. The bottom-right panel instead shows the fit for the case when only DES bands have been used to run HyperZ. We can see that the fit remains good, but as a very small portion of the spectrum is fitted (corresponding to the rest-frame $UV$~at these redshift) the physical solution differs from the one obtained when fitting all available bands. Interestingly, the photometric redshift remains high (although not as much), which reinforces the use of DES-only bands in D13 as a discriminator of high-$z$, massive galaxies.
This can be appreciated by looking at Figure~\ref{fig:GoldenSamplez_comparison}. Here, all best candidate galaxies' three types of redshifts ($z_{DES+VHS}$, $z_{DESonly}$, and $z_{BPZ}$) are shown as red, black, and blue histograms, respectively. In Section~\ref{sec:DESonlyVSDESVHS} we shall discuss in more detail how the addition of VHS data to the DES data affects our fitting results.

\begin{figure*}
\centering
\includegraphics[width=0.49\textwidth]{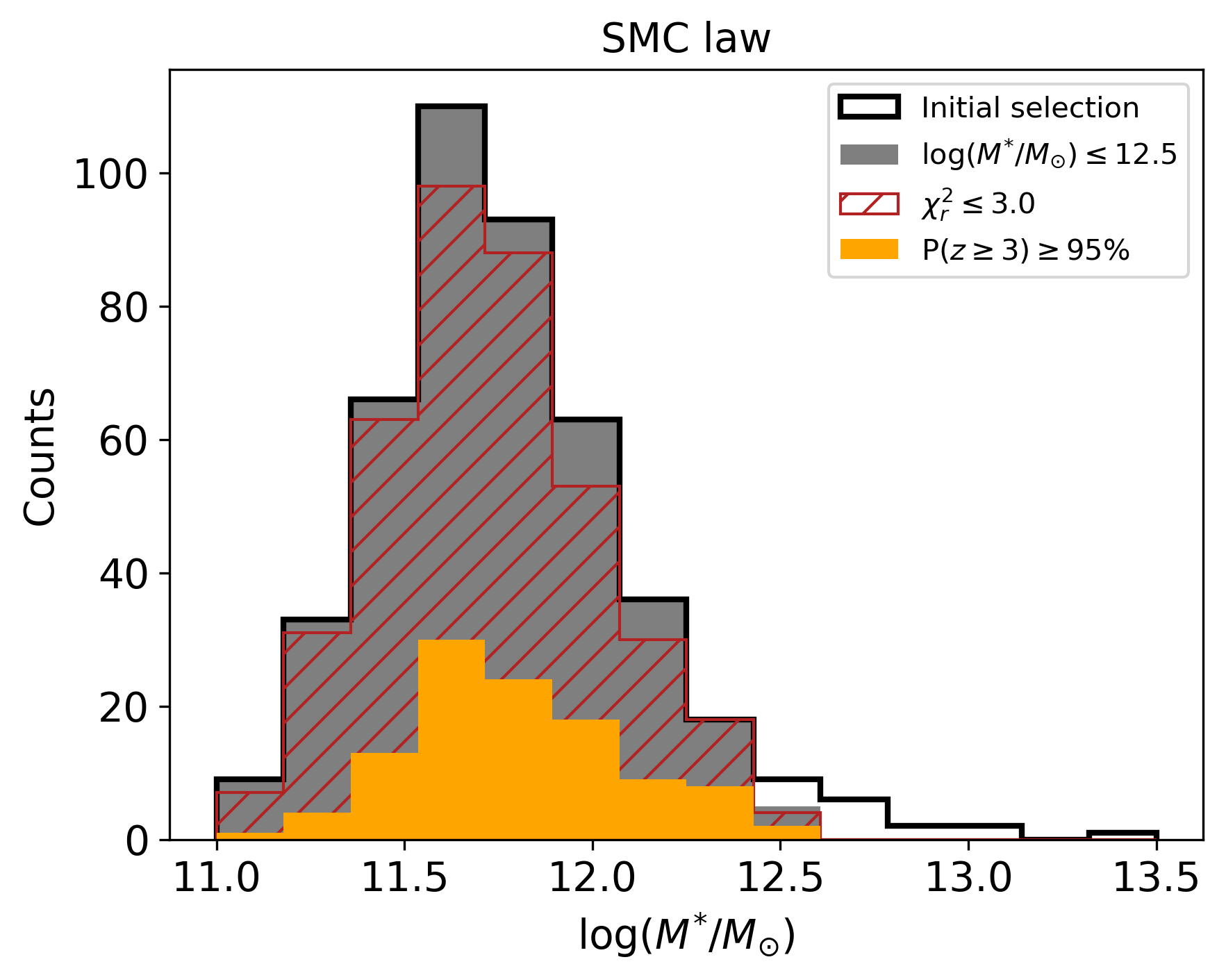}
\includegraphics[width=0.49\textwidth]{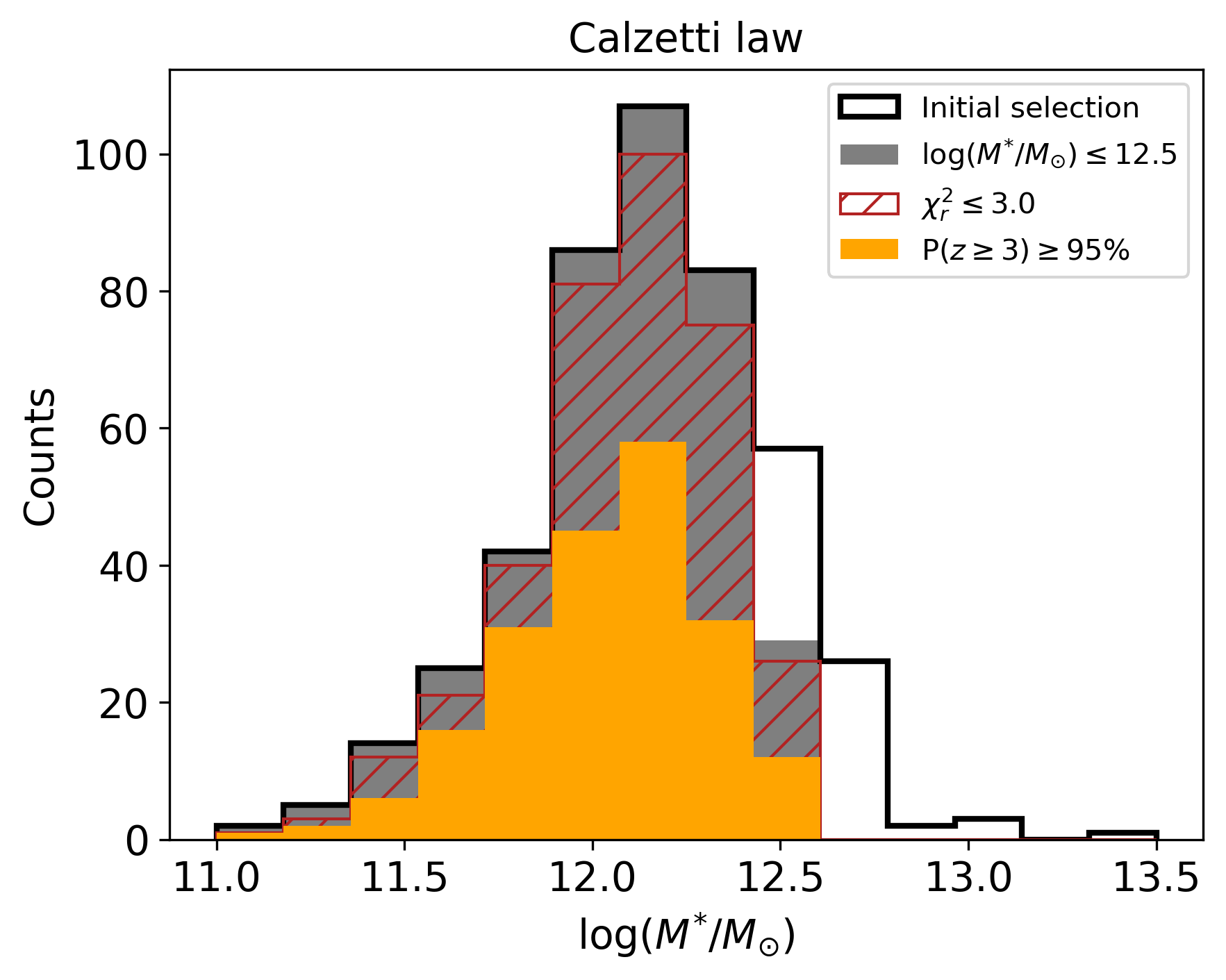}
\includegraphics[width=0.49\textwidth]{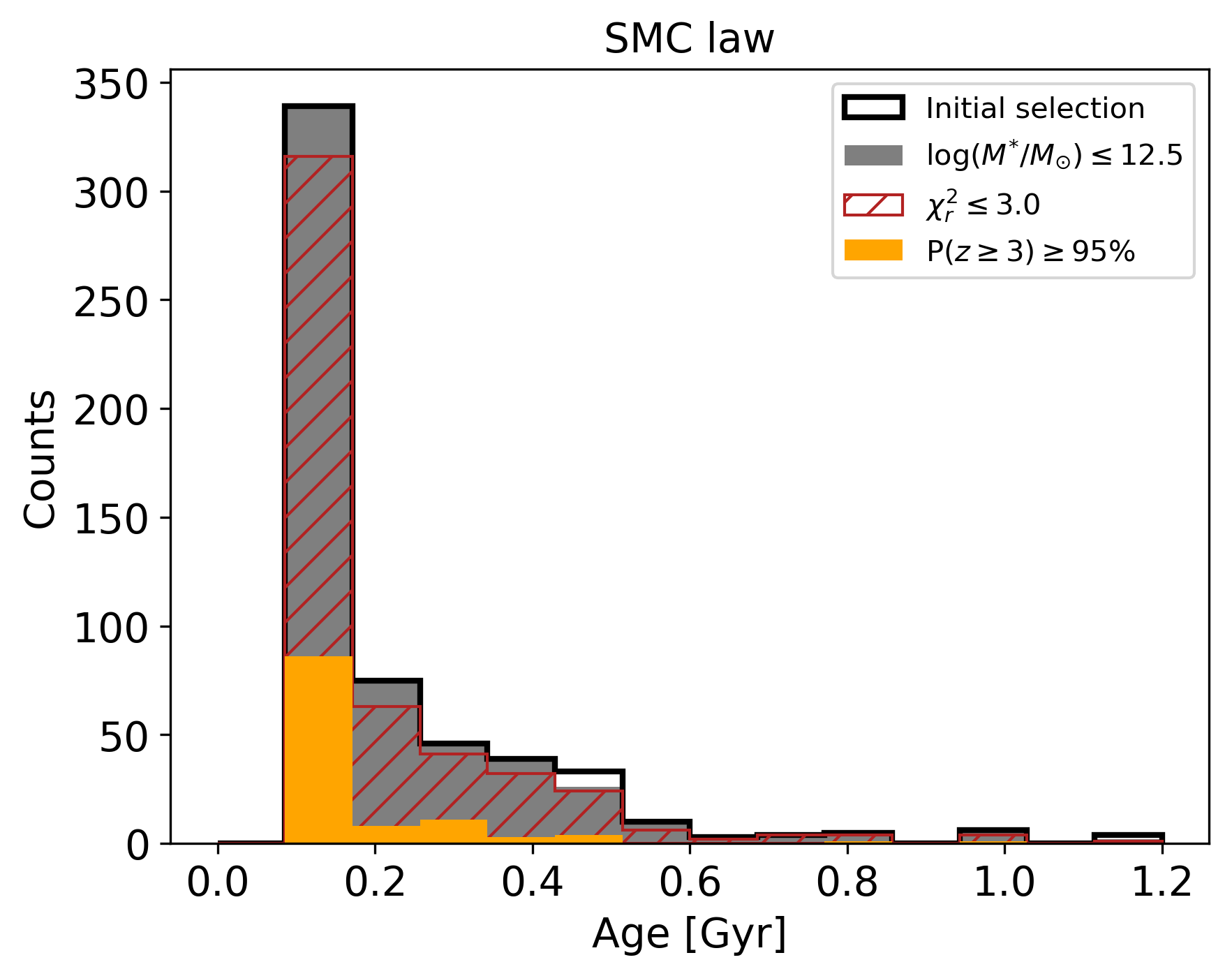}
\includegraphics[width=0.49\textwidth]{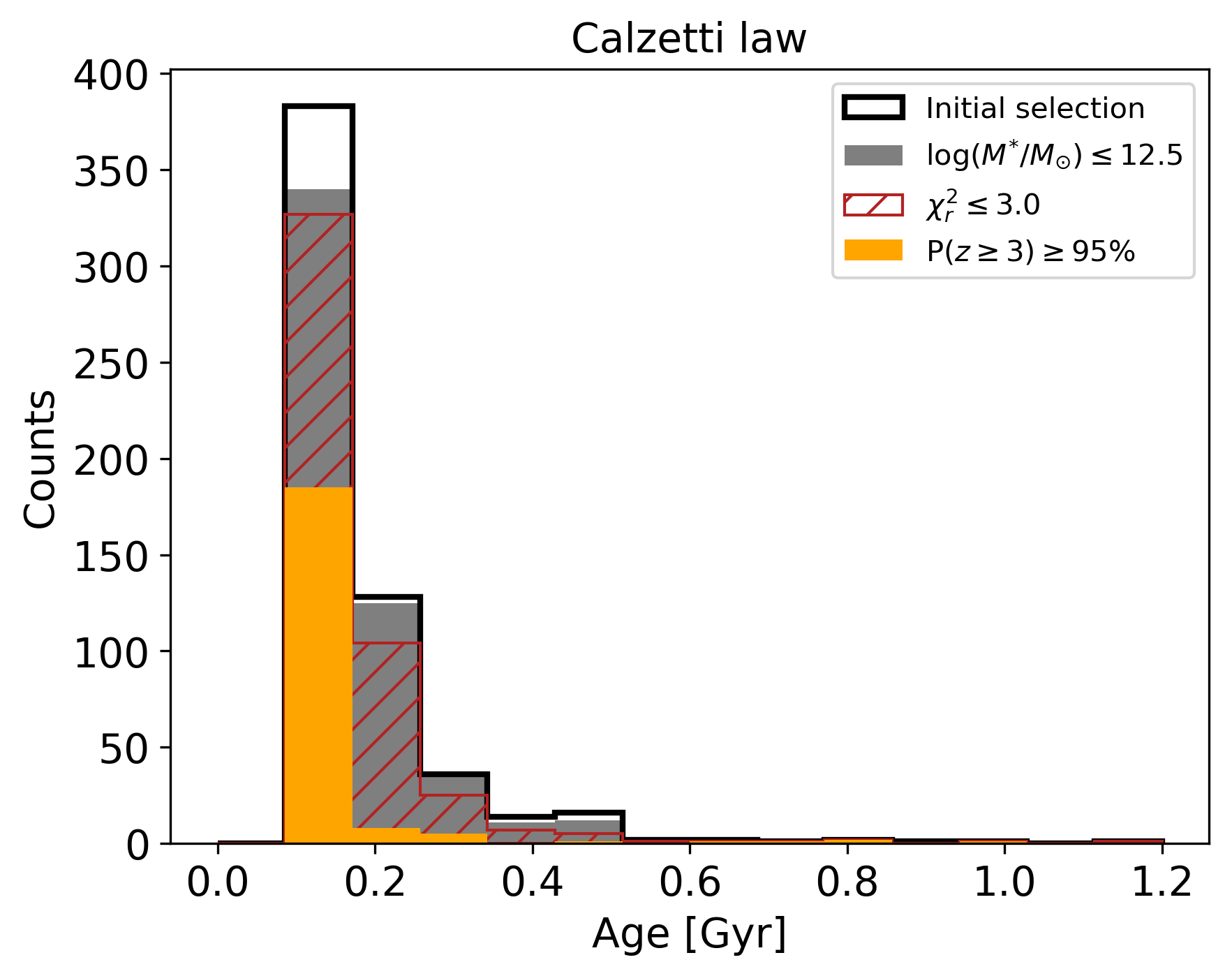}
\includegraphics[width=0.49\textwidth]{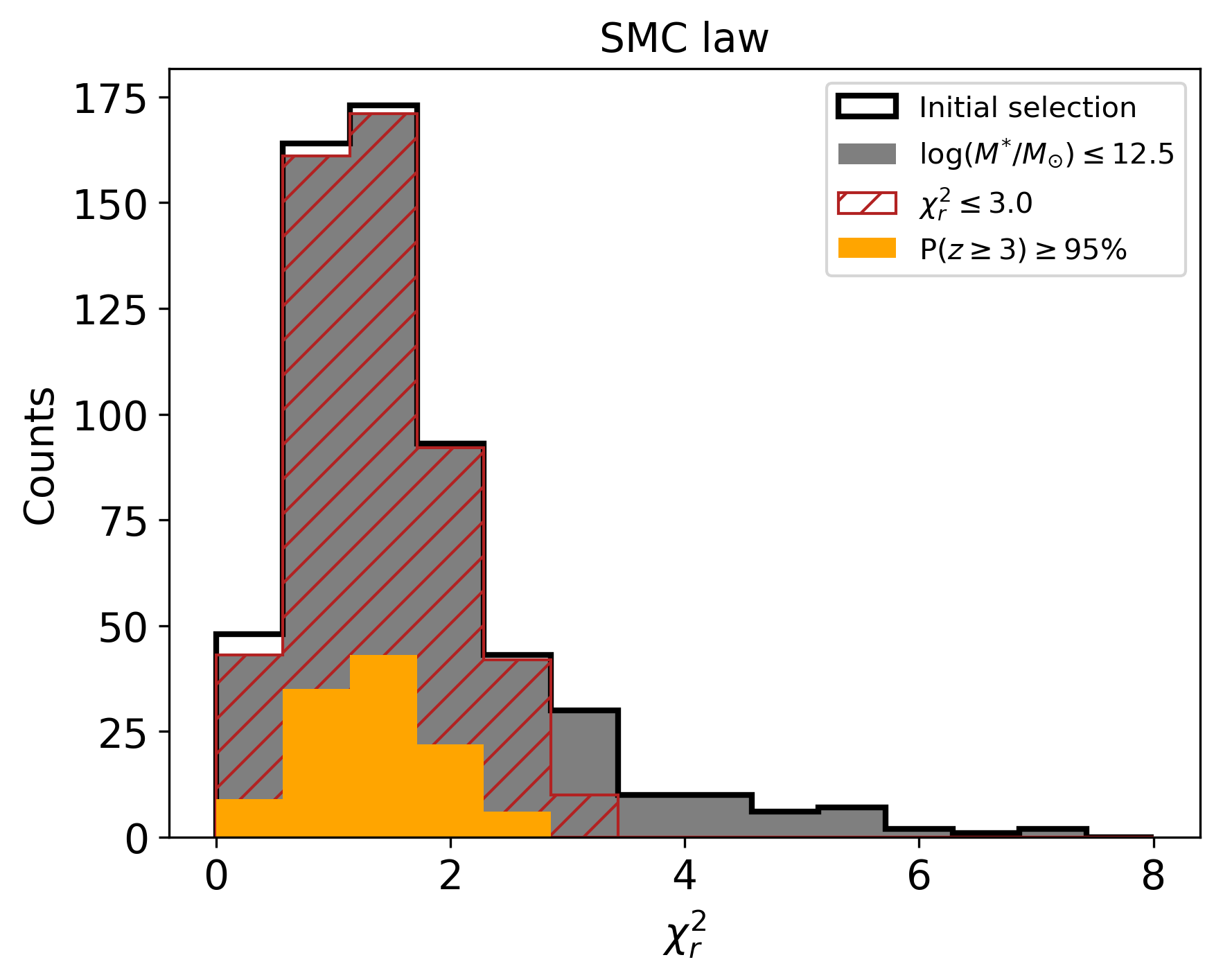}
\includegraphics[width=0.49\textwidth]{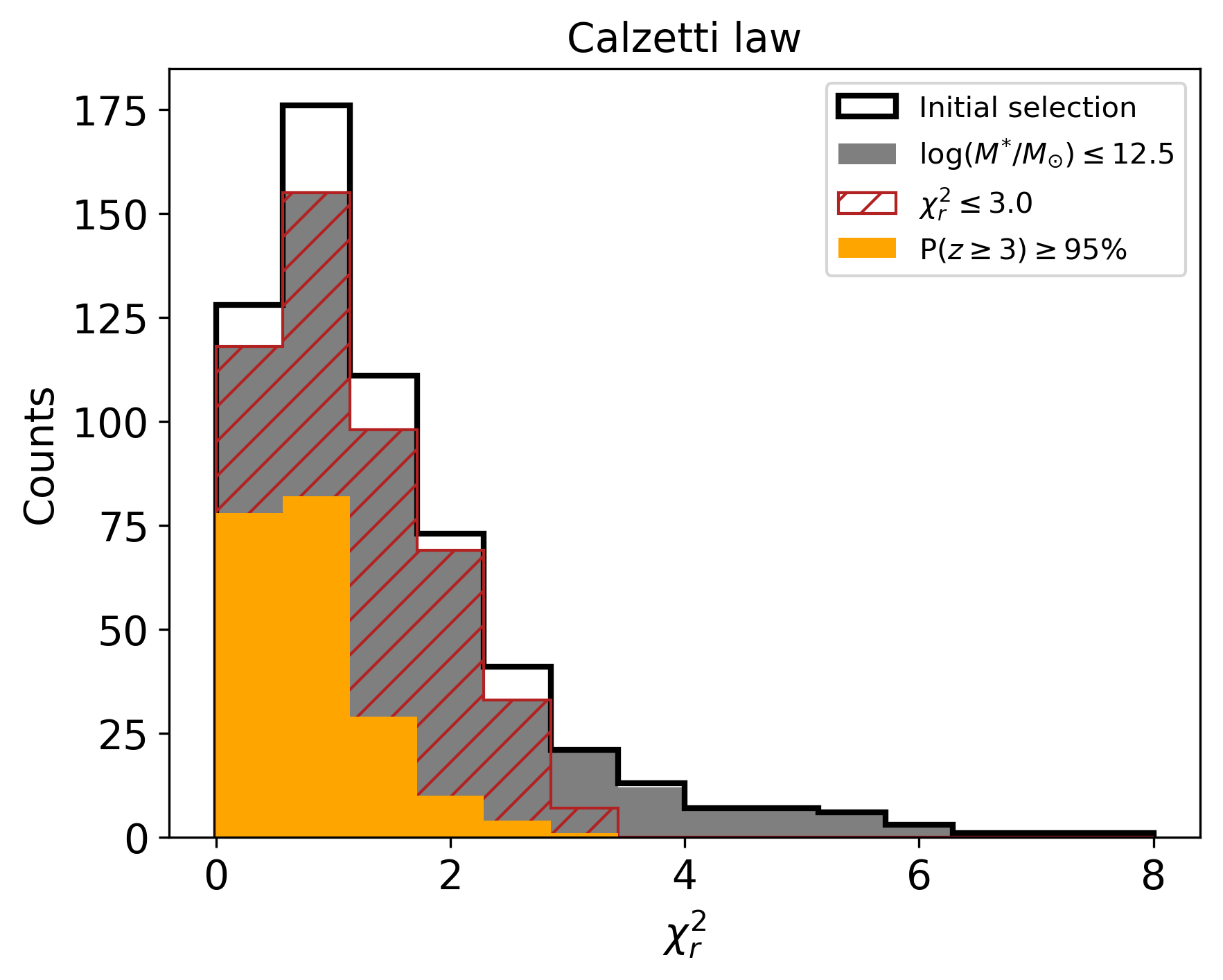}
\caption{$M^{*}$, age, and $\chi_{r}^{2}$ for the $z > 3$ candidates within the sample of 606 sources from the colour selection box. Legend as in Figure~\ref{fig:zDist78} (except for redshift cut (i.e.~$z\geq3$) applied to the initial selection). This is shown for all two reddening options (SMC law and Calzetti law) in the first and second columns, respectively.}
\label{fig:DistributionsOfTOP}
\end{figure*}

We summarise the properties of our best candidate galaxies in Figure~\ref{fig:DistributionsOfTOP}, in comparison with the initial sample of 606 galaxies (among these here we only show galaxies at $z \geq 3$). The solid orange bars show the distribution of stellar masses, ages, and the goodness of these fits (i.e.~$\chi_{r}^{2}$), for the best candidates. Objects from 606 at $z > 3.0$ are distributed according to the solid black line, with those having physical mass $\log_{10}(M^{*}/M_{\odot}\leq12.5$) shown by the filled grey bars. The cut in $\chi_{r}^{2} \leq 3.0$ is highlighted by hatched red bars. Furthermore, the PDF cut (probability of at least 95\% of being at $z \geq 3$) shrinks the distribution in what is shown with the filled orange bars which match also correspond to the best candidates. The final histograms show that the many quality cuts we apply do not distort the initial distribution of galaxies. More quantitatively, performing a K-S test, we find that we cannot reject the null hypothesis that the distributions are the same with p-values, on average for the two reddening options, of 0.27, 0.29, and 0.24, respectively for mass, age, and $\chi_{r}^{2}$. Even the objects with poorer fits trace similar galaxy physical properties. All the cuts are applied in series, one after the other.

Our best candidate galaxies have large stellar masses $10^{11.5} \leq M/M_{\odot} \leq 10^{12.5}$, ages $\sim 0.1$ Gyr, have been forming stars following a rapid mode generally consistent with an exponentially-declining mode with short $e$-folding time, $\tau = 0.1-0.3$ Gyr, or a truncated model with similarly short timescales. Metallicity is generally high ($Z > $ Z$_{\odot}$). Our derived ages tend to cluster around the lower age limit set in HyperZ (0.1 Gyr). This is to be expected as - in $\chi^2$~minimisation fitting of stellar populations containing young components - the best-fitting model tend to be associated to the youngest possible age, especially when only UV/optical bands are fitted. When this happens, as the energy emission of young stellar populations is order of magnitudes higher than that of older populations, the youngest models will dominate the fit, an effect that is known as `over-shining' (Maraston et al.~2010). Hence, if we allow the fitting procedure to use any value of age, rather than an age estimate for the whole galaxy, we tend to obtain the age of the last generation of stars, which may contribute very little by mass. This over-shining effect may hamper us to recover realistic galaxy stellar population properties, in particular making stellar mass estimates unreliable. In order to circumvent this problem, it is common to adopt a lower age limit in the fitting procedure (e.g.~Daddi et al.~2005), which is usually set to an age (0.1 Gyr) when the most massive stars are already dead (for passive models). It is not guaranteed that such a limit ensures fully realistic ages for all galaxies, which may still be biased low. If this is the case, our stellar mass estimates will be lower limits of the true stellar mass.

In Section~\ref{sec:AgeLimPhotFits} we shall assess the effect of this assumption on the two main properties we are interested in, namely the photometric redshift and the stellar mass. 

Results are summarised in Table~\ref{tab:app-SMC} and~\ref{tab:app-Calzetti}. The results of the model fitting are also consistent with the simulations of D13 and with being a younger version of the stellar populations inhabiting the most massive galaxies in the local universe~\citep{Thomas_etal2010}. 

\begin{figure*}
\centering
\includegraphics[width=0.32\textwidth]{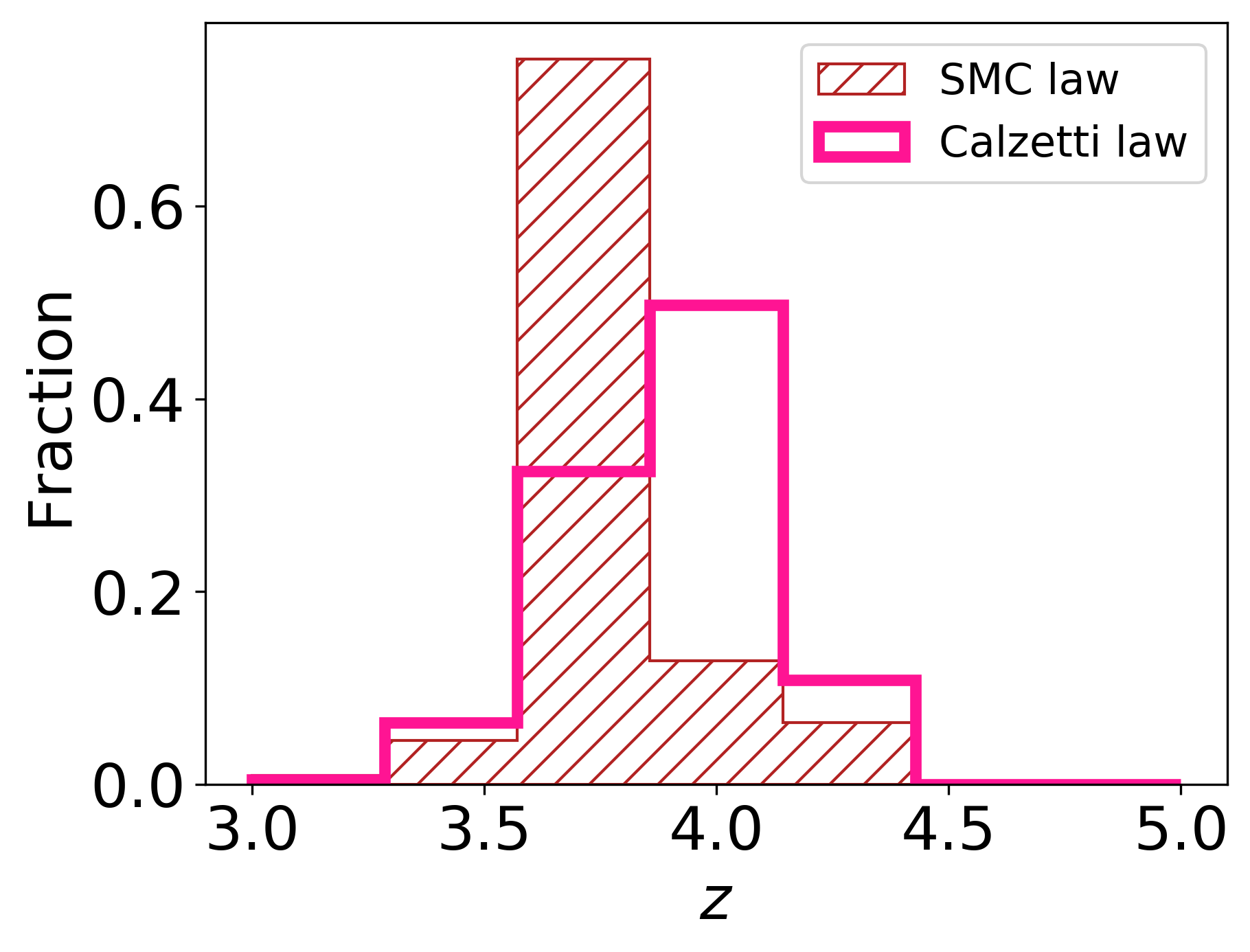}
\includegraphics[width=0.33\textwidth]{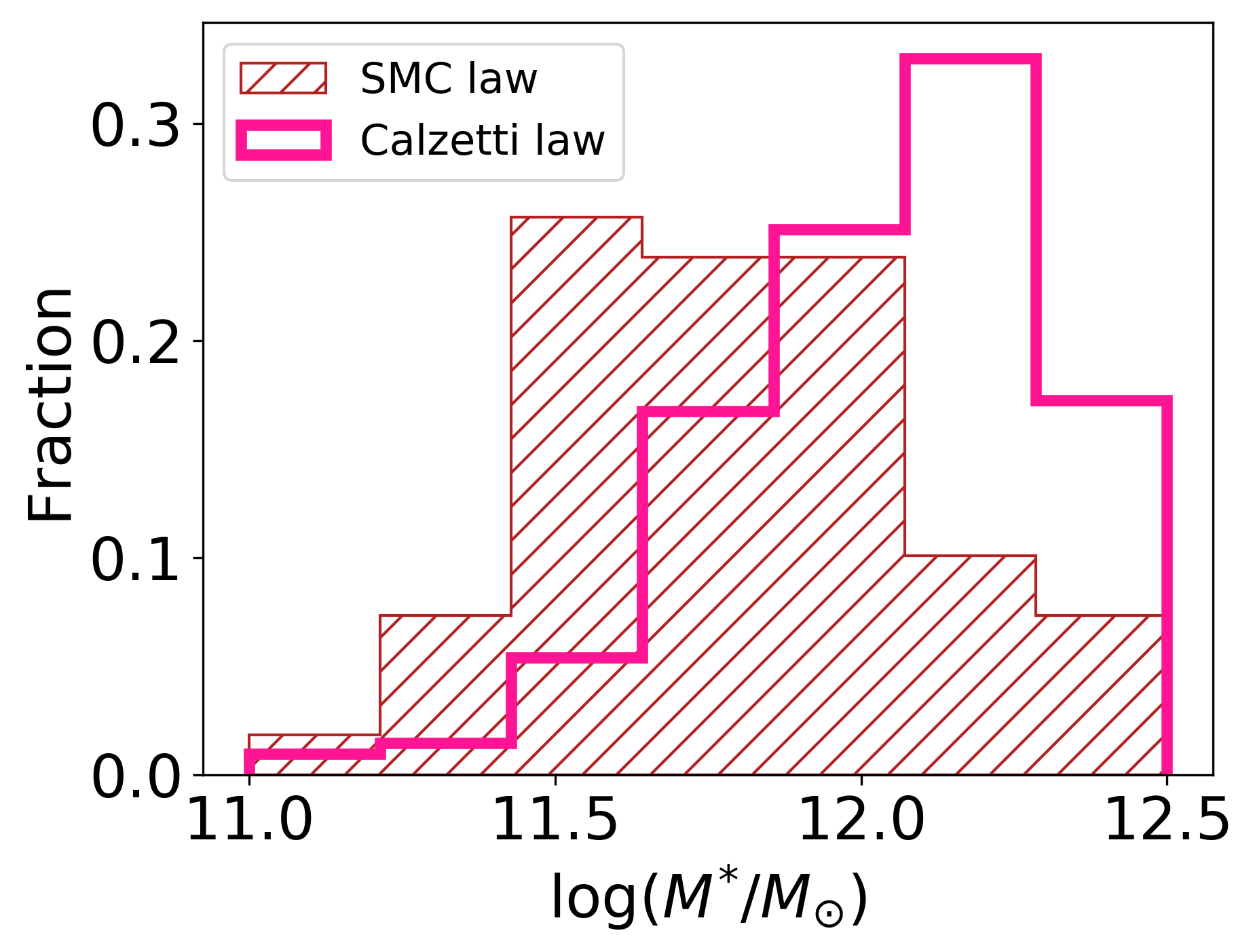}
\includegraphics[width=0.33\textwidth]{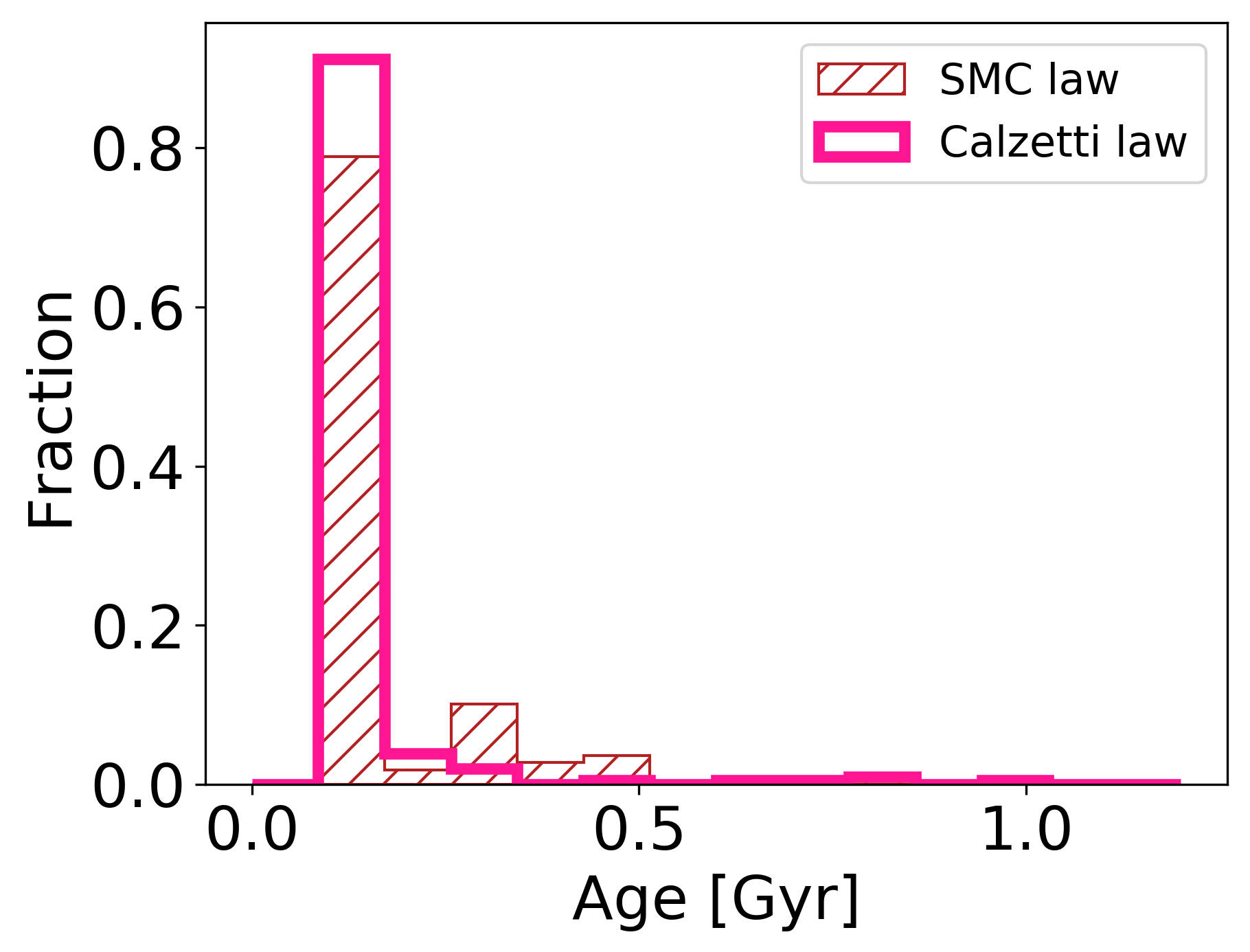}
\caption{Distributions of (from left to right) photometric redshift, stellar mass, and age in terms of fraction for the best candidates found with the two different reddening laws: SMC law (hatched dark red bars) and Calzetti law (solid deep pink line).}
\label{fig:3ReddeningDistributions}
\end{figure*}

Looking in more detail at the redshift and physical properties of the best candidates we can see that, compared to the assumption of an SMC law, the use of the Calzetti law often pushes redshifts and masses to higher values. 
All this information is summarised in Figure~\ref{fig:3ReddeningDistributions}, where we show fractional distributions for each reddening law (hatched dark red for SMC law and deep pink solid line for Calzetti law) in terms of redshift, mass, and age, respectively. 

\begin{figure*}
\centering
\includegraphics[width=0.49\textwidth]{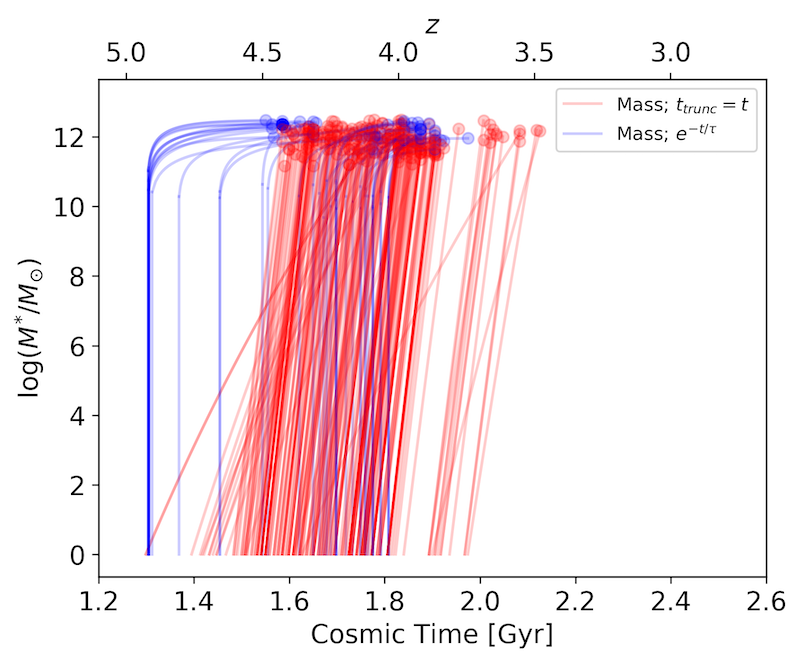}
\includegraphics[width=0.50\textwidth]{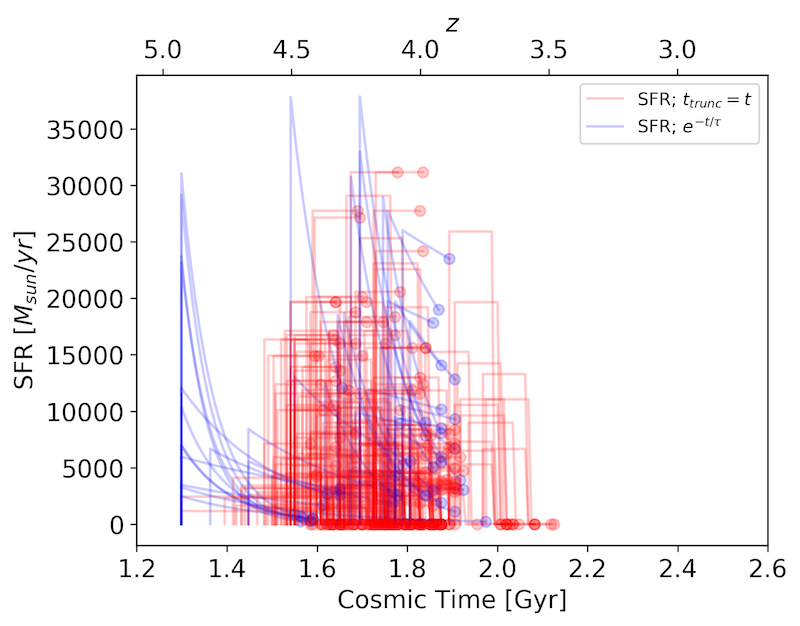}
\caption{Galaxy mass assembly derived for the best candidate galaxies. Typical SFHs as $t_{trunc} = t$ and $e^{-t/\tau}$ are plotted in red and blue respectively for the mass assembly (left panel) and for the SFR (right panel). A few candidates with SFHs different from those two laws have not been plotted. Candidates from all two reddening options have been plotted together.}
\label{fig:MassAssembly}
\end{figure*}

The SFHs of these candidates also allow us to create a visual representation of the mass assembly and SFR of such galaxies. These are shown for two different types of SFH in Figure~\ref{fig:MassAssembly}, which represent the vast majority of the best candidates. They are: $t_{trunc} = t$ and $e^{-t/\tau}$ (red and blue, respectively), plotted in the left panel for the mass assembly and in the right one for the SFR. These rapid timescales point to formation redshifts close to $z \sim 5$, which is in agreement with what is deduced from the fossil record of local high stellar mass galaxies (we shall return to this point in the discussion). Note that even if these SFRs appear extreme, values of tens of thousands of solar masses per year have already been recorded in the literature \citep[e.g.][]{Rowan-Robinson_etal2016}. The results from both reddening options are plotted together.

Let us conclude with a comment on the population model effect. Here we assume Maraston's models, while results in the literature are usually based on Bruzual \& Charlot (BC) models. As is well known, ages derived with Maraston models for high-$z$ galaxies are smaller than those based on BC models, which results in smaller stellar masses $M^{*}$~by e.g. $\sim 0.2$~dex (depending on the fitted ages). This is due to the different prescriptions for the thermally pulsing asymptotic giant branch (TP-AGB) and to the different onset age for the RGB assumed in the underlying stellar tracks \citep[see][]{Maraston_etal2006}. 
Hence, our estimates of age and $M^{*}$~lie likely on the lower side than what we would obtain had we used BC-type models.

In Section~\ref{sec:FittingStraatman14} we shall discuss the results we obtain when we re-fit some of the galaxies recently reported in the literature, including the~\citet{Glazebrook_etal2017} object, with Maraston models and our fitting procedure. 

\subsection{The Effects of Fitting DES-only vs.~DES+VHS Bands}
\label{sec:DESonlyVSDESVHS}

\begin{figure*}
\centering
\includegraphics[width=0.49\textwidth]{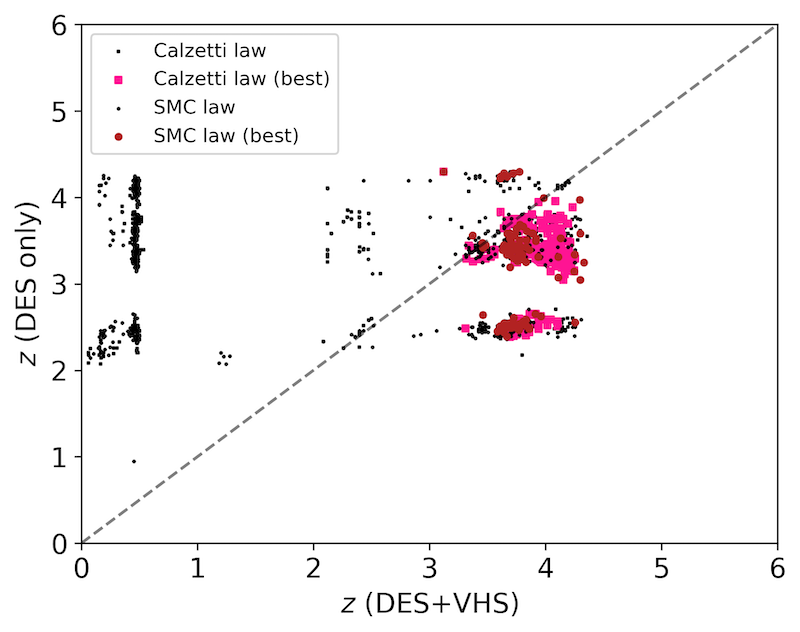}
\includegraphics[width=0.50\textwidth]{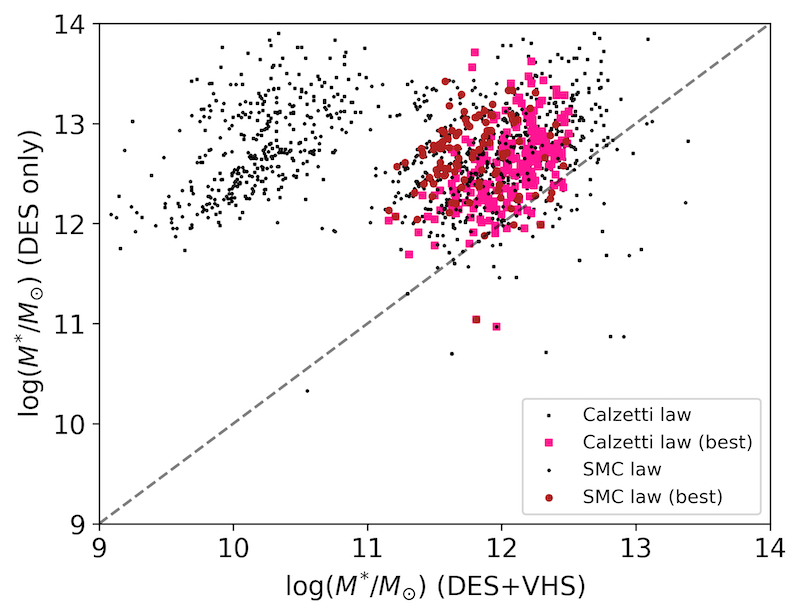}
\caption{Comparison of the results for photometric redshift (left panel) and stellar mass (right panel) of the 606 template-fitted galaxies depending on the use of DES+VHS photometry versus the DES only data option. This is shown for both the SMC law (circles) and the Calzetti law (squares). The best candidates are highlighted in dark red and deep pink for the two reddening types, respectively, with the other sources being plotted in black. It can be appreciated that near-IR data is essential to successfully recover the true photometric redshift and mass of lower-redshift galaxies.}
\label{fig:DESVHSvsVHS}
\end{figure*}

Given our database and modelling, we are in the best position of quantitatively assessing the effect of using only DES bands and DES+VHS bands on the results of template fitting. Figure~\ref{fig:DESVHSvsVHS} summarises the results. Here we plot photometric redshifts (left-hand panel) and stellar masses (right-hand panel) for all 606 sources as determined via template fittings using only DES bands (y-axis) or DES+VHS bands (x-axis). The best candidates are highlighted in dark red and deep pink, also distinguishing those obtained when assuming SMC (circles) or Calzetti (squares) reddening types, respectively.

The fitting performed with DES bands gives photometric redshifts that are always larger than 2 and very high, possibly unphysical stellar masses ($\log_{10}(M^{*}/M_{\odot}) > 12.5$), for $\sim 60\%$~of the galaxies. The inclusion of VHS bands in the fitting pushes the photometric redshift down to values below 2 for 25\% of the sample (150 galaxies, independently of the reddening law) and their stellar masses also become lower.
More generally, the addition of VHS bands results in a healthier distribution of stellar masses overall ($9 < \log_{10}(M^{*}/M_{\odot}) < 12.5$), with only $\sim2\%$ of objects having $\log_{10}(M^{*}/M_{\odot}) > 12.5$. This confirms the notion that by fitting a wider wavelength range in the data we recover the physical properties of galaxies~\citep[as shown, for example, by][]{Pforr_etal2012, Pforr_etal2013}.

More importantly for the present work, we find that the subsample of galaxies ($\sim233$) defining our best candidate sample remain at high redshift and with a large stellar mass independently of whether we fit only DES bands or DES+VHS bands.
This confirms that our selection cut based on DES-only bands~\citep[as in][]{Davies_etal_2013} plus the additional criteria put forward in this paper (goodness of fits, unimodal PDF of redshift and the addition of VHS bands) are effective at selecting a robust sample of very massive, high-redshift galaxies. The combination of DES and VHS allows us to obtain more robust galaxy properties and to limit the number of low-$z$~ contaminants.

\subsection{Caveats and Reliability Tests}

In this section we test the reliability of our method and results. In \ref{sec:ContaminationComment} we quantify the number of low-$z$~contaminants we found after performing spectral fitting and in \ref{sec:lowZscattering} we comment on the number of sources expected to scatter in and out of the colour map selection box because of photometric errors. In the remaining subsections we test the effect of various parameters on photometric fitting results.

\subsubsection{Contamination in the Selection Box}
\label{sec:ContaminationComment}

Colour selection based on models is a commonly adopted method to prioritise the galaxies one would like to obtain before performing the (time consuming) model fitting (e.g. Daddi et al.~2005). A certain level of contaminants of various kind is expected due to various types of objects having similar colours. 

D13 quantify and discuss the types of expected contaminants using DES galaxy simulations as well as real objects, and find that contamination in the $z \sim 4$ box is mostly due to intermediate-redshift galaxies (of similar colour because of their Balmer/$4000 \AA$ break) and quasars (D13, Figure~7). Using both DES galaxy simulations as well as real data from the MUltiwavelength Survey by Yale-Chile (MUSYC; Gawiser et al. 2006), D13 find (their Figure~11) that the $z < 2.5$ contamination in the $z \sim 4$ selection box should be ~ 38\%~\citep[consistent with the previous study by][]{Douglas+2009}.
Here we have the possibility of checking the D13 prediction with the real DES data after model fitting is performed. 
After fitting the 606 sources selected in the box, we find 34\% and 32\% lower-redshift galaxies for the SMC-type and Calzetti-type runs, respectively. These figures are in excellent agreement with the D13 estimates.

In conclusion, it seems unlikely that undetected lower redshift contaminants pass through the further best candidate selection cuts (in particular the redshift PDF and $\chi_{r}^{2}$ cuts), while the PSF cut should prune most pure quasars (see Figure~\ref{fig:psfvsmag}).

\subsubsection{Number of Sources Scattered Into the Selection Box}
\label{sec:lowZscattering}

Photometric errors may affect the position of galaxies on our colour selection plot (specifically of the $\sim2.7$ million sources without large errors and stellar contamination, see top-right panel of Figure~\ref{fig:CCmapsonly}). If we perturb the position (i.e. the colors) of the 3243 sources found within the $z \sim 4$ selection box according to their photometric errors in the $g$, $r$, and $i$ bands and assuming a normal distribution, we find that $\sim 29$\% of these sources scatter outside the box.

We can conclude therefore that of the 606 galaxies from the selection box, around 176 (or around 68 of the 233 best candidates) could have gone unselected purely because of their random error, while the rest sits well inside the box within their error. On the other hand, it is also possible that galaxies that would be truly outside the selection box have scattered inside, which is non-trivial to model because of the non-gaussianity of the distribution. Overall, however, we stress that the colour selection served purely as a way to prioritise the galaxies to fit using stellar population models. It would be unlikely that all fitted photometric bands for the 606 sources would be affected by their errors in such a way to spuriously produce high-$z$ massive galaxy candidates in such numbers (i.e.~233 candidates out of 606 fitted galaxies in total). Future simulation work could investigate these uncertainties quantitatively.

\subsubsection{Detections vs.~VHS Magnitude Limits}

\begin{figure}
	\centering
	\includegraphics[width=\columnwidth]{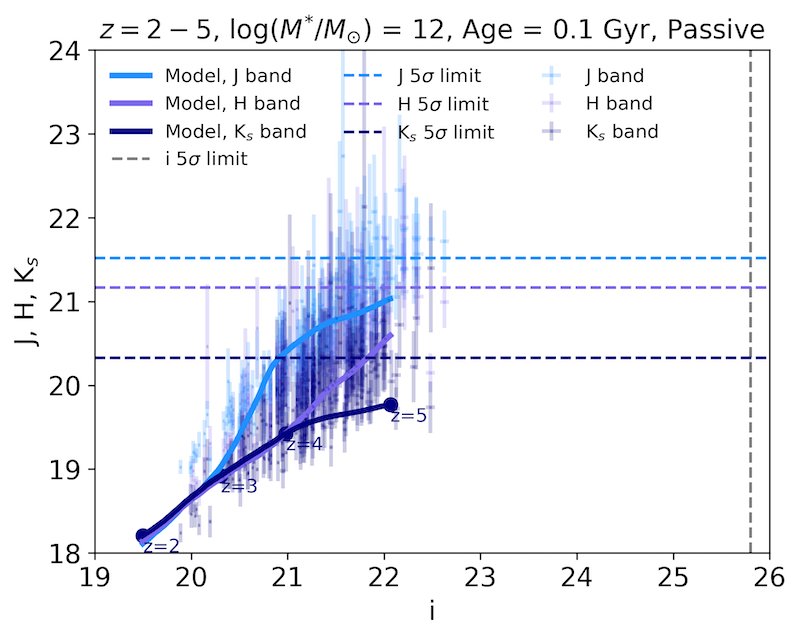}
	\caption{Expected $J, H, K_{s}$~VHS magnitude for given $i$~DES magnitude plotted as solid light blue, purple, and dark blue lines (with $5\sigma$ limits shown as dashed lines), respectively, for a passive M05 model with redshift between 2 and 5 and an age of 0.1 Gyr. Best candidate galaxies (for all reddening laws used) are plotted as dots for the available bands of each given source.}
	\label{fig:TableLims}
\end{figure}

As mentioned before, several objects do not have VHS detections, but this does not mean they do not lie at high-$z$. Figure~\ref{fig:TableLims} shows the expected VHS $J, H, K_{s}$ magnitudes, and the survey limits at $5\sigma$ confidence, for a given $i$~DES magnitude in the case of a passively evolving model with redshift between 2 and 5 and an age of 0.1 Gyr (to represent the most common value found for our objects). 
Our best candidate sources (when available for a given band) are shown as dots with errors. As expected, the majority of points lie below the survey detection limits. The $K_{s}$ points lying above the relative limit refer to sources with magnitudes in the other bands within the detection limits to which a $K_{s}$ magnitude was assigned via force-photometry. The difference between the points and the model lines can be accounted for by the varying age, stellar mass and SFH compared to the single model shown here, which should be taken as indicative.
We shall use upper limits during model fitting for these non-detections in a forthcoming paper exploiting the whole DES database. 

\subsubsection{Age Limits in Photometric Fittings}
\label{sec:AgeLimPhotFits}

\begin{figure}
	\centering
	\includegraphics[width=\columnwidth]{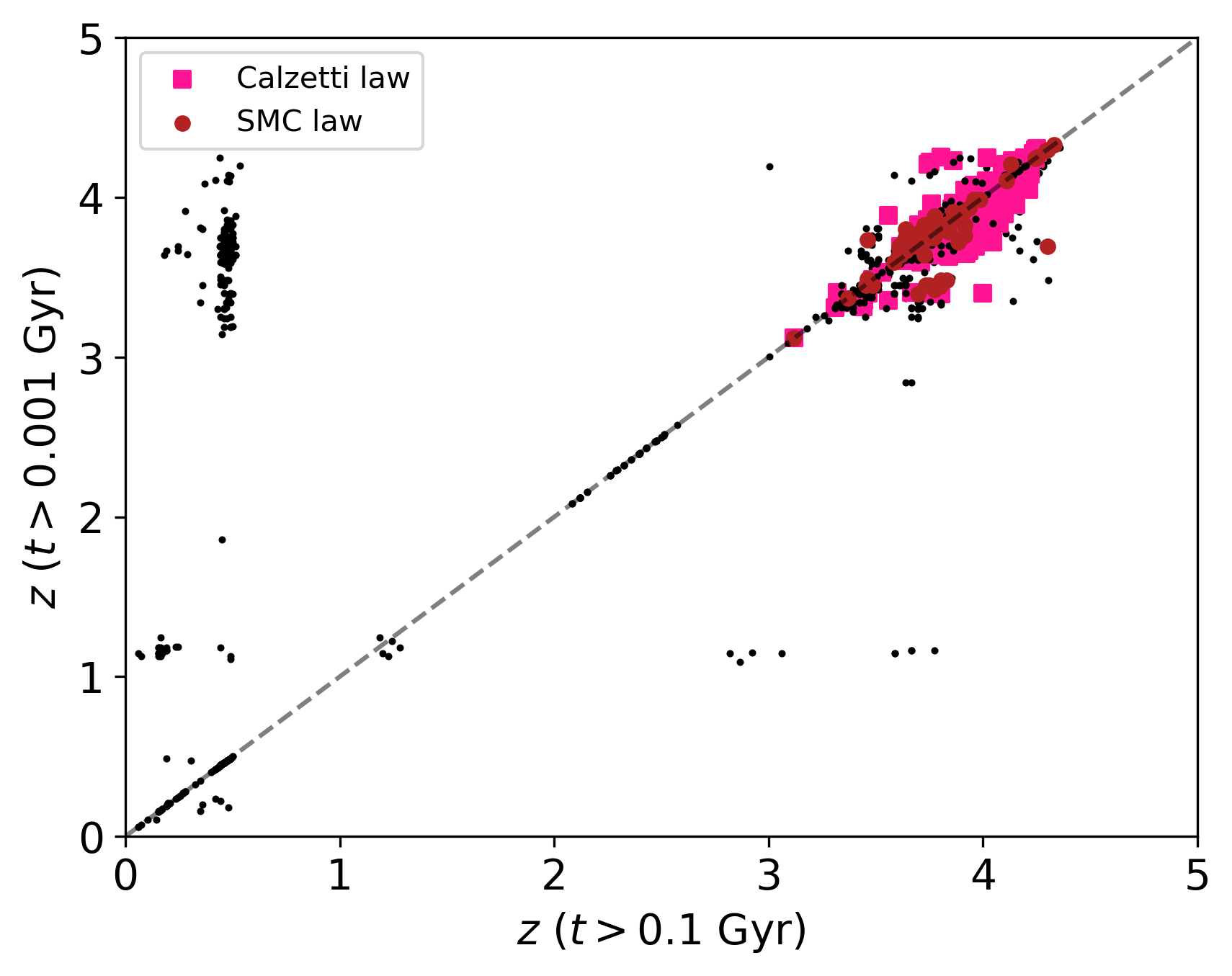}
	\includegraphics[width=\columnwidth]{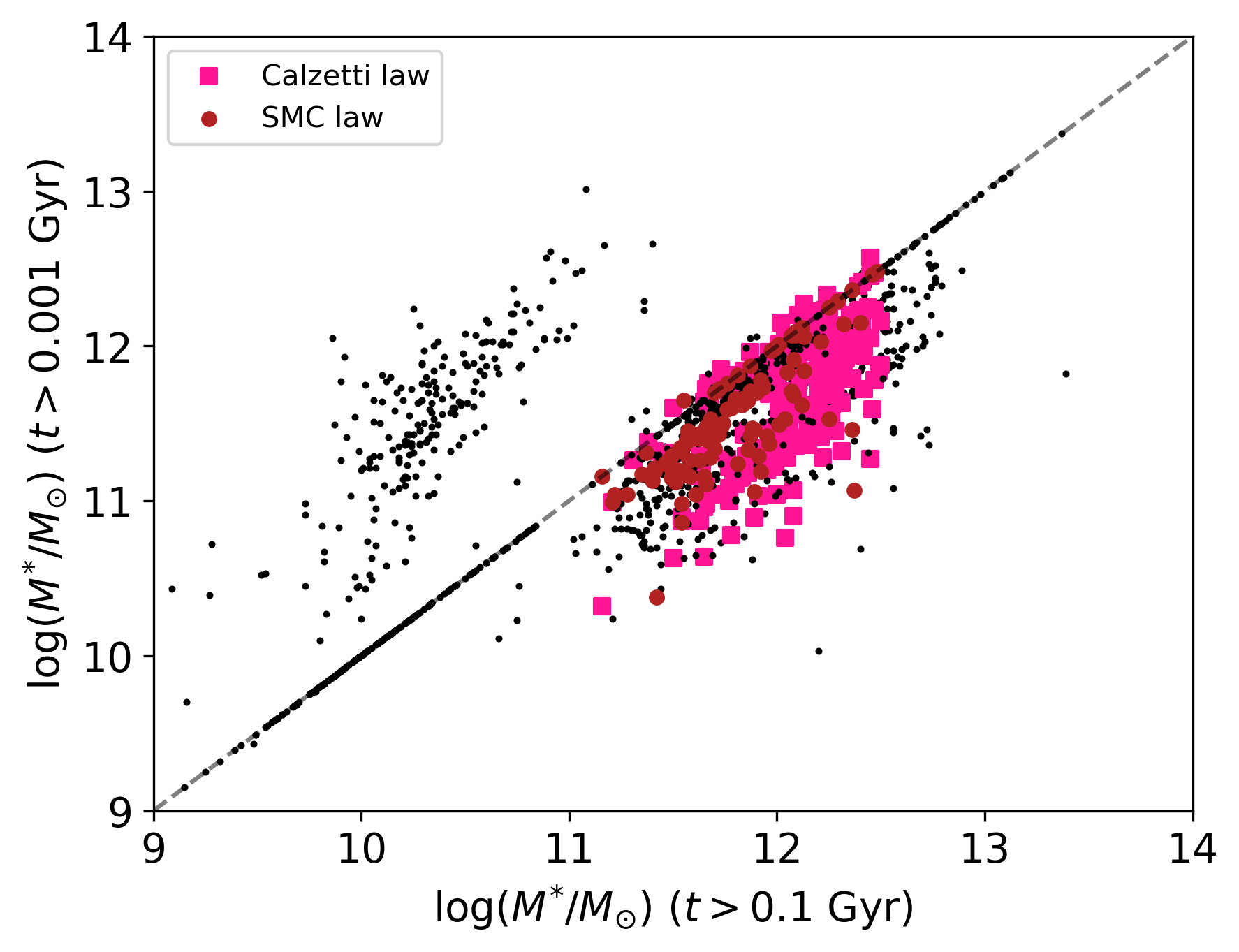}
    \caption{Effect of the minimum age allowed in the fitting (0.1 Gyr vs 1 Myr) on photometric redshifts (top panel) and stellar masses (bottom panel) of the 606 $z \sim 4$ selection box high-$z$~candidates (black dots). Best candidates are plotted as dark red circles and deep pink squares depending on the reddening option used in the fitting (SMC law and Calzetti law, respectively). Even though redshift and masses of objects may change, our robust candidates are relatively unaffected.}
    \label{fig:zANDmassAgeConANDnoAgeCon}
\end{figure}

In our model fitting we set a minimum age of 0.1 Gyr, which is larger than the minimum M05 model age (1 Myr). The setting of this limit is common practice in the literature \citep[e.g.][]{Daddi_etal2005, Maraston_etal2010} in order to contain the so-called `over-shining' effect \citep{Maraston_etal2010} by which low levels of star formation generate luminosities that are so high that any older populations remain undetected. This effect leads to very low fitted ages, which translates into huge star formation rates and a general underestimation of the stellar mass \citep[see discussion in][]{Maraston_etal2010}.
It should be noted that the over-shining effect is a threat mainly for a robust calculation of $M^{*}$~and not for the photometric redshift, which is regulated by the SED shape. This is exactly what we find when we compare the photometric redshift and stellar masses we derive from a fitting run without an age limit (Figure~\ref{fig:zANDmassAgeConANDnoAgeCon}) to our results. The plot includes results for all the 606 sources initially selected from the colour selection maps, as black dots, and for our best candidates, as larger symbols. The test shows that changing the age limit has no effect on the derived photometric redshifts (top panel) and little influence on $M^{*}$~($\sim-0.35$ dex on average; bottom panel) as far as our best candidates are concerned.

\section{Comparison With the Literature}
\label{sec:FittingStraatman14}

In Section~\ref{sec:StraatmanReFit} we re-fit data from \cite{Straatman_etal2014} to assess whether our fitting procedure recover their results. Moreover, we compare the results for our best candidates with recent literature in Section~\ref{sec:OtherLiteratureComparison}.
\subsection{Re-fitting Past Literature}
\label{sec:StraatmanReFit}

As is well known \citep[e.g.][]{Maraston_2005, Maraston_etal2006, Pforr_etal2012} the results of spectro-photometric model fitting to data depend on the adopted stellar population model, the fitting setup, the fitting code, and any assumptions made (e.g. the IMF, the fitting age grid, etc.). In order to compare our results with similar ones obtained in the literature, it is useful to also adopt our fitting framework on literature data. Here we present results for the galaxies from \cite{Straatman_etal2014}. One of these objects (ZF-COSMOS-13172) is the spectroscopically confirmed $z=3.717$ galaxy published in \cite{Glazebrook_etal2017}\footnote{This object is now named ZF-COSMOS-20115 in \cite{Glazebrook_etal2017}.}. Results are summarised in Table~\ref{tab:Straatman14Fitting}, which also reports the literature values. 

\renewcommand{\arraystretch}{2}
\begin{table*}
	\caption{Properties of the \citet{Straatman_etal2014}, `S14', sample of {\it quiescent} (according to their own classification) galaxies compared to the values we obtain (`us') using our fitting setup and those bands matching our DES+VHS photometry. For each object we obtain two results according to the assumed reddening option, whose $\chi_{r}^{2}$ is given in the last column. Note that the \citet{Straatman_etal2014} stellar masses refer to a Chabrier IMF, while ours to a Salpeter IMF. The latter are $\sim 0.2$~dex larger, hence the \citet{Straatman_etal2014} values should be increased by +0.2 dex to ensure a meaningful comparison with our derived values.}
	\centering
	\resizebox{\linewidth}{!}{%
	\begin{tabular}{ ccccccccc }
		\hline
		\textbf{ID} & \begin{tabular}{@{}c@{}}\textbf{$z_{phot}$} \\ \textbf{(S14)}\end{tabular} & \begin{tabular}{@{}c@{}}\textbf{$z_{phot}$} \\ \textbf{(us)}\end{tabular} & \begin{tabular}{@{}c@{}}\textbf{$\log_{10}(M^{*}/M_{\odot}$)} \\ \textbf{(S14)}\end{tabular} & \begin{tabular}{@{}c@{}}\textbf{$\log_{10}(M^{*}/M_{\odot}$)} \\ \textbf{(us)}\end{tabular} & \begin{tabular}{@{}c@{}}\textbf{Age [Gyr]} \\ \textbf{(S14)}\end{tabular} & \begin{tabular}{@{}c@{}}\textbf{Age [Gyr]} \\ \textbf{(us)}\end{tabular} & \begin{tabular}{@{}c@{}}\textbf{Reddening} \\ \textbf{(us)}\end{tabular} & \begin{tabular}{@{}c@{}}\textbf{$\chi_{r}^{2}$} \\ \textbf{(us)}\end{tabular} \\
		\hline
		ZF-COSMOS-13129 & $3.81 \pm 0.17$ & & 11.25 & & 1.58 & & & \\
		 & & 4.65 & & 11.39 & & 0.64 & SMC & 0.000 \\
		 & & 5.28 & & 12.27 & & 0.51 & Calzetti & 0.000 \\
		\hline
		\hline
		ZF-COSMOS-13172 & $3.55 \pm 0.06$ & & 11.16 & & 0.79 & & &  \\
		& & 2.88 & & 11.85 & & 0.20 & SMC & 1.693 \\
		& & 4.13 & & 11.82 & & 1.02 & Calzetti & 4.016 \\
		\hline
		\hline
		ZF-COSMOS-13414 & $3.57 \pm 0.19$ & & 10.64 & & 1.00 & & & \\
		& & 3.84 & & 10.77 & & 0.64 & SMC & 4.103 \\
		& & 3.80 & & 11.47 & & 1.28 & Calzetti & 0.702 \\
		\hline
		\hline
		ZF-CDFS-5657 & $3.56 \pm 0.07$ & & 10.88 & & 1.26 & & & \\
		& & 2.71 & & 10.50 & & 0.51 & SMC & 3.786 \\
		& & 2.81 & & 10.76 & & 0.51 & Calzetti & 3.489 \\
		\hline
		\hline
		ZF-CDFS-403 & 3.660 ($z_{spec}$) & & 11.06 & & 0.79 & & & \\
		& & 1.76 & & 10.62 & & 1.80 & SMC & 0.034 \\
		& & 2.42 & & 10.76 & & 0.14 & Calzetti & 0.000 \\
		\hline
		\hline
		ZF-CDFS-209 & $3.56 \pm 0.05$ & & 10.88 & & 0.63 & & & \\
		& & 2.01 & & 10.83 & & 2.60 & SMC & 0.054 \\
		& & 2.05 & & 10.53 & & 0.29 & Calzetti & 0.005 \\
		\hline
		\hline
		ZF-CDFS-4907 & $3.46 \pm 0.16$ & & 10.60 & & 0.40 & & & \\
		& & 2.89 & & 10.31 & & 0.32 & SMC & 1.297 \\
		& & 2.89 & & 10.31 & & 0.32 & Calzetti & 1.297 \\
		\hline
		\hline
		ZF-CDFS-4719 & $3.59 \pm 0.14$ & & 10.65 & & 1.00 & & & \\
		& & 2.92 & & 10.84 & & 1.68 & SMC & 0.093 \\
		& & 2.83 & & 10.90 & & 1.43 & Calzetti & 0.091 \\
		\hline
		\hline
		ZF-UDS-885 & $3.99 \pm 0.41$ & & 10.78 & & 0.40 & & & \\
		& & 5.42 & & 11.83 & & 0.81 & SMC & 2.411 \\
		& & 3.95 & & 12.02 & & 1.14 & Calzetti & 0.410 \\
		\hline
		\hline
		ZF-UDS-1236 & $3.58 \pm 0.08$ & & 10.78 & & 0.50 & & & \\
		& & 4.27 & & 11.41 & & 0.64 & SMC & 8.198 \\
		& & 3.45 & & 11.30 & & 0.51 & Calzetti & 6.397 \\
		\hline
	\end{tabular}}
	\label{tab:Straatman14Fitting}
\end{table*}
\renewcommand{\arraystretch}{1}
\renewcommand{\arraystretch}{2}
\begin{table*}
	\contcaption{}
	\centering
	\resizebox{\linewidth}{!}{%
	\begin{tabular}{ ccccccccc }
		\hline
		\textbf{ID} & \begin{tabular}{@{}c@{}}\textbf{$z_{phot}$} \\ \textbf{(S14)}\end{tabular} & \begin{tabular}{@{}c@{}}\textbf{$z_{phot}$} \\ \textbf{(us)}\end{tabular} & \begin{tabular}{@{}c@{}}\textbf{$\log_{10}(M^{*}/M_{\odot}$)} \\ \textbf{(S14)}\end{tabular} & \begin{tabular}{@{}c@{}}\textbf{$\log_{10}(M^{*}/M_{\odot}$)} \\ \textbf{(us)}\end{tabular} & \begin{tabular}{@{}c@{}}\textbf{Age [Gyr]} \\ \textbf{(S14)}\end{tabular} & \begin{tabular}{@{}c@{}}\textbf{Age [Gyr]} \\ \textbf{(us)}\end{tabular} & \begin{tabular}{@{}c@{}}\textbf{Reddening} \\ \textbf{(us)}\end{tabular} & \begin{tabular}{@{}c@{}}\textbf{$\chi_{r}^{2}$} \\ \textbf{(us)}\end{tabular} \\
		\hline
		ZF-UDS-2622 & $3.77 \pm 0.10$ & & 10.94 & & 0.63 & & & \\
		& & 3.81 & & 11.00 & & 0.72 & SMC & 1.611 \\
		& & 3.87 & & 11.41 & & 0.90 & Calzetti & 0.962 \\
		\hline
		\hline
		ZF-UDS-3112 & $3.53 \pm 0.06$ & & 10.63 & & 1.26 & & & \\
		& & 3.99 & & 10.95 & & 0.57 & SMC & 6.044 \\
		& & 3.99 & & 10.95 & & 0.57 & Calzetti & 6.044 \\
		\hline
		\hline
		ZF-UDS-5418 & $3.53 \pm 0.07$ & & 10.64 & & 0.79 & & & \\
		& & 3.37 & & 11.00 & & 1.61 & SMC & 4.306 \\
		& & 3.22 & & 10.91 & & 1.28 & Calzetti & 2.806 \\
		\hline
		\hline
		ZF-UDS-6119 & $4.05 \pm 0.27$ & & 10.74 & & 0.50 & & & \\
		& & 3.91 & & 10.84 & & 1.02 & SMC & 2.040 \\
		& & 4.13 & & 11.48 & & 1.02 & Calzetti & 1.368 \\
		\hline
		\hline
		ZF-UDS-9526 & $3.97 \pm 0.18$ & & 10.95 & & 0.63 & & & \\
		& & 2.67 & & 10.33 & & 1.43 & SMC & 0.128 \\
		& & 2.67 & & 10.33 & & 1.43 & Calzetti & 0.128 \\
		\hline
		\hline
		ZF-UDS-10401 & $3.91 \pm 0.38$ & & 10.58 & & 0.25 & & & \\
		& & 3.96 & & 10.62 & & 0.72 & SMC & 0.013 \\
		& & 3.25 & & 10.64 & & 0.57 & Calzetti & 0.000 \\
		\hline
		\hline
		ZF-UDS-10684 & $3.95 \pm 0.48$ & & 10.93 & & 1.26 & & & \\
		& & 3.78 & & 10.51 & & 0.64 & SMC & 4.681 \\
		& & 3.67 & & 10.90 & & 1.61 & Calzetti & 4.526 \\
		\hline
		\hline
		ZF-UDS-11483 & $3.63 \pm 0.32$ & & 11.01 & & 1.00 & & & \\
		& & 1.20 & & 10.25 & & 2.10 & SMC & 0.000 \\
		& & 1.80 & & 10.45 & & 1.80 & Calzetti & 0.000 \\
		\hline
	\end{tabular}}
	\label{tab:cont:Straatman14Fitting}
\end{table*}
\renewcommand{\arraystretch}{1}

It is important to note that the \citet{Straatman_etal2014} results have been obtained using 36-band photometry (with wavelength range of $3000-80000~\AA$) while our DES+VHS combination grants us 8 bands at most. Hence, in order to mimic our fitting, we selected the same bands from the \cite{Straatman_etal2014} database, and therefore the results of Table~\ref{tab:Straatman14Fitting} refer to fits with these bands.

We find results that are similar in both photometric redshift and mass. Note that the $\chi_{r}^{2}$~for most results lie slightly above our adopted cut of $\chi_{r}^{2}<3$ (on the other hand, one object has $\chi_{r}^{2} = 0$ as the available photometry is made up of 3 bands out of the 8 we considered and they align with the best fitted model). Overall, however, our method with only a few bands allows us to obtain a similar picture of galaxy evolution at high redshift.

Most interestingly, for the \cite{Glazebrook_etal2017} object we obtain a best solution which includes dust and a younger age than the one reported in their paper. Our result is consistent with \cite{Simpson_etal2017}, who find - using ALMA and SCUBA data - dust detection in the same source, thereby questioning the quiescent nature of that galaxy. On the other hand, \cite{Schreiber_etal2017} show that only 3.1 kpc away from that galaxy there is a massive, extremely obscured galaxy that they identify as the origin of the sub-mm emission observed by \cite{Simpson_etal2017}. In this paper, we have not applied any particular prior on the nature of the galaxy before fitting and we provide fitting solutions with and without dust. These will be useful to potentially perform ALMA follow-ups of our candidates.

\subsection{Further Comparisons with the Literature}
\label{sec:OtherLiteratureComparison}

\begin{figure*}
	\centering
	\includegraphics[width=\textwidth]{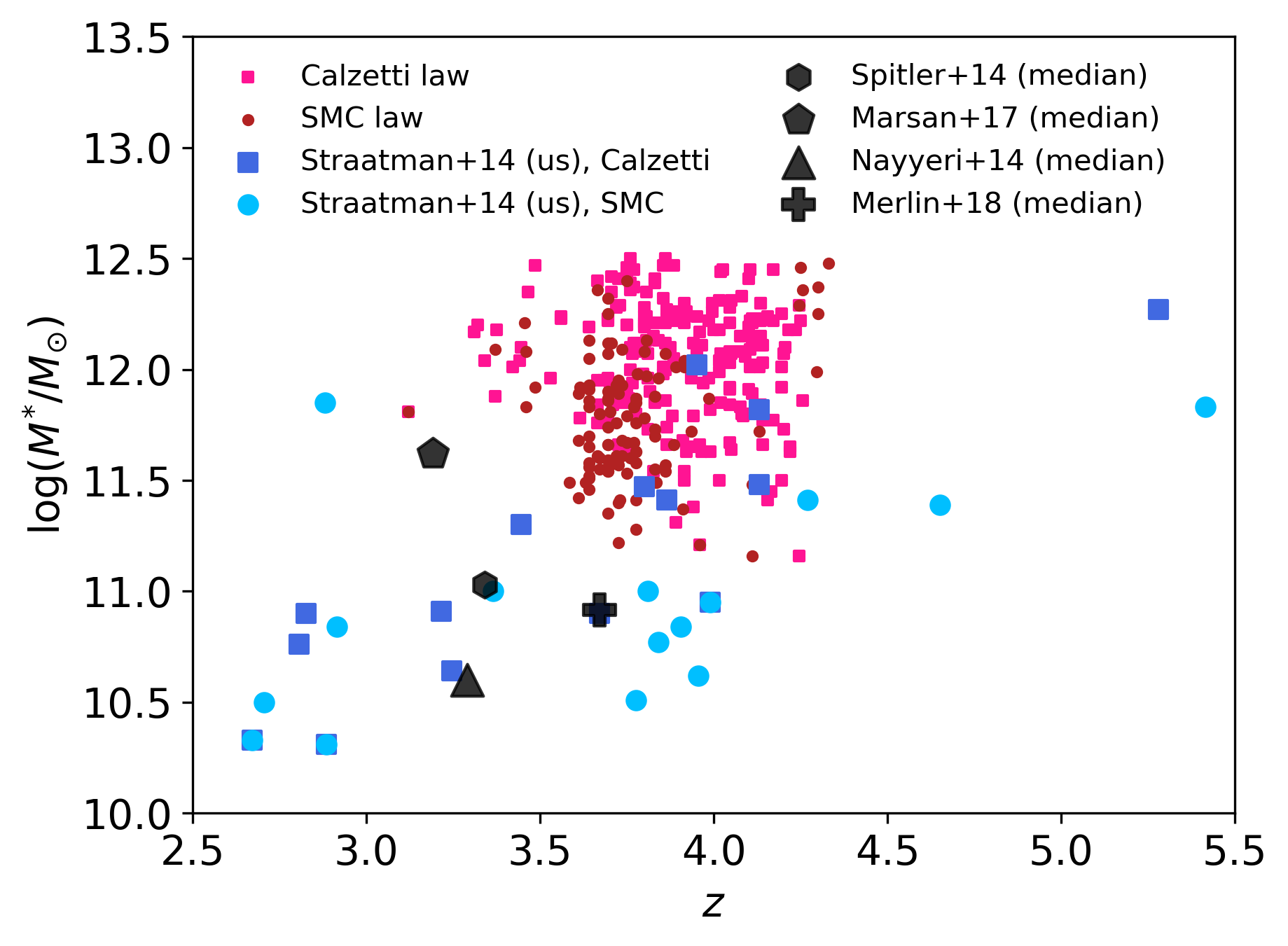}
    \caption{Stellar mass vs redshift for our candidates (large symbols for various reddening laws) compared to the median value for 57 galaxies by \citet{Spitler_etal2014} (black hexagon), 7 galaxies by \citet{Marsan_etal2017} (black pentagon), 16 galaxies by \citet{Nayyeri_etal2014} (black triangle), 10 galaxies by \citet{Merlin_etal2018} (black plus), and to the 18 values we obtain for the galaxies by \citet{Straatman_etal2014} with our fitting set-up and their data (blue symbols labelled as \citet{Straatman_etal2014}).} 
    \label{fig:SpitlerVSus}
\end{figure*}

In Figure~\ref{fig:SpitlerVSus} we show the median values of stellar mass and redshift of the 57 galaxies from \citet{Spitler_etal2014}, the 7 galaxies by \citet{Marsan_etal2017}, the 16 galaxies by \citet{Nayyeri_etal2014}, and the 10 galaxies by \citet{Merlin_etal2018}, as a black hexagon, pentagon, triangle and a plus, respectively. The latter two studies focus on what they call `passive galaxies', while we do not take any particular prior here. For comparison we also plot our best candidates (with the two usual shape and colour schemes for SMC and Calzetti laws) as well as the \cite{Straatman_etal2014} sources as re-fitted by us (as light blue circles and dark blue squares for SMC law and Calzetti law, respectively).

It can be appreciated that our candidates are of typically higher redshifts and stellar masses than these previous studies and expand the envelope of results for high-$z$ massive galaxies. 

\section{Comparison With Model Galaxies}
\label{sec:CompGalModels}
\begin{figure}
	\centering
	\includegraphics[width=\columnwidth]{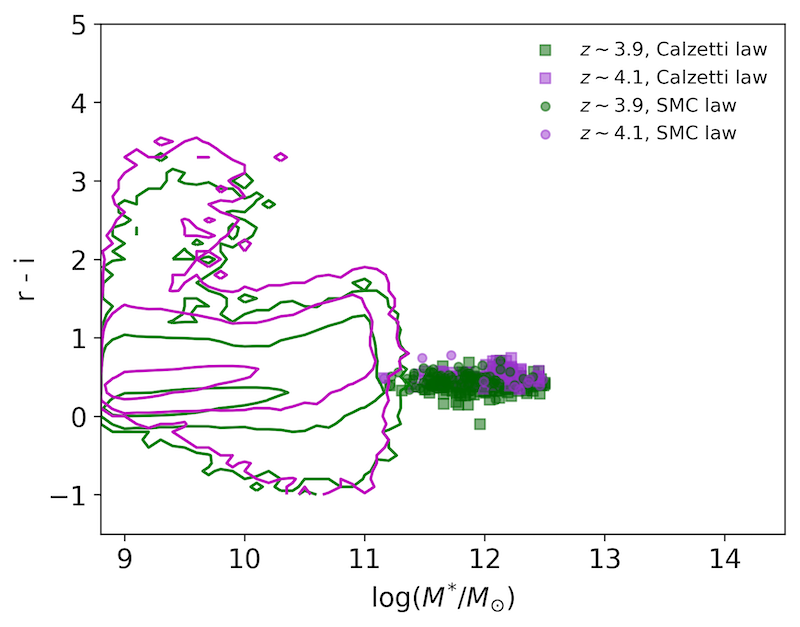}
    \caption{Observed-frame $(r-i)$ versus stellar mass for our best candidate galaxies, compared to galaxy distributions at $z \sim 3.9$ and $z \sim 4.1$ (colour-code in green and magenta respectively) from the \citet{GP14} model, assuming the stellar population synthesis model from \citet{Maraston_2005}. The model galaxies are shown as contours with the outermost one outlining the end of such a distribution, while the best candidates are plotted as scatter plot. Both observed and model galaxies include the effect of dust attenuation. The model stellar masses have been corrected to a Salpeter IMF \citep{Lagos_etal14, Lacey_etal2016} for consistency with our fitting assumptions. No model galaxies are found with masses above $\sim10^{11.5}$ at these redshifts.} 
    \label{fig:sam}
\end{figure}

We use the Millennium Simulation Database\footnote{\href{url}{http://virgo.dur.ac.uk}} (volume: $(500 Mpc/h)^{3}$) and extract the number density of galaxies with stellar masses above $\sim10^{11}$ from different semi-analytical models, at the snapshots corresponding to $z = 3.9$ and $z = 4.1$. We find co-moving densities of $1.6 \times 10^{-5}$Mpc$^{-3}$ for the model described in \cite{GP14}, $5.2 \times 10^{-7}$Mpc$^{-3}$ for model galaxies from~\cite{Henriques_etal2015} and $5 \times 10^{-6}$Mpc$^{-3}$ for the model described in \cite{Lacey_etal2016}. These densities compare to a value of $5 \times 10^{-5}$Mpc$^{-3}$ for \citet{DeLucia_etal2006}; it is important to note that large density variations for the massive end are found in the literature \citep[e.g.][]{Knebe+2018}.

At $z= 3.9$, no galaxies from the \cite{GP14} or \cite{Lacey_etal2016}  models have a mass above $10^{11.5}$. The same is true for the model described in \citet{Croton16}\footnote{\href{url}{https://tao.asvo.org.au/tao/}}. This lack of galaxies with masses above $10^{11.5}M_{\odot}$ at high redshift is shown in Figure~\ref{fig:sam} for \citet{GP14}, while most of our high-redshift candidates have masses above this limit.

Some galaxies from the \cite{Henriques_etal2015} model are indeed expected to have masses above $10^{11.5}$ at $z= 3.9$, but they are rare, with a density of $7.3 \times 10^{-9}$Mpc$^{-3}$. This last model was designed to match the evolution of the stellar mass function up to $z=3$ mainly by allowing the reincorporation timescale of wind ejecta to vary with cosmic time. The other two semi-analytical models do not reproduce the observed evolution of the galaxy stellar mass function above $z\sim 1$, predicting a too rapidly evolving high mass end. Moreover, models assuming warm dark matter produce more massive galaxies at high redshifts than standard cold dark matter cosmologies \citep{Wang_etal2016}. Thus, redshift and physical properties of galaxies in place above $z=2$ provide crucial information to help constrain our understanding of galaxy evolution within a cosmological context. We shall compare number densities in models and data when the DES observations are complete.

In this paper we take a first glance at the physical properties - namely, colours vs.~stellar mass - of massive galaxies in simulations compared to our best candidates. Figure~\ref{fig:sam} presents the comparison for the \citet{GP14} model run using the version based on the stellar population synthesis model from \citet{Maraston_2005}. This was chosen as, currently, it is the only model of the five explored that provides magnitudes for the DES bands. The observed-frame $r-i$ colour of the data are consistent with a subset of model galaxies at the appropriate redshift, possibly suggesting compatible star formation histories. The model galaxies have, however, too low a stellar mass with respect to those inferred from the observational data.

As stated above, very few massive model galaxies are found above $z=3.9$. Thus, these galaxies have the potential to change the way we understand galaxy formation, once their redshifts are spectroscopically confirmed and data for the whole DES survey are analysed. 

\section{Conclusions and Discussion}
\label{sec:Conclusions}
We have used data from the first three years of Dark Energy Survey (DES) operations together with the Science Verification (SV) catalogue, to probe the existence of high-redshift ($z\sim4$), massive ($M\sim10^{12} M_{\odot}$) galaxies down to the nominal full survey depth, and to study their nature. This analysis was motivated by our previous forecast (D13) according to which DES would be one of the very few surveys (the only one, at time of writing) that has the right combination of area (5000 deg.~sq.~at completion) and depth ($i = 24.3$) to allow for these rare objects to be detected, should they exist. In turn, these galaxies are key probes of the galaxy formation process within a cosmological framework, as they form from the largest primordial density fluctuations. Their number and characterisation vary substantially between cosmological simulations, and extrapolations of local and intermediate-redshift mass functions predict a large variation in the number density of these objects (see D13, Figure 1). In a similar fashion, DES data have been recently used to search for high-redshift quasars \citep{Banerji_etal2015b, Reed_etal2017}. 

We have applied a rigorous identification method, using theoretical maps as well as visual inspection of $\sim 600$~potential sources, along with data cleaning (artefacts, error cuts, and star/galaxy separation). We have then performed spectro-photometric template fitting, using a large variety of models and reddening assumptions, to minimise potential biases in the determination of photometric redshifts and galaxy properties. Critical to the model fitting, we extended our DES data $g,r,i,z,Y$~baseline with near-IR $J,H,K$~data from the VHS survey. After model fitting, we further applied other quality cuts in order to extract the most promising sample of high-redshift, massive galaxies, namely: goodness of fit ($\chi_{r}^{2}<3$); unimodal redshift probability distribution function showing a clear high-$z$ solution; physical mass.

Our final result is a sample of 233 galaxies that are $z\sim4$ candidates, with large stellar masses ($10^{11.5} \leq M/M_{\odot} \leq 10^{12.5}$) and ages of $\sim 0.1$ Gyr, some of which are on the verge of becoming passive. Their average star formation rates are of the order of thousands of solar masses per year and place them among the most extreme objects found so far (as in \cite{Rowan-Robinson_etal2016}). This number of objects should be regarded as a lower limit due to our very conservative selection. Our best candidates, when paired with our PSF comparison consideration, constitutes an excellent sample for spectroscopic and ALMA follow-ups.

The properties of our objects make them the most likely progenitors of the most massive elliptical galaxies studied in the local universe \citep{Thomas_etal2005, Thomas_etal2010}. The existence of such a class of mature, massive galaxies in the early universe pose unprecedented constraints to our understanding of galaxy formation and evolution in a cosmological context. An initial test of galaxy formation simulations reveals that galaxies with such large masses are absent in cold dark matter-based cosmologies.

Our work extends the hunt for massive galaxies at $z \sim 4$ by almost an order of magnitude in stellar mass, following the studies performed, for instance, in \citet{Straatman_etal2014}, \citet{Spitler_etal2014}, \citet{Marsan_etal2017}, \citet{Nayyeri_etal2014}, and \citet{Merlin_etal2018}. Our tests on the \citet{Straatman_etal2014} data also confirm that even though we use a limited amount of photometric bands for the fitting, we obtain results that are similar to those obtained with 36-band photometry or even spectroscopy. This is because in order to constrain a model matching galaxy spectra what is important is the wavelength baseline covered by the data rather than the density of bands over a narrower wavelength span.

We show we are able to recover the properties of one object with a spectroscopic redshift of 3.717 reported by \citep{Glazebrook_etal2017}, and in particular we provide solutions including the possible presence of dust. 

One is then tempted to push our method to even higher redshifts.
Photometric techniques similar to ours have also been used in studies based on the Hubble Frontier Fields (HFF) programme \citep[e.g.][]{Laporte_etal2016, Laporte_etal2016_inprep}. HFF is a survey aimed at imaging six massive clusters at moderate redshift in order to observe very high-redshift ($z > 7$) galaxies (or proto-galaxies) that are visible because they have been lensed by the foreground clusters. Studies include the discovery of several tens of Lyman-break galaxies at redshifts $z \sim 7-11$ \citep[e.g.][]{Zheng_etal2014, Jauzac_etal2015, Coe_etal2015}. 

Let us finally comment on the number counts of these extreme systems. D13 calculated the predicted galaxy number counts for a variety of approaches, including fully theoretical cosmological simulations (as per the Millennium Simulation), semi-empirical predictions obtained by passively evolving the $z = 0$ and $0.5 < z < 0.7$ published mass functions to higher redshift, available high-$z$~mass functions \citep{Marchesini_etal2010}, and also results from abundance-matching techniques \citep{Behroozi_etal2013}. For example, empirical mass functions do forecast the presence of $\sim7000$ galaxies with $\log(M^{*}/M_{\odot}) \geq 12.0$ within the whole DES. Of these, $\sim 100$ will represent true $\log(M^{*}/M_{\odot}) \geq 12.0$ systems, with the rest being misidentified due to Eddington bias \citep[e.g.][]{Maraston_etal2013}. In summary, the predicted number counts of D13 (Figure 1) are diverse and, at completion, a survey such as DES will help discriminate among these possibilities at the massive end of the stellar mass function.
As the SV-sized data we use here include only $\sim150$ deg.~sq and because we calculated photometric redshifts that will need to be confirmed spectroscopically, we refrain from probing the precise number counts from D13.

Finally, this manuscript outlines a new method that only uses optical bands (in the context of DES) to identify likely massive high-$z$ galaxies. This will work as a foundation to probe such galaxies using larger and larger photometric datasets that will be completed in the future (such as EUCLID, LSST, etc.).

\section*{Acknowledgements}
We would like to thank the MNRAS anonymous referee for his/her very useful and competent report, and Micol Bolzonella for a much appreciated support with HyperZ.
This paper has also gone through internal review by the DES collaboration and we would like to thank our internal referees Karl Glazebrook and Richard McMahon for their comments which greatly improved this manuscript. We are also grateful to the DES final reader Alex Kim who gave us very useful editorial suggestions. Thanks also go to Will Hartley and Manda Banerji for providing comments, to Brian Yanny for instructing us on how to get blending information for each source, to Sophie Reed for her suggestions regarding potential mismatch between DES and VHS data, to Carlotta Gruppioni for the discussions regarding potential AGN contamination, to Benjamin Mawdsley for discussions regarding the toy model simulations predicting spurious low-z objects scattering in the selection box, and to Daniel Goddard for important stylistic plotting suggestions. We also acknowledge discussions with Alvio Renzini, Emanuele Daddi, David Elbaz, and Johan Comparat. 

Funding for the DES Projects has been provided by the U.S. Department of Energy, the U.S. National Science Foundation, the Ministry of Science and Education of Spain, 
the Science and Technology Facilities Council of the United Kingdom, the Higher Education Funding Council for England, the National Center for Supercomputing 
Applications at the University of Illinois at Urbana-Champaign, the Kavli Institute of Cosmological Physics at the University of Chicago, 
the Center for Cosmology and Astro-Particle Physics at the Ohio State University,
the Mitchell Institute for Fundamental Physics and Astronomy at Texas A\&M University, Financiadora de Estudos e Projetos, 
Funda{\c c}{\~a}o Carlos Chagas Filho de Amparo {\`a} Pesquisa do Estado do Rio de Janeiro, Conselho Nacional de Desenvolvimento Cient{\'i}fico e Tecnol{\'o}gico and 
the Minist{\'e}rio da Ci{\^e}ncia, Tecnologia e Inova{\c c}{\~a}o, the Deutsche Forschungsgemeinschaft and the Collaborating Institutions in the Dark Energy Survey. 

The Collaborating Institutions are Argonne National Laboratory, the University of California at Santa Cruz, the University of Cambridge, Centro de Investigaciones Energ{\'e}ticas, 
Medioambientales y Tecnol{\'o}gicas-Madrid, the University of Chicago, University College London, the DES-Brazil Consortium, the University of Edinburgh, 
the Eidgen{\"o}ssische Technische Hochschule (ETH) Z{\"u}rich, 
Fermi National Accelerator Laboratory, the University of Illinois at Urbana-Champaign, the Institut de Ci{\`e}ncies de l'Espai (IEEC/CSIC), 
the Institut de F{\'i}sica d'Altes Energies, Lawrence Berkeley National Laboratory, the Ludwig-Maximilians Universit{\"a}t M{\"u}nchen and the associated Excellence Cluster Universe, 
the University of Michigan, the National Optical Astronomy Observatory, the University of Nottingham, The Ohio State University, the University of Pennsylvania, the University of Portsmouth, 
SLAC National Accelerator Laboratory, Stanford University, the University of Sussex, Texas A\&M University, and the OzDES Membership Consortium.

Based in part on observations at Cerro Tololo Inter-American Observatory, National Optical Astronomy Observatory, which is operated by the Association of 
Universities for Research in Astronomy (AURA) under a cooperative agreement with the National Science Foundation.

The DES data management system is supported by the National Science Foundation under Grant Numbers AST-1138766 and AST-1536171.
The DES participants from Spanish institutions are partially supported by MINECO under grants AYA2015-71825, ESP2015-66861, FPA2015-68048, SEV-2016-0588, SEV-2016-0597, and MDM-2015-0509, 
some of which include ERDF funds from the European Union. IFAE is partially funded by the CERCA program of the Generalitat de Catalunya.
Research leading to these results has received funding from the European Research
Council under the European Union's Seventh Framework Program (FP7/2007-2013) including ERC grant agreements 240672, 291329, and 306478.
We  acknowledge support from the Australian Research Council Centre of Excellence for All-sky Astrophysics (CAASTRO), through project number CE110001020, and the Brazilian Instituto Nacional de Ci\^encia
e Tecnologia (INCT) e-Universe (CNPq grant 465376/2014-2).

This manuscript has been authored by Fermi Research Alliance, LLC under Contract No. DE-AC02-07CH11359 with the U.S. Department of Energy, Office of Science, Office of High Energy Physics. The United States Government retains and the publisher, by accepting the article for publication, acknowledges that the United States Government retains a non-exclusive, paid-up, irrevocable, world-wide license to publish or reproduce the published form of this manuscript, or allow others to do so, for United States Government purposes.




\bibliographystyle{mnras}
\bibliography{DESpaper} 

\begin{thebibliography}{}
\makeatletter
\relax
\def\mn@urlcharsother{\let\do\@makeother \do\$\do\&\do\#\do\^\do\_\do\%\do\~}
\def\mn@doi{\begingroup\mn@urlcharsother \@ifnextchar [ {\mn@doi@}
  {\mn@doi@[]}}
\def\mn@doi@[#1]#2{\def\@tempa{#1}\ifx\@tempa\@empty \href
  {http://dx.doi.org/#2} {doi:#2}\else \href {http://dx.doi.org/#2} {#1}\fi
  \endgroup}
\def\mn@eprint#1#2{\mn@eprint@#1:#2::\@nil}
\def\mn@eprint@arXiv#1{\href {http://arxiv.org/abs/#1} {{\tt arXiv:#1}}}
\def\mn@eprint@dblp#1{\href {http://dblp.uni-trier.de/rec/bibtex/#1.xml}
  {dblp:#1}}
\def\mn@eprint@#1:#2:#3:#4\@nil{\def\@tempa {#1}\def\@tempb {#2}\def\@tempc
  {#3}\ifx \@tempc \@empty \let \@tempc \@tempb \let \@tempb \@tempa \fi \ifx
  \@tempb \@empty \def\@tempb {arXiv}\fi \@ifundefined
  {mn@eprint@\@tempb}{\@tempb:\@tempc}{\expandafter \expandafter \csname
  mn@eprint@\@tempb\endcsname \expandafter{\@tempc}}}

\bibitem[\protect\citeauthoryear{{Abbott} et~al.}{{Abbott}
  et~al.}{2018}]{Abbott+2018}
{Abbott} T.~M.~C.,  et~al., 2018, preprint, \href
  {http://adsabs.harvard.edu/abs/2018arXiv180103181A} {} (\mn@eprint {arXiv}
  {1801.03181})

\bibitem[\protect\citeauthoryear{{Banerji} et~al.}{{Banerji}
  et~al.}{2008}]{Banerji_etal2008}
{Banerji} M.,  et~al., 2008, \mn@doi [\mnras]
  {10.1111/j.1365-2966.2008.13095.x}, \href
  {http://adsabs.harvard.edu/abs/2008MNRAS.386.1219B} {386, 1219}

\bibitem[\protect\citeauthoryear{{Banerji} et~al.}{{Banerji}
  et~al.}{2015a}]{Banerji_etal2015}
{Banerji} M.,  et~al., 2015a, \mn@doi [\mnras] {10.1093/mnras/stu2261}, \href
  {http://adsabs.harvard.edu/abs/2015MNRAS.446.2523B} {446, 2523}

\bibitem[\protect\citeauthoryear{{Banerji} et~al.}{{Banerji}
  et~al.}{2015b}]{Banerji_etal2015b}
{Banerji} M.,  et~al., 2015b, \mn@doi [\mnras] {10.1093/mnras/stu2649}, \href
  {http://adsabs.harvard.edu/abs/2015MNRAS.447.3368B} {447, 3368}

\bibitem[\protect\citeauthoryear{{Behroozi} et~al.}{{Behroozi}
  et~al.}{2013}]{Behroozi_etal2013}
{Behroozi} P.~S.,  et~al., 2013, \mn@doi [\apj] {10.1088/0004-637X/770/1/57},
  \href {http://adsabs.harvard.edu/abs/2013ApJ...770...57B} {770, 57}

\bibitem[\protect\citeauthoryear{{Bolzonella}, {Miralles}  \&
  {Pell{\'o}}}{{Bolzonella} et~al.}{2000}]{HyperZ}
{Bolzonella} M.,  {Miralles} J.-M.,   {Pell{\'o}} R.,  2000, \aap, \href
  {http://adsabs.harvard.edu/abs/2000A%26A...363..476B} {363, 476}

\bibitem[\protect\citeauthoryear{{Bouchet} et~al.}{{Bouchet}
  et~al.}{1985}]{SMC_bouchet}
{Bouchet} P.,  et~al., 1985, \aap, \href
  {http://adsabs.harvard.edu/abs/1985A%26A...149..330B} {149, 330}

\bibitem[\protect\citeauthoryear{{Calzetti} et~al.}{{Calzetti}
  et~al.}{2000}]{CalzettiLaw}
{Calzetti} D.,  et~al., 2000, \mn@doi [\apj] {10.1086/308692}, \href
  {http://adsabs.harvard.edu/abs/2000ApJ...533..682C} {533, 682}

\bibitem[\protect\citeauthoryear{{Caputi} et~al.}{{Caputi}
  et~al.}{2012}]{Caputi_etal2012}
{Caputi} K.~I.,  et~al., 2012, \mn@doi [\apjl] {10.1088/2041-8205/750/1/L20},
  \href {http://adsabs.harvard.edu/abs/2012ApJ...750L..20C} {750, L20}

\bibitem[\protect\citeauthoryear{{Caputi} et~al.}{{Caputi}
  et~al.}{2015}]{Caputi_etal2015}
{Caputi} K.~I.,  et~al., 2015, \mn@doi [\apj] {10.1088/0004-637X/810/1/73},
  \href {http://adsabs.harvard.edu/abs/2015ApJ...810...73C} {810, 73}

\bibitem[\protect\citeauthoryear{{Cimatti} et~al.}{{Cimatti}
  et~al.}{2004}]{Cimatti_etal2004}
{Cimatti} A.,  et~al., 2004, \mn@doi [\nat] {10.1038/nature02668}, \href
  {http://adsabs.harvard.edu/abs/2004Natur.430..184C} {430, 184}

\bibitem[\protect\citeauthoryear{{Cimatti}, {Daddi}  \& {Renzini}}{{Cimatti}
  et~al.}{2006}]{CimattiDaddiRenzini2006}
{Cimatti} A.,  {Daddi} E.,   {Renzini} A.,  2006, \mn@doi [\aap]
  {10.1051/0004-6361:20065155}, \href
  {http://adsabs.harvard.edu/abs/2006A%26A...453L..29C} {453, L29}

\bibitem[\protect\citeauthoryear{{Coe}, {Bradley}  \& {Zitrin}}{{Coe}
  et~al.}{2015}]{Coe_etal2015}
{Coe} D.,  {Bradley} L.,   {Zitrin} A.,  2015, \mn@doi [\apj]
  {10.1088/0004-637X/800/2/84}, \href
  {http://adsabs.harvard.edu/abs/2015ApJ...800...84C} {800, 84}

\bibitem[\protect\citeauthoryear{{Conselice} et~al.}{{Conselice}
  et~al.}{2007}]{Conselice_etal2007}
{Conselice} C.~J.,  et~al., 2007, \mn@doi [\mnras]
  {10.1111/j.1365-2966.2007.12316.x}, \href
  {http://adsabs.harvard.edu/abs/2007MNRAS.381..962C} {381, 962}

\bibitem[\protect\citeauthoryear{{Cowie} \& {Barger}}{{Cowie} \&
  {Barger}}{2008}]{CoBa_2008}
{Cowie} L.~L.,  {Barger} A.~J.,  2008, \mn@doi [\apj] {10.1086/591176}, \href
  {http://adsabs.harvard.edu/abs/2008ApJ...686...72C} {686, 72}

\bibitem[\protect\citeauthoryear{{Cowie}, {Songaila}  \& {Barger}}{{Cowie}
  et~al.}{1999}]{CoSoBa_1999}
{Cowie} L.~L.,  {Songaila} A.,   {Barger} A.~J.,  1999, \mn@doi [\aj]
  {10.1086/300959}, \href {http://adsabs.harvard.edu/abs/1999AJ....118..603C}
  {118, 603}

\bibitem[\protect\citeauthoryear{{Cross} et~al.}{{Cross}
  et~al.}{2012}]{Cross+2012}
{Cross} N.~J.~G.,  et~al., 2012, \mn@doi [\aap] {10.1051/0004-6361/201219505},
  \href {http://adsabs.harvard.edu/abs/2012A%26A...548A.119C} {548, A119}

\bibitem[\protect\citeauthoryear{{Croton} et~al.}{{Croton}
  et~al.}{2016}]{Croton16}
{Croton} D.~J.,  et~al., 2016, \mn@doi [\apjs] {10.3847/0067-0049/222/2/22},
  \href {http://adsabs.harvard.edu/abs/2016ApJS..222...22C} {222, 22}

\bibitem[\protect\citeauthoryear{{Daddi} et~al.}{{Daddi}
  et~al.}{2005}]{Daddi_etal2005}
{Daddi} E.,  et~al., 2005, \mn@doi [\apj] {10.1086/430104}, \href
  {http://adsabs.harvard.edu/abs/2005ApJ...626..680D} {626, 680}

\bibitem[\protect\citeauthoryear{{Davies} et~al.}{{Davies}
  et~al.}{2013}]{Davies_etal_2013}
{Davies} L.~J.~M.,  et~al., 2013, \mn@doi [\mnras] {10.1093/mnras/stt1018},
  \href {http://adsabs.harvard.edu/abs/2013MNRAS.434..296D} {434, 296}

\bibitem[\protect\citeauthoryear{{De Lucia} et~al.}{{De Lucia}
  et~al.}{2006}]{DeLucia_etal2006}
{De Lucia} G.,  et~al., 2006, \mn@doi [\mnras]
  {10.1111/j.1365-2966.2005.09879.x}, \href
  {http://adsabs.harvard.edu/abs/2006MNRAS.366..499D} {366, 499}

\bibitem[\protect\citeauthoryear{{Douglas} et~al.}{{Douglas}
  et~al.}{2009}]{Douglas+2009}
{Douglas} L.~S.,  et~al., 2009, \mn@doi [\mnras]
  {10.1111/j.1365-2966.2009.15482.x}, \href
  {http://adsabs.harvard.edu/abs/2009MNRAS.400..561D} {400, 561}

\bibitem[\protect\citeauthoryear{{Etherington} et~al.}{{Etherington}
  et~al.}{2017}]{Etherington_etal2017}
{Etherington} J.,  et~al., 2017, \mn@doi [\mnras] {10.1093/mnras/stw3069},
  \href {http://adsabs.harvard.edu/abs/2017MNRAS.466..228E} {466, 228}

\bibitem[\protect\citeauthoryear{{Glazebrook} et~al.}{{Glazebrook}
  et~al.}{2017}]{Glazebrook_etal2017}
{Glazebrook} K.,  et~al., 2017, preprint, \href
  {http://adsabs.harvard.edu/abs/2017arXiv170201751G} {} (\mn@eprint {arXiv}
  {1702.01751})

\bibitem[\protect\citeauthoryear{{Goddard} et~al.}{{Goddard}
  et~al.}{2017}]{Goddard_etal2017}
{Goddard} D.,  et~al., 2017, \mn@doi [\mnras] {10.1093/mnras/stw3371}, \href
  {http://adsabs.harvard.edu/abs/2017MNRAS.466.4731G} {466, 4731}

\bibitem[\protect\citeauthoryear{{Gonzalez-Perez} et~al.}{{Gonzalez-Perez}
  et~al.}{2009}]{Violeta_etal2009}
{Gonzalez-Perez} V.,  et~al., 2009, \mn@doi [\mnras]
  {10.1111/j.1365-2966.2009.14397.x}, \href
  {http://adsabs.harvard.edu/abs/2009MNRAS.398..497G} {398, 497}

\bibitem[\protect\citeauthoryear{{Gonzalez-Perez} et~al.}{{Gonzalez-Perez}
  et~al.}{2014}]{GP14}
{Gonzalez-Perez} V.,  et~al., 2014, \mn@doi [\mnras] {10.1093/mnras/stt2410},
  \href {http://adsabs.harvard.edu/abs/2014MNRAS.439..264G} {439, 264}

\bibitem[\protect\citeauthoryear{{Guo}}{{Guo}}{2013}]{Guo_2013}
{Guo} Y.,  2013, {UV Snapshot of Low-redshift Massive Star-forming Galaxies:
  Searching for the Analogs of High-redshift Clumpy Galaxies}, HST Proposal

\bibitem[\protect\citeauthoryear{{Henriques}, {White}, {Thomas}, {Angulo},
  {Guo}, {Lemson}, {Springel}  \& {Overzier}}{{Henriques}
  et~al.}{2015}]{Henriques_etal2015}
{Henriques} B.~M.~B.,  {White} S.~D.~M.,  {Thomas} P.~A.,  {Angulo} R.,  {Guo}
  Q.,  {Lemson} G.,  {Springel} V.,   {Overzier} R.,  2015, \mn@doi [\mnras]
  {10.1093/mnras/stv705}, \href
  {http://adsabs.harvard.edu/abs/2015MNRAS.451.2663H} {451, 2663}

\bibitem[\protect\citeauthoryear{{Hoyle} et~al.}{{Hoyle}
  et~al.}{2017}]{Hoyle_etal2017}
{Hoyle} B.,  et~al., 2017, preprint, \href
  {http://adsabs.harvard.edu/abs/2017arXiv170801532H} {} (\mn@eprint {arXiv}
  {1708.01532})

\bibitem[\protect\citeauthoryear{{Ilbert} et~al.}{{Ilbert}
  et~al.}{2013}]{Ilbert_etal2013}
{Ilbert} O.,  et~al., 2013, in {Cambresy} L.,  {Martins} F.,  {Nuss} E.,
  {Palacios} A.,  eds, SF2A-2013: Proceedings of the Annual meeting of the
  French Society of Astronomy and Astrophysics. pp 545--548

\bibitem[\protect\citeauthoryear{{Jauzac} et~al.}{{Jauzac}
  et~al.}{2015}]{Jauzac_etal2015}
{Jauzac} M.,  et~al., 2015, \mn@doi [\mnras] {10.1093/mnras/stv1402}, \href
  {http://adsabs.harvard.edu/abs/2015MNRAS.452.1437J} {452, 1437}

\bibitem[\protect\citeauthoryear{{Kim} et~al.}{{Kim}
  et~al.}{2015}]{Kim_etal2015}
{Kim} E.~J.,  et~al., 2015, \mn@doi [\mnras] {10.1093/mnras/stv1608}, \href
  {http://adsabs.harvard.edu/abs/2015MNRAS.453..507K} {453, 507}

\bibitem[\protect\citeauthoryear{{Knebe} et~al.}{{Knebe}
  et~al.}{2018}]{Knebe+2018}
{Knebe} A.,  et~al., 2018, \mn@doi [\mnras] {10.1093/mnras/stx3274}, \href
  {http://adsabs.harvard.edu/abs/2018MNRAS.475.2936K} {475, 2936}

\bibitem[\protect\citeauthoryear{{Kriek} \& {Conroy}}{{Kriek} \&
  {Conroy}}{2013}]{KriekConroy2013}
{Kriek} M.,  {Conroy} C.,  2013, \mn@doi [\apjl] {10.1088/2041-8205/775/1/L16},
  \href {http://adsabs.harvard.edu/abs/2013ApJ...775L..16K} {775, L16}

\bibitem[\protect\citeauthoryear{{Kriek} et~al.}{{Kriek}
  et~al.}{2016}]{Kriek_etal2016}
{Kriek} M.,  et~al., 2016, \mn@doi [\nat] {10.1038/nature20570}, \href
  {http://adsabs.harvard.edu/abs/2016Natur.540..248K} {540, 248}

\bibitem[\protect\citeauthoryear{{Lacey} et~al.}{{Lacey}
  et~al.}{2016}]{Lacey_etal2016}
{Lacey} C.~G.,  et~al., 2016, \mn@doi [\mnras] {10.1093/mnras/stw1888}, \href
  {http://adsabs.harvard.edu/abs/2016MNRAS.462.3854L} {462, 3854}

\bibitem[\protect\citeauthoryear{{Lagos} et~al.}{{Lagos}
  et~al.}{2014}]{Lagos_etal14}
{Lagos} C.~D.~P.,  et~al., 2014, \mn@doi [\mnras] {10.1093/mnras/stu266}, \href
  {http://adsabs.harvard.edu/abs/2014MNRAS.440..920L} {440, 920}

\bibitem[\protect\citeauthoryear{{Laporte} et~al.}{{Laporte}
  et~al.}{2016a}]{Laporte_etal2016_inprep}
{Laporte} N.,  et~al., 2016a, in {Reyl{\'e}} C.,  {Richard} J.,  {Cambr{\'e}sy}
  L.,  {Deleuil} M.,  {P{\'e}contal} E.,  {Tresse} L.,   {Vauglin} I.,  eds,
  SF2A-2016: Proceedings of the Annual meeting of the French Society of
  Astronomy and Astrophysics. pp 411--415

\bibitem[\protect\citeauthoryear{{Laporte} et~al.}{{Laporte}
  et~al.}{2016b}]{Laporte_etal2016}
{Laporte} N.,  et~al., 2016b, \mn@doi [\apj] {10.3847/0004-637X/820/2/98},
  \href {http://adsabs.harvard.edu/abs/2016ApJ...820...98L} {820, 98}

\bibitem[\protect\citeauthoryear{{Lonoce} et~al.}{{Lonoce}
  et~al.}{2015}]{Lonoce_etal2015}
{Lonoce} I.,  et~al., 2015, \mn@doi [\mnras] {10.1093/mnras/stv2150}, \href
  {http://adsabs.harvard.edu/abs/2015MNRAS.454.3912L} {454, 3912}

\bibitem[\protect\citeauthoryear{{Mancini} et~al.}{{Mancini}
  et~al.}{2009}]{Mancini_etal2009}
{Mancini} C.,  et~al., 2009, \mn@doi [\aap] {10.1051/0004-6361/200810630},
  \href {http://adsabs.harvard.edu/abs/2009A%26A...500..705M} {500, 705}

\bibitem[\protect\citeauthoryear{{Maraston}}{{Maraston}}{2005}]{Maraston_2005}
{Maraston} C.,  2005, \mn@doi [\mnras] {10.1111/j.1365-2966.2005.09270.x},
  \href {http://adsabs.harvard.edu/abs/2005MNRAS.362..799M} {362, 799}

\bibitem[\protect\citeauthoryear{{Maraston} et~al.}{{Maraston}
  et~al.}{2006}]{Maraston_etal2006}
{Maraston} C.,  et~al., 2006, \mn@doi [\apj] {10.1086/508143}, \href
  {http://adsabs.harvard.edu/abs/2006ApJ...652...85M} {652, 85}

\bibitem[\protect\citeauthoryear{{Maraston} et~al.}{{Maraston}
  et~al.}{2010}]{Maraston_etal2010}
{Maraston} C.,  et~al., 2010, \mn@doi [\mnras]
  {10.1111/j.1365-2966.2010.16973.x}, \href
  {http://adsabs.harvard.edu/abs/2010MNRAS.407..830M} {407, 830}

\bibitem[\protect\citeauthoryear{{Maraston} et~al.}{{Maraston}
  et~al.}{2013}]{Maraston_etal2013}
{Maraston} C.,  et~al., 2013, \mn@doi [\mnras] {10.1093/mnras/stt1424}, \href
  {http://adsabs.harvard.edu/abs/2013MNRAS.435.2764M} {435, 2764}

\bibitem[\protect\citeauthoryear{{Marchesini} et~al.}{{Marchesini}
  et~al.}{2010}]{Marchesini_etal2010}
{Marchesini} D.,  et~al., 2010, {Measuring the Rest-Frame UV Properties and the
  Number Density of Massive Galaxies at $3 < z < 4$}, NOAO Proposal

\bibitem[\protect\citeauthoryear{{Marsan} et~al.}{{Marsan}
  et~al.}{2017}]{Marsan_etal2017}
{Marsan} Z.~C.,  et~al., 2017, \mn@doi [\apj] {10.3847/1538-4357/aa7206}, \href
  {http://adsabs.harvard.edu/abs/2017ApJ...842...21M} {842, 21}

\bibitem[\protect\citeauthoryear{{McMahon}}{{McMahon}}{2012}]{VHS}
{McMahon} R.,  2012, in Science from the Next Generation Imaging and
  Spectroscopic Surveys. p.~37

\bibitem[\protect\citeauthoryear{{Mehlert}, {Thomas}, {Saglia}, {Bender}  \&
  {Wegner}}{{Mehlert} et~al.}{2003}]{Mehlert_etal2003}
{Mehlert} D.,  {Thomas} D.,  {Saglia} R.~P.,  {Bender} R.,   {Wegner} G.,
  2003, \mn@doi [\aap] {10.1051/0004-6361:20030886}, \href
  {http://adsabs.harvard.edu/abs/2003A%26A...407..423M} {407, 423}

\bibitem[\protect\citeauthoryear{{Melchior} et~al.}{{Melchior}
  et~al.}{2015}]{Melchior+2015}
{Melchior} P.,  et~al., 2015, \mn@doi [\mnras] {10.1093/mnras/stv398}, \href
  {http://adsabs.harvard.edu/abs/2015MNRAS.449.2219M} {449, 2219}

\bibitem[\protect\citeauthoryear{{Merlin} et~al.}{{Merlin}
  et~al.}{2018}]{Merlin_etal2018}
{Merlin} E.,  et~al., 2018, \mn@doi [\mnras] {10.1093/mnras/stx2385}, \href
  {http://adsabs.harvard.edu/abs/2018MNRAS.473.2098M} {473, 2098}

\bibitem[\protect\citeauthoryear{{Mortlock} et~al.}{{Mortlock}
  et~al.}{2015}]{Mortlock_etal2015}
{Mortlock} A.,  et~al., 2015, \mn@doi [\mnras] {10.1093/mnras/stu2403}, \href
  {http://adsabs.harvard.edu/abs/2015MNRAS.447....2M} {447, 2}

\bibitem[\protect\citeauthoryear{{Muzzin} et~al.}{{Muzzin}
  et~al.}{2013}]{Muzzin_etal2013}
{Muzzin} A.,  et~al., 2013, \mn@doi [\apj] {10.1088/0004-637X/777/1/18}, \href
  {http://adsabs.harvard.edu/abs/2013ApJ...777...18M} {777, 18}

\bibitem[\protect\citeauthoryear{{Nayyeri} et~al.}{{Nayyeri}
  et~al.}{2014}]{Nayyeri_etal2014}
{Nayyeri} H.,  et~al., 2014, \mn@doi [\apj] {10.1088/0004-637X/794/1/68}, \href
  {http://adsabs.harvard.edu/abs/2014ApJ...794...68N} {794, 68}

\bibitem[\protect\citeauthoryear{{Onodera} et~al.}{{Onodera}
  et~al.}{2012}]{Onodera_etal2012}
{Onodera} M.,  et~al., 2012, \mn@doi [\apj] {10.1088/0004-637X/755/1/26}, \href
  {http://adsabs.harvard.edu/abs/2012ApJ...755...26O} {755, 26}

\bibitem[\protect\citeauthoryear{{Pforr} et~al.}{{Pforr}
  et~al.}{2012}]{Pforr_etal2012}
{Pforr} J.,  et~al., 2012, \mn@doi [\mnras] {10.1111/j.1365-2966.2012.20848.x},
  \href {http://adsabs.harvard.edu/abs/2012MNRAS.422.3285P} {422, 3285}

\bibitem[\protect\citeauthoryear{{Pforr} et~al.}{{Pforr}
  et~al.}{2013}]{Pforr_etal2013}
{Pforr} J.,  et~al., 2013, \mn@doi [\mnras] {10.1093/mnras/stt1382}, \href
  {http://adsabs.harvard.edu/abs/2013MNRAS.435.1389P} {435, 1389}

\bibitem[\protect\citeauthoryear{{Pipino} et~al.}{{Pipino}
  et~al.}{2007}]{Pipino_etal2007}
{Pipino} A.,  et~al., 2007, \mn@doi [\apj] {10.1086/519546}, \href
  {http://adsabs.harvard.edu/abs/2007ApJ...665..295P} {665, 295}

\bibitem[\protect\citeauthoryear{{Pozzetti} et~al.}{{Pozzetti}
  et~al.}{2010}]{Pozzetti_etal2010}
{Pozzetti} L.,  et~al., 2010, \mn@doi [\aap] {10.1051/0004-6361/200913020},
  \href {http://adsabs.harvard.edu/abs/2010A%26A...523A..13P} {523, A13}

\bibitem[\protect\citeauthoryear{{Prevot} et~al.}{{Prevot}
  et~al.}{1984}]{SMC_prevot}
{Prevot} M.~L.,  et~al., 1984, \aap, \href
  {http://adsabs.harvard.edu/abs/1984A%26A...132..389P} {132, 389}

\bibitem[\protect\citeauthoryear{{Reed} et~al.}{{Reed}
  et~al.}{2017}]{Reed_etal2017}
{Reed} S.~L.,  et~al., 2017, \mn@doi [\mnras] {10.1093/mnras/stx728}, \href
  {http://adsabs.harvard.edu/abs/2017MNRAS.468.4702R} {468, 4702}

\bibitem[\protect\citeauthoryear{{Ricciardelli} \&
  {Franceschini}}{{Ricciardelli} \&
  {Franceschini}}{2010}]{RicciardelliANDFranceschini_2010}
{Ricciardelli} E.,  {Franceschini} A.,  2010, \mn@doi [\aap]
  {10.1051/0004-6361/200913374}, \href
  {http://adsabs.harvard.edu/abs/2010A%26A...518A..14R} {518, A14}

\bibitem[\protect\citeauthoryear{{Rossetto} et~al.}{{Rossetto}
  et~al.}{2011}]{Rossetto_etl2011}
{Rossetto} B.~M.,  et~al., 2011, \mn@doi [\aj] {10.1088/0004-6256/141/6/185},
  \href {http://adsabs.harvard.edu/abs/2011AJ....141..185R} {141, 185}

\bibitem[\protect\citeauthoryear{{Rowan-Robinson} et~al.}{{Rowan-Robinson}
  et~al.}{2016}]{Rowan-Robinson_etal2016}
{Rowan-Robinson} M.,  et~al., 2016, \mn@doi [\mnras] {10.1093/mnras/stw1169},
  \href {http://adsabs.harvard.edu/abs/2016MNRAS.461.1100R} {461, 1100}

\bibitem[\protect\citeauthoryear{{S{\'a}nchez} et~al.}{{S{\'a}nchez}
  et~al.}{2012}]{CALIFA}
{S{\'a}nchez} S.~F.,  et~al., 2012, \mn@doi [\aap]
  {10.1051/0004-6361/201117353}, \href
  {http://adsabs.harvard.edu/abs/2012A%26A...538A...8S} {538, A8}

\bibitem[\protect\citeauthoryear{{S{\'a}nchez} et~al.}{{S{\'a}nchez}
  et~al.}{2014}]{NeuralNet}
{S{\'a}nchez} C.,  et~al., 2014, \mn@doi [\mnras] {10.1093/mnras/stu1836},
  \href {http://adsabs.harvard.edu/abs/2014MNRAS.445.1482S} {445, 1482}

\bibitem[\protect\citeauthoryear{{Santini} et~al.}{{Santini}
  et~al.}{2009}]{Santini_etal2009}
{Santini} P.,  et~al., 2009, \mn@doi [\aap] {10.1051/0004-6361/200811434},
  \href {http://adsabs.harvard.edu/abs/2009A%26A...504..751S} {504, 751}

\bibitem[\protect\citeauthoryear{{Schreiber} et~al.}{{Schreiber}
  et~al.}{2018}]{Schreiber_etal2017}
{Schreiber} C.,  et~al., 2018, \mn@doi [\aap] {10.1051/0004-6361/201731917},
  \href {http://adsabs.harvard.edu/abs/2018A%26A...611A..22S} {611, A22}

\bibitem[\protect\citeauthoryear{{Simpson} et~al.}{{Simpson}
  et~al.}{2017}]{Simpson_etal2017}
{Simpson} J.~M.,  et~al., 2017, \mn@doi [\apjl] {10.3847/2041-8213/aa7cf2},
  \href {http://adsabs.harvard.edu/abs/2017ApJ...844L..10S} {844, L10}

\bibitem[\protect\citeauthoryear{{Spitler} et~al.}{{Spitler}
  et~al.}{2014}]{Spitler_etal2014}
{Spitler} L.~R.,  et~al., 2014, \mn@doi [\apjl] {10.1088/2041-8205/787/2/L36},
  \href {http://adsabs.harvard.edu/abs/2014ApJ...787L..36S} {787, L36}

\bibitem[\protect\citeauthoryear{{Stefanon} et~al.}{{Stefanon}
  et~al.}{2013}]{Stefanon_etal2013}
{Stefanon} M.,  et~al., 2013, \mn@doi [\apj] {10.1088/0004-637X/768/1/92},
  \href {http://adsabs.harvard.edu/abs/2013ApJ...768...92S} {768, 92}

\bibitem[\protect\citeauthoryear{{Straatman} et~al.}{{Straatman}
  et~al.}{2014}]{Straatman_etal2014}
{Straatman} C.~M.~S.,  et~al., 2014, \mn@doi [\apjl]
  {10.1088/2041-8205/783/1/L14}, \href
  {http://adsabs.harvard.edu/abs/2014ApJ...783L..14S} {783, L14}

\bibitem[\protect\citeauthoryear{{The Dark Energy Survey Collaboration}}{{The
  Dark Energy Survey Collaboration}}{2005}]{DEScollaboration}
{The Dark Energy Survey Collaboration} 2005, ArXiv Astrophysics e-prints, \href
  {http://adsabs.harvard.edu/abs/2005astro.ph.10346T} {}

\bibitem[\protect\citeauthoryear{{Thomas} et~al.}{{Thomas}
  et~al.}{2005}]{Thomas_etal2005}
{Thomas} D.,  et~al., 2005, \mn@doi [\apj] {10.1086/426932}, \href
  {http://adsabs.harvard.edu/abs/2005ApJ...621..673T} {621, 673}

\bibitem[\protect\citeauthoryear{{Thomas} et~al.}{{Thomas}
  et~al.}{2010}]{Thomas_etal2010}
{Thomas} D.,  et~al., 2010, \mn@doi [\mnras]
  {10.1111/j.1365-2966.2010.16427.x}, \href
  {http://adsabs.harvard.edu/abs/2010MNRAS.404.1775T} {404, 1775}

\bibitem[\protect\citeauthoryear{{Wake} et~al.}{{Wake}
  et~al.}{2006}]{Wake_etal2006}
{Wake} D.~A.,  et~al., 2006, \mn@doi [\mnras]
  {10.1111/j.1365-2966.2006.10831.x}, \href
  {http://adsabs.harvard.edu/abs/2006MNRAS.372..537W} {372, 537}

\bibitem[\protect\citeauthoryear{{Wang} et~al.}{{Wang}
  et~al.}{2016}]{Wang_etal2016}
{Wang} L.,  et~al., 2016, preprint, \href
  {http://adsabs.harvard.edu/abs/2016arXiv161204540W} {} (\mn@eprint {arXiv}
  {1612.04540})

\bibitem[\protect\citeauthoryear{{Whitaker} et~al.}{{Whitaker}
  et~al.}{2013}]{Whitaker_etal2013}
{Whitaker} K.~E.,  et~al., 2013, \mn@doi [\apjl] {10.1088/2041-8205/770/2/L39},
  \href {http://adsabs.harvard.edu/abs/2013ApJ...770L..39W} {770, L39}

\bibitem[\protect\citeauthoryear{{White} \& {Rees}}{{White} \&
  {Rees}}{1978}]{WhiteRees1978}
{White} S.~D.~M.,  {Rees} M.~J.,  1978, \mn@doi [\mnras]
  {10.1093/mnras/183.3.341}, \href
  {http://adsabs.harvard.edu/abs/1978MNRAS.183..341W} {183, 341}

\bibitem[\protect\citeauthoryear{{Yan} et~al.}{{Yan}
  et~al.}{2004}]{Yan_etal2004}
{Yan} H.,  et~al., 2004, \mn@doi [\apj] {10.1086/424898}, \href
  {http://adsabs.harvard.edu/abs/2004ApJ...616...63Y} {616, 63}

\bibitem[\protect\citeauthoryear{{Zheng} et~al.}{{Zheng}
  et~al.}{2014}]{Zheng_etal2014}
{Zheng} W.,  et~al., 2014, \mn@doi [\apj] {10.1088/0004-637X/795/1/93}, \href
  {http://adsabs.harvard.edu/abs/2014ApJ...795...93Z} {795, 93}

\makeatother
\end{thebibliography}




\appendix
%
%
%

\section{Fitting Results for the Best Candidates}
\label{app:FullGoldenSampleResults}

Here we provide the stellar population properties and photometric redshift for all best candidate galaxies. As usual, fits are performed for two reddening options: SMC law and Calzetti law.

\renewcommand{\arraystretch}{2}
\begin{table*}
	\caption{Properties of the best candidates for the SMC law case, from left to right: object ID, $z_{phot}$, $\chi_{r}^{2}$, stellar mass $M^{*}$, absolute magnitude, age, star formation history, metallicity, the number $\sigma$'s used to estimate potential AGN contamination, photometric redshift found by fitting DES only bands, redshift from the DES pipeline, extinction as $E(B-V))$. Errors refer to the 99\% confidence level.}
	\centering
	\resizebox{\linewidth}{!}{%
	\begin{tabular}{ ccccccccccccc }
		\hline
		\textbf{ID} & \textbf{\textit{$z_{phot}$}} & \textbf{$\chi_r^2$} & \textbf{$\log_{10}(M^{*}/M_{\odot}$)} & \begin{tabular}{@{}c@{}}\textbf{Abs.} \\ \textbf{Mag. $(i)$}\end{tabular} & \begin{tabular}{@{}c@{}}\textbf{Age} \\ \textbf{(Gyr)}\end{tabular} & \textbf{SFH} & \begin{tabular}{@{}c@{}}\textbf{[Z/H]} \\ \textbf{(Z$_{\odot}$)}\end{tabular} & \textbf{$\sigma_{AGN}$} & \textbf{$z_{DESonly}$} & \textbf{\textit{$z_{BPZ}$}} & \textbf{E (B-V)} \\
		\hline
		100600870 & $3.67_{-0.23}^{+0.16}$ & 1.224 & $11.8_{-0.0}^{+0.03}$ & -27.02 & 0.1 & $t_{trunc} = 1.0$ & 1 & 0.16 & 3.37 & 0.38 & 0.18 \\ 
102002089 & $3.77_{-0.3}^{+0.47}$ & 2.019 & $11.28_{-0.06}^{+0.71}$ & -25.63 & 0.1 & $t_{trunc} = 1.0$ & 2 & 5.59 & 3.37 & 0.51 & 0.18 \\ 
102009403 & $3.64_{-0.19}^{+0.19}$ & 1.578 & $11.7_{-0.01}^{+0.02}$ & -26.69 & 0.1 & $t_{trunc} = 1.0$ & 2 & 1.71 & 4.25 & 0.35 & 0.18 \\ 
102009835 & $3.77_{-0.1}^{+0.17}$ & 1.632 & $11.83_{-0.05}^{+0.0}$ & -27.01 & 0.1 & $t_{trunc} = 0.1$ & 2 & 10.86 & 2.47 & 0.47 & 0.18 \\ 
102009849 & $3.73_{-0.14}^{+0.23}$ & 1.007 & $11.93_{-0.25}^{+0.04}$ & -27.24 & 0.1 & $t_{trunc} = 0.1$ & 2 & 12.22 & 3.58 & 0.47 & 0.18 \\ 
102031864 & $3.77_{-0.14}^{+0.16}$ & 1.978 & $11.67_{-0.07}^{+0.01}$ & -26.6 & 0.1 & $t_{trunc} = 0.1$ & 2 & 3.34 & 2.48 & 0.5 & 0.18 \\ 
132987082 & $3.7_{-0.26}^{+0.14}$ & 1.522 & $11.57_{-0.02}^{+0.42}$ & -26.54 & 0.1 & CONSTANT & 1/2 & 7.21 & 2.52 & 0.37 & 0.18 \\ 
133572897 & $3.46_{-0.08}^{+0.12}$ & 1.792 & $12.21_{-0.04}^{+0.05}$ & -27.52 & 0.11 & $t_{trunc} = 0.1$ & 2 & 1.77 & 3.48 & 0.49 & 0.18 \\ 
136067262 & $3.73_{-0.16}^{+0.12}$ & 2.073 & $12.09_{-0.24}^{+0.0}$ & -27.66 & 0.1 & $t_{trunc} = 0.1$ & 2 & 15.27 & 2.54 & 0.4 & 0.18 \\ 
137806706 & $3.78_{-0.28}^{+0.55}$ & 0.338 & $11.98_{-0.17}^{+0.21}$ & -26.6 & 0.4 & $t_{trunc} = 0.3$ & 1 & 0.06 & 3.69 & 0.49 & 0.0 \\ 
164738198 & $3.67_{-0.19}^{+0.16}$ & 2.642 & $12.36_{-0.01}^{+0.28}$ & -27.76 & 0.29 & $t_{trunc} = 1.0$ & 2 & 19.48 & 2.39 & 0.49 & 0.18 \\ 
287114376 & $3.73_{-0.32}^{+0.26}$ & 1.54 & $11.89_{-0.13}^{+0.27}$ & -26.76 & 0.13 & $e^{-t/0.1~\textrm{Gyr}}$ & 2 & 2.52 & 3.4 & 0.41 & 0.18 \\ 
396223342 & $3.69_{-0.26}^{+0.11}$ & 0.891 & $11.59_{-0.02}^{+0.04}$ & -26.59 & 0.1 & $t_{trunc} = 1.0$ & 1/2 & 4.44 & 3.34 & 0.37 & 0.18 \\ 
396276124 & $3.69_{-0.26}^{+0.18}$ & 1.022 & $11.66_{-0.23}^{+0.02}$ & -26.66 & 0.1 & $t_{trunc} = 0.1$ & 1 & 2.61 & 3.45 & 0.48 & 0.18 \\ 
397300605 & $3.83_{-0.3}^{+0.09}$ & 2.084 & $11.73_{-0.0}^{+0.02}$ & -26.93 & 0.1 & CONSTANT & 1/2 & 7.77 & 3.25 & 0.47 & 0.18 \\ 
397303505 & $3.69_{-0.32}^{+0.15}$ & 1.604 & $11.54_{-0.27}^{+0.01}$ & -26.29 & 0.11 & $t_{trunc} = 1.0$ & 1 & 4.8 & 3.4 & 0.49 & 0.18 \\ 
397554368 & $3.71_{-0.28}^{+0.35}$ & 1.366 & $12.12_{-0.44}^{+0.0}$ & -27.23 & 0.29 & SSP & 1/5 & 3.19 & 2.5 & 0.37 & 0.0 \\ 
397764328 & $3.77_{-0.27}^{+0.13}$ & 2.321 & $11.58_{-0.04}^{+0.02}$ & -26.4 & 0.1 & CONSTANT & 2 & 4.27 & 4.3 & 0.54 & 0.18 \\ 
397885462 & $3.8_{-0.56}^{+0.2}$ & 0.709 & $11.78_{-0.04}^{+0.4}$ & -26.18 & 0.18 & $t_{trunc} = 0.1$ & 2 & 1.74 & 3.34 & 0.5 & 0.0 \\ 
398107560 & $3.67_{-0.2}^{+0.25}$ & 1.0 & $11.61_{-0.2}^{+0.03}$ & -26.46 & 0.1 & $t_{trunc} = 0.1$ & 2 & 3.3 & 3.34 & 0.45 & 0.18 \\ 
399804681 & $3.86_{-0.48}^{+0.21}$ & 1.764 & $11.54_{-0.3}^{+0.02}$ & -26.28 & 0.1 & $t_{trunc} = 0.1$ & 2 & 1.88 & 2.49 & 0.5 & 0.18 \\ 
399842053 & $3.72_{-0.25}^{+0.16}$ & 0.946 & $11.61_{-0.03}^{+0.01}$ & -26.48 & 0.1 & CONSTANT & 2 & 5.7 & 4.28 & 0.43 & 0.18 \\ 
399842613 & $4.11_{-0.48}^{+0.15}$ & 1.443 & $11.48_{-0.38}^{+0.0}$ & -26.15 & 0.1 & $t_{trunc} = 1.0$ & 2 & 0.36 & 3.31 & 0.59 & 0.18 \\ 
400998781 & $3.64_{-0.17}^{+0.23}$ & 0.863 & $11.52_{-0.08}^{+0.04}$ & -26.23 & 0.1 & $t_{trunc} = 0.1$ & 2 & 2.2 & 2.53 & 0.47 & 0.18 \\ 
401003476 & $3.86_{-0.41}^{+0.18}$ & 0.666 & $12.07_{-0.71}^{+0.07}$ & -26.72 & 1.02 & $t_{trunc} = 1.0$ & 2 & 1.37 & 2.48 & 0.39 & 0.0 \\ 
404788215 & $3.61_{-0.2}^{+0.19}$ & 1.098 & $11.68_{-0.02}^{+0.03}$ & -26.65 & 0.1 & $t_{trunc} = 1.0$ & 2 & 4.63 & 2.51 & 0.43 & 0.18 \\ 
404798117 & $3.46_{-0.13}^{+0.38}$ & 0.929 & $12.08_{-0.13}^{+0.26}$ & -27.31 & 0.11 & $t_{trunc} = 0.1$ & 1/5 & 16.39 & 3.42 & 0.48 & 0.18 \\ 
404886634 & $3.72_{-0.3}^{+0.15}$ & 1.837 & $11.76_{-0.21}^{+0.02}$ & -26.92 & 0.1 & $t_{trunc} = 0.1$ & 1 & 6.63 & 3.45 & 0.47 & 0.18 \\ 
404907811 & $3.61_{-0.26}^{+0.32}$ & 1.542 & $11.89_{-0.04}^{+0.54}$ & -26.86 & 0.23 & $e^{-t/1.0~\textrm{Gyr}}$ & 1/2 & 6.14 & 4.22 & 0.36 & 0.18 \\ 
405937444 & $3.64_{-0.2}^{+0.19}$ & 1.272 & $11.83_{-0.02}^{+0.01}$ & -27.02 & 0.1 & $t_{trunc} = 1.0$ & 2 & 11.74 & 3.42 & 0.42 & 0.18 \\ 
408135057 & $3.77_{-0.17}^{+0.23}$ & 1.411 & $11.41_{-0.07}^{+0.04}$ & -25.97 & 0.1 & $t_{trunc} = 0.1$ & 2 & 0.64 & 3.5 & 0.48 & 0.18 \\ 
408311797 & $3.83_{-0.19}^{+0.1}$ & 2.208 & $11.88_{-0.01}^{+0.05}$ & -27.15 & 0.1 & CONSTANT & 2 & 9.19 & 2.52 & 0.48 & 0.18 \\ 
409127588 & $3.77_{-0.48}^{+0.21}$ & 1.367 & $11.85_{-0.09}^{+0.41}$ & -26.38 & 0.11 & SSP & 2 & 4.41 & 3.45 & 0.52 & 0.0 \\ 
411491335 & $3.69_{-0.17}^{+0.23}$ & 1.521 & $12.32_{-0.08}^{+0.19}$ & -27.84 & 0.13 & $e^{-t/0.1~\textrm{Gyr}}$ & 2 & 27.57 & 3.19 & 0.43 & 0.18 \\ 
411500732 & $3.73_{-0.24}^{+0.31}$ & 2.43 & $11.4_{-0.03}^{+0.41}$ & -25.93 & 0.1 & $t_{trunc} = 1.0$ & 2 & 0.7 & 3.4 & 0.46 & 0.18 \\ 
412637681 & $3.61_{-0.25}^{+0.19}$ & 2.309 & $11.42_{-0.01}^{+0.4}$ & -26.06 & 0.1 & CONSTANT & 1 & 0.49 & 2.48 & 0.46 & 0.18 \\ 
\hline 
\end{tabular}} 
\label{tab:app-SMC} 
\end{table*} 
\renewcommand{\arraystretch}{1}
\renewcommand{\arraystretch}{2}
\begin{table*}
\contcaption{} 
\centering
\resizebox{\linewidth}{!}{%
\begin{tabular}{ ccccccccccccc }
\hline 
\textbf{ID} & \textbf{\textit{$z_{phot}$}} & \textbf{$\chi_r^2$} & \textbf{$\log_{10}(M^{*}/M_{\odot}$)} & \begin{tabular}{@{}c@{}}\textbf{Abs.} \\ \textbf{Mag. $(i)$}\end{tabular} & \begin{tabular}{@{}c@{}}\textbf{Age} \\ \textbf{(Gyr)}\end{tabular} & \textbf{SFH} & \begin{tabular}{@{}c@{}}\textbf{[Z/H]} \\ \textbf{(Z$_{\odot}$)}\end{tabular} & \textbf{$\sigma_{AGN}$} & \textbf{$z_{DESonly}$} & \textbf{\textit{$z_{BPZ}$}} & \textbf{E (B-V)} \\
\hline 
414233666 & $3.8_{-0.3}^{+0.15}$ & 2.121 & $12.08_{-0.21}^{+0.01}$ & -27.71 & 0.1 & $t_{trunc} = 0.1$ & 1 & 11.09 & 3.29 & 0.46 & 0.18 \\ 
414237423 & $3.73_{-0.29}^{+0.41}$ & 1.102 & $11.61_{-0.22}^{+0.77}$ & -26.45 & 0.1 & $t_{trunc} = 0.1$ & 2 & 2.3 & 3.58 & 0.43 & 0.18 \\ 
417565185 & $3.64_{-0.23}^{+0.22}$ & 1.051 & $11.51_{-0.27}^{+0.04}$ & -26.22 & 0.1 & $t_{trunc} = 1.0$ & 2 & 1.49 & 2.5 & 0.39 & 0.18 \\ 
431455424 & $3.81_{-0.24}^{+0.2}$ & 1.217 & $12.13_{-0.09}^{+0.12}$ & -27.04 & 0.13 & SSP & 2 & 14.07 & 3.67 & 0.64 & 0.0 \\ 
431827017 & $3.77_{-0.16}^{+0.16}$ & 0.764 & $11.83_{-0.28}^{+0.04}$ & -27.01 & 0.1 & $t_{trunc} = 0.1$ & 2 & 9.63 & 3.34 & 0.49 & 0.18 \\ 
434401854 & $3.69_{-0.29}^{+0.14}$ & 1.31 & $12.25_{-0.06}^{+0.08}$ & -27.82 & 0.16 & $e^{-t/1.0~\textrm{Gyr}}$ & 1 & 29.43 & 3.42 & 0.44 & 0.18 \\ 
444147103 & $3.69_{-0.36}^{+0.2}$ & 1.289 & $11.56_{-0.38}^{+0.35}$ & -26.41 & 0.1 & $t_{trunc} = 0.1$ & 1 & 4.9 & 3.42 & 0.46 & 0.18 \\ 
470611726 & $3.69_{-0.47}^{+0.23}$ & 1.081 & $11.35_{-0.29}^{+0.04}$ & -25.8 & 0.1 & $t_{trunc} = 0.1$ & 2 & 7.27 & 2.5 & 0.5 & 0.18 \\ 
470971747 & $4.25_{-0.89}^{+0.11}$ & 0.739 & $12.36_{-0.85}^{+0.04}$ & -27.54 & 0.45 & $e^{-t/0.1~\textrm{Gyr}}$ & 2 & 2.36 & 2.56 & 0.28 & 0.0 \\ 
471600124 & $3.94_{-0.37}^{+0.49}$ & 0.544 & $11.72_{-0.42}^{+0.36}$ & -26.17 & 0.32 & $t_{trunc} = 0.3$ & 2 & 2.32 & 3.31 & 0.49 & 0.0 \\ 
471985468 & $3.64_{-0.21}^{+0.14}$ & 2.193 & $12.13_{-0.1}^{+0.07}$ & -27.32 & 0.11 & $t_{trunc} = 0.1$ & 2 & 10.29 & 2.54 & 0.54 & 0.18 \\ 
473133985 & $3.67_{-0.22}^{+0.22}$ & 2.112 & $11.6_{-0.01}^{+0.43}$ & -26.52 & 0.1 & $t_{trunc} = 1.0$ & 1 & 7.56 & 3.58 & 0.38 & 0.18 \\ 
473136272 & $3.64_{-0.18}^{+0.17}$ & 1.491 & $11.91_{-0.06}^{+0.03}$ & -27.16 & 0.1 & $e^{-t/0.3~\textrm{Gyr}}$ & 2 & 11.56 & 2.52 & 0.41 & 0.18 \\ 
473140970 & $3.69_{-0.27}^{+0.11}$ & 1.047 & $11.74_{-0.01}^{+0.03}$ & -26.95 & 0.1 & $t_{trunc} = 1.0$ & 1/2 & 9.56 & 3.34 & 0.42 & 0.18 \\ 
473404298 & $4.3_{-0.14}^{+0.11}$ & 0.554 & $12.37_{-0.2}^{+0.04}$ & -27.57 & 0.45 & $e^{-t/0.1~\textrm{Gyr}}$ & 2 & 0.43 & 3.58 & 0.47 & 0.0 \\ 
473408311 & $3.75_{-0.28}^{+0.16}$ & 1.53 & $11.79_{-0.04}^{+0.39}$ & -27.08 & 0.1 & $t_{trunc} = 0.1$ & 1/2 & 8.66 & 3.65 & 0.44 & 0.18 \\ 
473411673 & $3.69_{-0.24}^{+0.22}$ & 1.016 & $11.54_{-0.26}^{+0.02}$ & -26.28 & 0.1 & $t_{trunc} = 1.0$ & 2 & 5.69 & 3.58 & 0.45 & 0.18 \\ 
473496203 & $4.3_{-0.84}^{+0.08}$ & 0.53 & $12.25_{-0.73}^{+0.37}$ & -26.98 & 1.28 & $e^{-t/1.0~\textrm{Gyr}}$ & 2 & 0.37 & 3.05 & 0.38 & 0.0 \\ 
473498930 & $3.64_{-0.18}^{+0.21}$ & 0.992 & $11.93_{-0.01}^{+0.01}$ & -27.26 & 0.1 & $t_{trunc} = 1.0$ & 2 & 20.72 & 3.42 & 0.41 & 0.18 \\ 
473503196 & $3.83_{-0.18}^{+0.28}$ & 0.835 & $11.49_{-0.26}^{+0.03}$ & -26.15 & 0.1 & $t_{trunc} = 0.1$ & 2 & 3.37 & 3.5 & 0.59 & 0.18 \\ 
473511031 & $3.64_{-0.16}^{+0.2}$ & 1.643 & $11.58_{-0.07}^{+0.05}$ & -26.39 & 0.1 & $t_{trunc} = 0.1$ & 2 & 1.92 & 2.44 & 0.41 & 0.18 \\ 
473512115 & $3.89_{-0.34}^{+0.2}$ & 2.122 & $12.01_{-0.37}^{+0.33}$ & -26.91 & 0.32 & $t_{trunc} = 0.3$ & 2 & 3.44 & 3.58 & 0.43 & 0.0 \\ 
473515263 & $3.63_{-0.2}^{+0.22}$ & 1.616 & $11.49_{-0.01}^{+0.02}$ & -26.18 & 0.1 & $t_{trunc} = 1.0$ & 2 & 2.27 & 3.4 & 0.45 & 0.18 \\ 
473519025 & $3.75_{-0.35}^{+0.41}$ & 1.318 & $11.53_{-0.36}^{+0.76}$ & -26.25 & 0.1 & $t_{trunc} = 0.1$ & 2 & 0.81 & 3.4 & 0.5 & 0.18 \\ 
473520285 & $3.64_{-0.21}^{+0.2}$ & 1.199 & $11.56_{-0.02}^{+0.03}$ & -26.35 & 0.1 & $t_{trunc} = 1.0$ & 2 & 3.35 & 3.42 & 0.38 & 0.18 \\ 
473521671 & $3.48_{-0.13}^{+0.62}$ & 1.08 & $11.92_{-0.16}^{+0.35}$ & -26.8 & 0.11 & $t_{trunc} = 0.1$ & 2 & 2.5 & 3.45 & 0.58 & 0.18 \\ 
473528868 & $3.88_{-0.48}^{+0.26}$ & 0.686 & $11.66_{-0.12}^{+0.36}$ & -26.03 & 0.32 & $t_{trunc} = 0.3$ & 2 & 1.9 & 3.59 & 0.5 & 0.0 \\ 
473530252 & $3.92_{-0.31}^{+0.27}$ & 1.431 & $12.04_{-0.36}^{+0.24}$ & -26.97 & 0.32 & $t_{trunc} = 0.3$ & 2 & 0.76 & 3.5 & 0.46 & 0.0 \\ 
477008049 & $4.25_{-0.17}^{+0.08}$ & 0.681 & $12.46_{-0.22}^{+0.03}$ & -27.78 & 0.45 & $e^{-t/0.1~\textrm{Gyr}}$ & 2 & 4.79 & 3.34 & 0.51 & 0.0 \\ 
477008438 & $3.77_{-0.21}^{+0.19}$ & 2.679 & $11.63_{-0.02}^{+0.01}$ & -26.52 & 0.1 & $t_{trunc} = 1.0$ & 2 & 4.99 & 3.31 & 0.45 & 0.18 \\ 
480339250 & $3.86_{-0.53}^{+0.25}$ & 1.089 & $11.57_{-0.27}^{+0.03}$ & -26.37 & 0.1 & $t_{trunc} = 0.1$ & 2 & 6.05 & 3.4 & 0.55 & 0.18 \\ 
480995070 & $3.75_{-0.16}^{+0.16}$ & 1.823 & $11.67_{-0.34}^{+0.0}$ & -26.62 & 0.1 & $t_{trunc} = 0.1$ & 2 & 7.03 & 2.52 & 0.43 & 0.18 \\ 
481350973 & $3.67_{-0.26}^{+0.28}$ & 0.967 & $11.55_{-0.32}^{+0.86}$ & -26.48 & 0.1 & $t_{trunc} = 1.0$ & 1/2 & 5.37 & 4.25 & 0.4 & 0.18 \\ 
482208365 & $3.69_{-0.21}^{+0.16}$ & 1.481 & $11.86_{-0.08}^{+0.04}$ & -27.03 & 0.1 & $e^{-t/0.3~\textrm{Gyr}}$ & 2 & 13.71 & 2.5 & 0.44 & 0.18 \\ 
483918716 & $3.64_{-0.48}^{+0.45}$ & 1.28 & $11.65_{-0.02}^{+0.47}$ & -26.21 & 0.16 & $e^{-t/0.1~\textrm{Gyr}}$ & 1/2 & 0.5 & 3.45 & 0.43 & 0.18 \\ 
489254835 & $3.83_{-0.49}^{+0.37}$ & 1.73 & $11.55_{-0.29}^{+0.75}$ & -26.31 & 0.1 & $t_{trunc} = 0.1$ & 2 & 4.53 & 3.42 & 0.48 & 0.18 \\ 
\hline 
\end{tabular}} 
\label{tab:cont:app-SMC} 
\end{table*} 
\renewcommand{\arraystretch}{1}
\renewcommand{\arraystretch}{2}
\begin{table*}
\contcaption{} 
\centering
\resizebox{\linewidth}{!}{%
\begin{tabular}{ ccccccccccccc }
\hline 
\textbf{ID} & \textbf{\textit{$z_{phot}$}} & \textbf{$\chi_r^2$} & \textbf{$\log_{10}(M^{*}/M_{\odot}$)} & \begin{tabular}{@{}c@{}}\textbf{Abs.} \\ \textbf{Mag. $(i)$}\end{tabular} & \begin{tabular}{@{}c@{}}\textbf{Age} \\ \textbf{(Gyr)}\end{tabular} & \textbf{SFH} & \begin{tabular}{@{}c@{}}\textbf{[Z/H]} \\ \textbf{(Z$_{\odot}$)}\end{tabular} & \textbf{$\sigma_{AGN}$} & \textbf{$z_{DESonly}$} & \textbf{\textit{$z_{BPZ}$}} & \textbf{E (B-V)} \\
\hline 
490704656 & $3.73_{-0.35}^{+0.68}$ & 1.209 & $11.22_{-0.27}^{+1.12}$ & -25.49 & 0.1 & $t_{trunc} = 1.0$ & 2 & 0.75 & 3.4 & 0.51 & 0.18 \\ 
494789087 & $3.96_{-0.38}^{+0.48}$ & 1.445 & $11.21_{-0.1}^{+0.55}$ & -25.38 & 0.11 & $t_{trunc} = 0.1$ & 2 & 0.73 & 2.63 & 0.4 & 0.0 \\ 
494790027 & $3.83_{-0.19}^{+0.31}$ & 1.281 & $11.7_{-0.08}^{+0.6}$ & -26.22 & 0.14 & $t_{trunc} = 0.1$ & 2 & 3.18 & 2.58 & 0.52 & 0.0 \\ 
494790169 & $4.29_{-0.62}^{+0.12}$ & 0.598 & $11.99_{-0.11}^{+0.35}$ & -26.76 & 0.29 & $e^{-t/0.1~\textrm{Gyr}}$ & 2 & 1.53 & 3.98 & 0.42 & 0.0 \\ 
494792459 & $4.11_{-0.34}^{+0.36}$ & 1.167 & $11.16_{-0.65}^{+1.68}$ & -25.27 & 0.11 & $t_{trunc} = 0.1$ & 2 & 0.36 & 3.08 & 0.44 & 0.0 \\ 
494793098 & $3.98_{-0.25}^{+0.29}$ & 0.383 & $11.87_{-0.11}^{+0.25}$ & -26.55 & 0.32 & $t_{trunc} = 0.3$ & 2 & 0.33 & 4.0 & 0.48 & 0.0 \\ 
494800805 & $4.25_{-1.22}^{+0.1}$ & 0.196 & $12.29_{-0.5}^{+0.26}$ & -27.19 & 0.81 & $e^{-t/0.3~\textrm{Gyr}}$ & 2 & 0.64 & 3.15 & 0.39 & 0.0 \\ 
494801634 & $3.37_{-0.3}^{+0.5}$ & 0.437 & $12.09_{-0.01}^{+0.1}$ & -27.08 & 0.45 & $t_{trunc} = 0.3$ & 1/5 & 0.95 & 3.56 & 0.35 & 0.0 \\ 
495323159 & $3.91_{-0.31}^{+0.42}$ & 1.903 & $11.37_{-0.08}^{+0.68}$ & -25.66 & 0.14 & $t_{trunc} = 0.1$ & 1/5 & 0.54 & 2.65 & 0.41 & 0.0 \\ 
495342175 & $3.64_{-0.23}^{+0.2}$ & 2.408 & $11.46_{-0.27}^{+0.38}$ & -26.09 & 0.1 & $t_{trunc} = 1.0$ & 2 & 0.81 & 2.46 & 0.4 & 0.18 \\ 
495566911 & $3.12_{-0.15}^{+0.21}$ & 0.438 & $11.81_{-0.11}^{+0.03}$ & -26.12 & 0.16 & SSP & 2 & 0.56 & 4.29 & 0.54 & 0.0 \\ 
497171956 & $3.72_{-0.11}^{+0.24}$ & 1.868 & $11.93_{-0.05}^{+0.04}$ & -27.26 & 0.1 & $t_{trunc} = 0.1$ & 2 & 6.9 & 3.56 & 0.44 & 0.18 \\ 
501217876 & $3.69_{-0.22}^{+0.16}$ & 1.175 & $11.87_{-0.03}^{+0.01}$ & -27.11 & 0.1 & CONSTANT & 2 & 14.42 & 2.5 & 0.42 & 0.18 \\ 
501218097 & $3.84_{-0.29}^{+0.11}$ & 1.961 & $11.96_{-0.35}^{+0.09}$ & -26.78 & 0.32 & $t_{trunc} = 0.3$ & 2 & 5.96 & 3.61 & 0.38 & 0.0 \\ 
501524910 & $3.92_{-0.46}^{+0.25}$ & 1.41 & $12.01_{-0.12}^{+0.36}$ & -26.91 & 0.32 & $t_{trunc} = 0.3$ & 2 & 4.02 & 3.53 & 0.4 & 0.0 \\ 
501577492 & $3.69_{-0.14}^{+0.15}$ & 1.59 & $12.12_{-0.07}^{+0.01}$ & -27.73 & 0.1 & $t_{trunc} = 0.1$ & 2 & 34.62 & 2.5 & 0.45 & 0.18 \\ 
503984762 & $3.77_{-0.28}^{+0.27}$ & 0.479 & $11.76_{-0.1}^{+0.36}$ & -26.28 & 0.32 & $t_{trunc} = 0.3$ & 2 & 0.27 & 2.54 & 0.38 & 0.0 \\ 
504038042 & $3.73_{-0.12}^{+0.22}$ & 1.172 & $11.95_{-0.06}^{+0.04}$ & -27.3 & 0.1 & $t_{trunc} = 0.1$ & 2 & 10.92 & 3.29 & 0.48 & 0.18 \\ 
504056183 & $3.76_{-0.53}^{+0.18}$ & 1.519 & $11.6_{-0.32}^{+0.0}$ & -26.51 & 0.1 & $t_{trunc} = 0.1$ & 1 & 5.42 & 3.27 & 0.43 & 0.18 \\ 
504194446 & $3.77_{-0.09}^{+0.25}$ & 2.009 & $11.87_{-0.05}^{+0.03}$ & -27.1 & 0.1 & $t_{trunc} = 0.1$ & 2 & 9.77 & 3.4 & 0.48 & 0.18 \\ 
504330828 & $3.58_{-0.21}^{+0.11}$ & 2.148 & $11.49_{-0.03}^{+0.03}$ & -26.32 & 0.1 & CONSTANT & 1/2 & 4.25 & 2.45 & 0.35 & 0.18 \\ 
505028285 & $3.69_{-0.13}^{+0.26}$ & 1.048 & $11.9_{-0.06}^{+0.04}$ & -27.18 & 0.1 & $t_{trunc} = 0.1$ & 2 & 17.04 & 3.56 & 0.47 & 0.18 \\ 
506017320 & $3.77_{-0.14}^{+0.19}$ & 0.719 & $11.63_{-0.07}^{+0.03}$ & -26.5 & 0.1 & $t_{trunc} = 0.1$ & 2 & 4.07 & 3.37 & 0.49 & 0.18 \\ 
506345182 & $3.73_{-0.47}^{+0.44}$ & 1.044 & $11.41_{-0.31}^{+0.78}$ & -25.97 & 0.1 & $t_{trunc} = 0.1$ & 2 & 1.09 & 3.4 & 0.53 & 0.18 \\ 
506534457 & $3.64_{-0.13}^{+0.18}$ & 1.484 & $11.86_{-0.06}^{+0.03}$ & -27.08 & 0.1 & $t_{trunc} = 0.1$ & 2 & 8.97 & 2.49 & 0.4 & 0.18 \\ 
506537406 & $3.73_{-0.24}^{+0.24}$ & 1.615 & $11.57_{-0.25}^{+0.03}$ & -26.35 & 0.1 & $t_{trunc} = 0.1$ & 2 & 4.12 & 2.61 & 0.5 & 0.18 \\ 
507681715 & $3.81_{-0.41}^{+0.62}$ & 1.395 & $11.97_{-0.04}^{+0.88}$ & -26.65 & 0.36 & $t_{trunc} = 0.3$ & 1 & 5.25 & 2.56 & 0.41 & 0.0 \\ 
507785363 & $3.46_{-0.18}^{+0.72}$ & 0.872 & $11.83_{-0.08}^{+0.33}$ & -26.56 & 0.13 & $t_{trunc} = 0.1$ & 1/2 & 0.67 & 2.64 & 0.48 & 0.18 \\ 
507791066 & $3.62_{-0.21}^{+0.17}$ & 1.08 & $11.92_{-0.1}^{+0.16}$ & -27.13 & 0.11 & $e^{-t/0.1~\textrm{Gyr}}$ & 1/2 & 10.98 & 4.25 & 0.47 & 0.18 \\ 
507791331 & $3.75_{-0.21}^{+0.19}$ & 1.514 & $12.4_{-0.07}^{+0.12}$ & -27.95 & 0.14 & $e^{-t/0.1~\textrm{Gyr}}$ & 2 & 30.31 & 2.5 & 0.48 & 0.18 \\ 
507810919 & $3.69_{-0.31}^{+0.15}$ & 1.282 & $11.66_{-0.27}^{+0.01}$ & -26.67 & 0.1 & $t_{trunc} = 0.1$ & 1 & 9.6 & 2.55 & 0.41 & 0.18 \\ 
507820438 & $3.64_{-0.2}^{+0.19}$ & 0.78 & $12.05_{-0.08}^{+0.04}$ & -27.47 & 0.11 & $e^{-t/1.0~\textrm{Gyr}}$ & 2 & 18.39 & 4.28 & 0.42 & 0.18 \\ 
508601732 & $4.13_{-0.15}^{+0.16}$ & 1.847 & $11.72_{-0.07}^{+0.05}$ & -26.74 & 0.1 & $t_{trunc} = 0.1$ & 2 & 5.08 & 3.53 & 0.65 & 0.18 \\ 
618652137 & $3.7_{-0.24}^{+0.17}$ & 0.918 & $11.81_{-0.27}^{+0.0}$ & -27.05 & 0.1 & $t_{trunc} = 0.1$ & 1 & 4.1 & 3.5 & 0.45 & 0.18 \\ 
618663972 & $3.69_{-0.2}^{+0.21}$ & 1.575 & $12.07_{-0.03}^{+0.13}$ & -27.31 & 0.11 & $e^{-t/0.1~\textrm{Gyr}}$ & 2 & 8.83 & 3.53 & 0.52 & 0.18 \\ 
618664093 & $3.73_{-0.31}^{+0.13}$ & 1.132 & $11.68_{-0.25}^{+0.0}$ & -26.81 & 0.1 & $t_{trunc} = 0.1$ & 1/2 & 1.29 & 3.4 & 0.47 & 0.18 \\ 
\hline 
\end{tabular}} 
\label{tab:cont:app-SMC} 
\end{table*} 
\renewcommand{\arraystretch}{1}
\renewcommand{\arraystretch}{2}
\begin{table*}
\contcaption{} 
\centering
\resizebox{\linewidth}{!}{%
\begin{tabular}{ ccccccccccccc }
\hline 
\textbf{ID} & \textbf{\textit{$z_{phot}$}} & \textbf{$\chi_r^2$} & \textbf{$\log_{10}(M^{*}/M_{\odot}$)} & \begin{tabular}{@{}c@{}}\textbf{Abs.} \\ \textbf{Mag. $(i)$}\end{tabular} & \begin{tabular}{@{}c@{}}\textbf{Age} \\ \textbf{(Gyr)}\end{tabular} & \textbf{SFH} & \begin{tabular}{@{}c@{}}\textbf{[Z/H]} \\ \textbf{(Z$_{\odot}$)}\end{tabular} & \textbf{$\sigma_{AGN}$} & \textbf{$z_{DESonly}$} & \textbf{\textit{$z_{BPZ}$}} & \textbf{E (B-V)} \\
\hline 
618664306 & $4.33_{-0.12}^{+0.08}$ & 1.867 & $12.48_{-0.16}^{+0.23}$ & -27.93 & 0.36 & $e^{-t/0.1~\textrm{Gyr}}$ & 2 & 1.62 & 3.25 & 0.41 & 0.0 \\ 
		\hline
	\end{tabular}}
	\label{tab:cont:app-SMC}
\end{table*}
\renewcommand{\arraystretch}{1}

\renewcommand{\arraystretch}{2}
\begin{table*}
	\caption{As in Table~\ref{tab:app-SMC}, but for the Calzetti-type reddening.}
	\centering
	\resizebox{\linewidth}{!}{%
	\begin{tabular}{ ccccccccccccc }
		\hline
		\textbf{ID} & \textbf{\textit{$z_{phot}$}} & \textbf{$\chi_r^2$} & \textbf{$\log_{10}(M^{*}/M_{\odot}$)} & \begin{tabular}{@{}c@{}}\textbf{Abs.} \\ \textbf{Mag. $(i)$}\end{tabular} & \begin{tabular}{@{}c@{}}\textbf{Age} \\ \textbf{(Gyr)}\end{tabular} & \textbf{SFH} & \begin{tabular}{@{}c@{}}\textbf{[Z/H]} \\ \textbf{(Z$_{\odot}$)}\end{tabular} & \textbf{$\sigma_{AGN}$} & \textbf{$z_{DESonly}$} & \textbf{\textit{$z_{BPZ}$}} & \textbf{E (B-V)} \\
		\hline
		100669215 & $3.87_{-0.38}^{+0.39}$ & 0.863 & $11.74_{-0.28}^{+0.25}$ & -26.5 & 0.1 & $t_{trunc} = 0.1$ & 2 & 0.04 & 2.65 & 0.47 & 0.25 \\ 
102002089 & $4.22_{-0.44}^{+0.12}$ & 1.209 & $11.65_{-0.02}^{+0.04}$ & -26.26 & 0.1 & $t_{trunc} = 0.1$ & 2 & 5.59 & 3.31 & 0.51 & 0.25 \\ 
102009403 & $4.05_{-0.38}^{+0.13}$ & 2.106 & $12.07_{-0.33}^{+0.02}$ & -27.33 & 0.1 & $t_{trunc} = 0.1$ & 2 & 1.71 & 3.75 & 0.35 & 0.25 \\ 
102031864 & $4.01_{-0.19}^{+0.28}$ & 2.487 & $12.31_{-0.01}^{+0.4}$ & -27.6 & 0.1 & $t_{trunc} = 1.0$ & 2 & 3.34 & 3.58 & 0.5 & 0.37 \\ 
105765488 & $4.1_{-0.47}^{+0.24}$ & 1.199 & $12.19_{-0.28}^{+0.44}$ & -27.29 & 0.1 & $t_{trunc} = 0.1$ & 2 & 2.41 & 2.56 & 0.51 & 0.37 \\ 
115286147 & $3.96_{-0.32}^{+0.31}$ & 0.35 & $11.63_{-0.0}^{+0.15}$ & -26.23 & 0.1 & $t_{trunc} = 1.0$ & 2 & 0.66 & 3.25 & 0.43 & 0.25 \\ 
132987082 & $3.98_{-0.21}^{+0.22}$ & 0.733 & $11.96_{-0.04}^{+0.08}$ & -27.03 & 0.1 & $t_{trunc} = 0.1$ & 2 & 7.21 & 3.64 & 0.37 & 0.25 \\ 
133076071 & $3.56_{-0.25}^{+0.75}$ & 0.926 & $12.23_{-0.02}^{+0.6}$ & -26.75 & 0.1 & SSP & 2 & 1.9 & 3.4 & 0.49 & 0.25 \\ 
133575827 & $4.11_{-0.22}^{+0.18}$ & 1.165 & $12.23_{-0.05}^{+0.33}$ & -27.4 & 0.1 & $t_{trunc} = 0.1$ & 2 & 11.06 & 3.31 & 0.5 & 0.37 \\ 
133592684 & $3.81_{-0.48}^{+0.65}$ & 0.023 & $12.07_{-0.2}^{+0.62}$ & -26.65 & 0.13 & $t_{trunc} = 0.1$ & 2 & 1.18 & 3.71 & 0.51 & 0.25 \\ 
133755647 & $3.73_{-0.45}^{+0.63}$ & 0.404 & $12.41_{-0.18}^{+0.13}$ & -27.14 & 0.18 & $e^{-t/0.1~\textrm{Gyr}}$ & 2 & 0.38 & 3.75 & 0.33 & 0.37 \\ 
133779875 & $4.13_{-0.51}^{+0.2}$ & 0.978 & $12.11_{-0.25}^{+0.33}$ & -27.08 & 0.1 & $t_{trunc} = 0.1$ & 2 & 4.32 & 3.64 & 0.52 & 0.37 \\ 
133785852 & $3.69_{-0.38}^{+0.57}$ & 0.773 & $11.77_{-0.03}^{+0.47}$ & -26.26 & 0.1 & CONSTANT & 2 & 1.44 & 2.46 & 0.46 & 0.37 \\ 
134036466 & $3.96_{-0.34}^{+0.3}$ & 0.829 & $12.17_{-0.26}^{+0.48}$ & -27.24 & 0.1 & $t_{trunc} = 0.1$ & 2 & 1.73 & 3.73 & 0.51 & 0.37 \\ 
134797801 & $4.04_{-0.76}^{+0.29}$ & 1.2 & $11.67_{-0.17}^{+0.37}$ & -26.33 & 0.1 & $t_{trunc} = 0.1$ & 2 & 0.82 & 3.15 & 0.48 & 0.25 \\ 
135449486 & $3.73_{-0.54}^{+0.56}$ & 0.707 & $11.9_{-0.12}^{+0.25}$ & -26.57 & 0.1 & $t_{trunc} = 1.0$ & 2 & 2.37 & 3.25 & 0.4 & 0.37 \\ 
135756581 & $4.02_{-0.2}^{+0.19}$ & 0.85 & $12.44_{-0.01}^{+0.18}$ & -27.93 & 0.1 & $t_{trunc} = 1.0$ & 2 & 2.51 & 3.35 & 0.53 & 0.37 \\ 
135760809 & $3.75_{-0.31}^{+0.19}$ & 0.731 & $12.2_{-0.08}^{+0.3}$ & -27.32 & 0.1 & $t_{trunc} = 0.1$ & 2 & 1.05 & 3.69 & 0.45 & 0.37 \\ 
135856576 & $3.64_{-0.32}^{+0.31}$ & 1.148 & $12.19_{-0.03}^{+0.54}$ & -27.31 & 0.1 & $e^{-t/0.3~\textrm{Gyr}}$ & 1 & 0.4 & 3.64 & 0.42 & 0.37 \\ 
135857162 & $3.86_{-0.43}^{+0.32}$ & 1.097 & $12.0_{-0.06}^{+0.09}$ & -26.82 & 0.1 & CONSTANT & 2 & 0.05 & 3.45 & 0.39 & 0.37 \\ 
136034648 & $4.01_{-0.53}^{+0.37}$ & 0.488 & $11.5_{-0.07}^{+0.52}$ & -25.88 & 0.1 & $t_{trunc} = 0.1$ & 2 & 0.39 & 3.79 & 0.42 & 0.25 \\ 
137552954 & $3.8_{-0.36}^{+0.31}$ & 1.028 & $12.21_{-0.02}^{+0.19}$ & -27.36 & 0.1 & CONSTANT & 2 & 1.72 & 3.4 & 0.41 & 0.37 \\ 
137650861 & $3.44_{-0.08}^{+0.61}$ & 0.733 & $12.1_{-0.04}^{+0.66}$ & -26.95 & 0.11 & $t_{trunc} = 0.1$ & 2 & 1.76 & 3.4 & 0.42 & 0.25 \\ 
137806706 & $3.94_{-0.43}^{+0.53}$ & 0.234 & $12.12_{-0.3}^{+0.27}$ & -26.84 & 0.32 & $t_{trunc} = 0.3$ & 2 & 0.06 & 3.94 & 0.49 & 0.12 \\ 
164738777 & $3.69_{-0.53}^{+0.53}$ & 0.571 & $12.22_{-0.46}^{+0.15}$ & -26.8 & 0.23 & $e^{-t/0.1~\textrm{Gyr}}$ & 2 & 1.85 & 2.38 & 0.51 & 0.25 \\ 
285308599 & $3.81_{-0.52}^{+0.6}$ & 0.176 & $11.9_{-0.01}^{+0.58}$ & -26.46 & 0.11 & $t_{trunc} = 0.1$ & 2 & 0.37 & 2.58 & 0.4 & 0.25 \\ 
287114376 & $4.0_{-0.27}^{+0.31}$ & 1.173 & $12.3_{-0.01}^{+0.31}$ & -27.57 & 0.1 & $t_{trunc} = 1.0$ & 2 & 2.52 & 3.33 & 0.41 & 0.37 \\ 
287127591 & $3.67_{-0.42}^{+0.64}$ & 0.138 & $11.95_{-0.09}^{+0.41}$ & -26.68 & 0.1 & $t_{trunc} = 0.1$ & 2 & 1.94 & 3.73 & 0.4 & 0.37 \\ 
289328303 & $4.01_{-0.39}^{+0.29}$ & 0.6 & $12.04_{-0.1}^{+0.37}$ & -26.91 & 0.1 & $t_{trunc} = 0.1$ & 2 & 0.3 & 2.52 & 0.45 & 0.37 \\ 
289329064 & $4.2_{-0.45}^{+0.18}$ & 0.744 & $12.01_{-0.23}^{+0.35}$ & -26.86 & 0.1 & $t_{trunc} = 1.0$ & 2 & 2.4 & 3.69 & 0.54 & 0.37 \\ 
290792079 & $3.42_{-0.2}^{+0.85}$ & 1.864 & $12.01_{-0.14}^{+0.69}$ & -26.74 & 0.11 & $t_{trunc} = 0.1$ & 2 & 2.83 & 3.42 & 0.36 & 0.25 \\ 
395017226 & $3.69_{-0.33}^{+0.35}$ & 0.435 & $12.23_{-0.07}^{+0.39}$ & -27.34 & 0.1 & $e^{-t/0.3~\textrm{Gyr}}$ & 2 & 2.96 & 3.4 & 0.41 & 0.37 \\ 
395746810 & $4.04_{-0.29}^{+0.26}$ & 0.244 & $12.28_{-0.0}^{+0.35}$ & -27.54 & 0.1 & $t_{trunc} = 1.0$ & 2 & 8.31 & 3.25 & 0.45 & 0.37 \\ 
396223342 & $4.04_{-0.18}^{+0.17}$ & 0.557 & $11.92_{-0.34}^{+0.32}$ & -26.95 & 0.1 & $t_{trunc} = 0.3$ & 2 & 4.44 & 3.44 & 0.37 & 0.25 \\ 
396276124 & $3.85_{-0.18}^{+0.32}$ & 0.701 & $12.32_{-0.01}^{+0.36}$ & -27.63 & 0.1 & $t_{trunc} = 1.0$ & 2 & 2.61 & 3.42 & 0.48 & 0.37 \\ 
396551822 & $3.77_{-0.38}^{+0.54}$ & 0.231 & $12.07_{-0.03}^{+0.06}$ & -27.01 & 0.1 & $t_{trunc} = 1.0$ & 2 & 1.27 & 3.29 & 0.47 & 0.37 \\ 
\hline 
\end{tabular}} 
\label{tab:app-Calzetti} 
\end{table*} 
\renewcommand{\arraystretch}{1}
\renewcommand{\arraystretch}{2}
\begin{table*}
\contcaption{} 
\centering
\resizebox{\linewidth}{!}{%
\begin{tabular}{ ccccccccccccc }
\hline 
\textbf{ID} & \textbf{\textit{$z_{phot}$}} & \textbf{$\chi_r^2$} & \textbf{$\log_{10}(M^{*}/M_{\odot}$)} & \begin{tabular}{@{}c@{}}\textbf{Abs.} \\ \textbf{Mag. $(i)$}\end{tabular} & \begin{tabular}{@{}c@{}}\textbf{Age} \\ \textbf{(Gyr)}\end{tabular} & \textbf{SFH} & \begin{tabular}{@{}c@{}}\textbf{[Z/H]} \\ \textbf{(Z$_{\odot}$)}\end{tabular} & \textbf{$\sigma_{AGN}$} & \textbf{$z_{DESonly}$} & \textbf{\textit{$z_{BPZ}$}} & \textbf{E (B-V)} \\
\hline 
397300605 & $4.14_{-0.13}^{+0.15}$ & 1.197 & $12.03_{-0.02}^{+0.4}$ & -27.23 & 0.1 & $t_{trunc} = 1.0$ & 2 & 7.77 & 3.29 & 0.47 & 0.25 \\ 
397303505 & $3.8_{-0.34}^{+0.34}$ & 1.263 & $12.19_{-0.02}^{+0.18}$ & -27.29 & 0.1 & CONSTANT & 2 & 4.8 & 3.4 & 0.49 & 0.37 \\ 
397554368 & $3.77_{-0.33}^{+0.54}$ & 0.717 & $12.12_{-0.05}^{+0.41}$ & -27.09 & 0.11 & $t_{trunc} = 0.1$ & 1 & 3.19 & 3.61 & 0.37 & 0.25 \\ 
397885462 & $4.14_{-0.67}^{+0.22}$ & 0.569 & $12.11_{-0.18}^{+0.53}$ & -27.08 & 0.1 & $t_{trunc} = 0.1$ & 2 & 1.74 & 3.29 & 0.5 & 0.37 \\ 
398107560 & $3.91_{-0.21}^{+0.29}$ & 0.699 & $12.25_{-0.04}^{+0.13}$ & -27.46 & 0.1 & CONSTANT & 2 & 3.3 & 3.4 & 0.45 & 0.37 \\ 
399842613 & $4.25_{-0.16}^{+0.15}$ & 1.231 & $12.22_{-0.02}^{+0.45}$ & -27.37 & 0.1 & $t_{trunc} = 0.1$ & 2 & 0.36 & 3.32 & 0.59 & 0.37 \\ 
401003476 & $3.86_{-0.53}^{+0.42}$ & 0.44 & $12.0_{-0.34}^{+0.19}$ & -26.54 & 0.16 & $e^{-t/0.1~\textrm{Gyr}}$ & 2 & 1.37 & 3.4 & 0.39 & 0.25 \\ 
401582291 & $3.7_{-0.46}^{+0.6}$ & 0.259 & $11.91_{-0.04}^{+0.35}$ & -26.49 & 0.11 & $t_{trunc} = 0.1$ & 2 & 2.58 & 2.46 & 0.4 & 0.25 \\ 
404760121 & $3.88_{-0.28}^{+0.33}$ & 0.224 & $11.79_{-0.3}^{+0.43}$ & -26.63 & 0.1 & $t_{trunc} = 0.1$ & 2 & 4.56 & 3.61 & 0.4 & 0.25 \\ 
404798494 & $3.92_{-0.1}^{+0.17}$ & 0.823 & $12.02_{-0.05}^{+0.04}$ & -27.2 & 0.1 & $t_{trunc} = 0.1$ & 2 & 7.32 & 3.34 & 0.44 & 0.25 \\ 
404907811 & $3.72_{-0.32}^{+0.42}$ & 0.664 & $12.28_{-0.02}^{+0.21}$ & -27.52 & 0.1 & $e^{-t/1.0~\textrm{Gyr}}$ & 2 & 6.14 & 3.4 & 0.36 & 0.37 \\ 
405529691 & $3.71_{-0.31}^{+0.43}$ & 0.264 & $12.42_{-0.12}^{+0.12}$ & -27.72 & 0.11 & $e^{-t/0.3~\textrm{Gyr}}$ & 2 & 7.91 & 3.42 & 0.41 & 0.37 \\ 
405537460 & $4.2_{-0.84}^{+0.18}$ & 0.301 & $12.07_{-0.11}^{+0.52}$ & -26.99 & 0.1 & $t_{trunc} = 0.1$ & 2 & 7.74 & 3.39 & 0.59 & 0.37 \\ 
405539533 & $3.8_{-0.37}^{+0.35}$ & 0.246 & $12.2_{-0.03}^{+0.19}$ & -27.33 & 0.1 & $t_{trunc} = 1.0$ & 2 & 2.12 & 3.45 & 0.46 & 0.37 \\ 
405686502 & $3.99_{-0.38}^{+0.32}$ & 0.789 & $11.63_{-0.34}^{+0.15}$ & -26.22 & 0.1 & $t_{trunc} = 1.0$ & 2 & 1.33 & 3.25 & 0.41 & 0.25 \\ 
405937444 & $4.04_{-0.14}^{+0.12}$ & 1.5 & $12.21_{-0.35}^{+0.03}$ & -27.67 & 0.1 & $t_{trunc} = 0.1$ & 2 & 11.74 & 3.4 & 0.42 & 0.25 \\ 
406039218 & $3.76_{-0.17}^{+0.37}$ & 1.012 & $12.5_{-0.0}^{+0.15}$ & -28.09 & 0.1 & $t_{trunc} = 1.0$ & 2 & 22.07 & 3.42 & 0.39 & 0.37 \\ 
406366767 & $3.44_{-0.25}^{+0.86}$ & 0.957 & $12.04_{-0.0}^{+0.25}$ & -26.82 & 0.11 & $t_{trunc} = 0.1$ & 2 & 0.03 & 3.29 & 0.43 & 0.25 \\ 
407630148 & $3.88_{-0.34}^{+0.38}$ & 0.985 & $12.05_{-0.02}^{+0.16}$ & -26.95 & 0.1 & CONSTANT & 2 & 2.07 & 3.37 & 0.53 & 0.37 \\ 
408132796 & $3.76_{-0.33}^{+0.4}$ & 1.048 & $12.36_{-0.09}^{+0.48}$ & -27.71 & 0.1 & $t_{trunc} = 0.1$ & 2 & 4.01 & 3.34 & 0.46 & 0.37 \\ 
408135057 & $4.08_{-0.24}^{+0.24}$ & 1.633 & $12.08_{-0.02}^{+0.42}$ & -27.02 & 0.1 & $t_{trunc} = 1.0$ & 2 & 0.64 & 3.75 & 0.48 & 0.37 \\ 
409127588 & $4.11_{-0.63}^{+0.21}$ & 1.095 & $12.14_{-0.23}^{+0.35}$ & -27.18 & 0.1 & $t_{trunc} = 1.0$ & 2 & 4.41 & 3.48 & 0.52 & 0.37 \\ 
410163990 & $4.11_{-0.49}^{+0.22}$ & 0.959 & $12.22_{-0.23}^{+0.26}$ & -27.38 & 0.1 & $t_{trunc} = 0.1$ & 2 & 0.52 & 2.51 & 0.54 & 0.37 \\ 
411500732 & $4.17_{-0.28}^{+0.13}$ & 1.938 & $11.77_{-0.31}^{+0.03}$ & -26.57 & 0.1 & $t_{trunc} = 0.1$ & 2 & 0.7 & 3.37 & 0.46 & 0.25 \\ 
411502452 & $3.94_{-0.42}^{+0.28}$ & 0.828 & $11.96_{-0.32}^{+0.46}$ & -27.06 & 0.1 & $t_{trunc} = 0.1$ & 2 & 2.72 & 3.4 & 0.41 & 0.25 \\ 
412637681 & $3.94_{-0.47}^{+0.26}$ & 1.938 & $11.79_{-0.29}^{+0.44}$ & -26.61 & 0.1 & $t_{trunc} = 0.1$ & 2 & 0.49 & 2.48 & 0.46 & 0.25 \\ 
414173316 & $3.67_{-0.2}^{+0.5}$ & 0.421 & $11.76_{-0.0}^{+0.31}$ & -26.56 & 0.1 & CONSTANT & 2 & 2.24 & 3.4 & 0.34 & 0.25 \\ 
414235028 & $4.04_{-0.61}^{+0.29}$ & 0.151 & $12.08_{-0.19}^{+0.46}$ & -27.07 & 0.16 & CONSTANT & 2 & 8.39 & 3.29 & 0.35 & 0.25 \\ 
414237423 & $4.0_{-0.3}^{+0.31}$ & 0.331 & $12.26_{-0.01}^{+0.17}$ & -27.48 & 0.1 & $t_{trunc} = 1.0$ & 2 & 2.3 & 3.46 & 0.43 & 0.37 \\ 
414248322 & $3.8_{-0.27}^{+0.48}$ & 0.397 & $12.08_{-0.04}^{+0.12}$ & -27.06 & 0.1 & $t_{trunc} = 0.3$ & 1/5 & 4.73 & 3.67 & 0.45 & 0.37 \\ 
415246403 & $3.79_{-0.35}^{+0.37}$ & 1.017 & $11.98_{-0.02}^{+0.35}$ & -26.79 & 0.1 & CONSTANT & 2 & 0.17 & 2.46 & 0.49 & 0.37 \\ 
417446833 & $3.86_{-0.46}^{+0.51}$ & 0.283 & $12.12_{-0.23}^{+0.39}$ & -26.77 & 0.13 & $t_{trunc} = 0.1$ & 2 & 0.43 & 3.65 & 0.57 & 0.25 \\ 
417565001 & $3.83_{-0.34}^{+0.13}$ & 2.192 & $11.86_{-0.28}^{+0.03}$ & -26.8 & 0.1 & $t_{trunc} = 0.1$ & 2 & 2.01 & 2.42 & 0.37 & 0.25 \\ 
417565185 & $3.83_{-0.33}^{+0.41}$ & 0.801 & $12.21_{-0.02}^{+0.08}$ & -27.35 & 0.1 & CONSTANT & 2 & 1.49 & 3.77 & 0.39 & 0.37 \\ 
417579802 & $3.83_{-0.48}^{+0.41}$ & 0.281 & $12.41_{-0.04}^{+0.46}$ & -27.52 & 0.1 & $t_{trunc} = 0.1$ & 2 & 6.89 & 3.5 & 0.51 & 0.49 \\ 
429617726 & $3.92_{-0.26}^{+0.33}$ & 0.289 & $12.3_{-0.29}^{+0.44}$ & -27.56 & 0.1 & $t_{trunc} = 0.1$ & 2 & 5.91 & 3.34 & 0.46 & 0.37 \\ 
\hline 
\end{tabular}} 
\label{tab:cont:app-Calzetti} 
\end{table*} 
\renewcommand{\arraystretch}{1}
\renewcommand{\arraystretch}{2}
\begin{table*}
\contcaption{} 
\centering
\resizebox{\linewidth}{!}{%
\begin{tabular}{ ccccccccccccc }
\hline 
\textbf{ID} & \textbf{\textit{$z_{phot}$}} & \textbf{$\chi_r^2$} & \textbf{$\log_{10}(M^{*}/M_{\odot}$)} & \begin{tabular}{@{}c@{}}\textbf{Abs.} \\ \textbf{Mag. $(i)$}\end{tabular} & \begin{tabular}{@{}c@{}}\textbf{Age} \\ \textbf{(Gyr)}\end{tabular} & \textbf{SFH} & \begin{tabular}{@{}c@{}}\textbf{[Z/H]} \\ \textbf{(Z$_{\odot}$)}\end{tabular} & \textbf{$\sigma_{AGN}$} & \textbf{$z_{DESonly}$} & \textbf{\textit{$z_{BPZ}$}} & \textbf{E (B-V)} \\
\hline 
431449768 & $3.76_{-0.37}^{+0.4}$ & 0.636 & $12.1_{-0.03}^{+0.07}$ & -27.07 & 0.1 & $t_{trunc} = 1.0$ & 2 & 3.77 & 3.4 & 0.34 & 0.37 \\ 
431455424 & $3.81_{-0.24}^{+0.53}$ & 1.217 & $12.13_{-0.09}^{+0.27}$ & -27.04 & 0.13 & SSP & 2 & 14.07 & 3.42 & 0.64 & 0.0 \\ 
444147103 & $3.87_{-0.39}^{+0.34}$ & 0.609 & $12.23_{-0.04}^{+0.18}$ & -27.39 & 0.1 & $t_{trunc} = 1.0$ & 2 & 4.9 & 3.44 & 0.46 & 0.37 \\ 
444182193 & $3.89_{-0.41}^{+0.3}$ & 0.76 & $12.26_{-0.22}^{+0.39}$ & -27.47 & 0.1 & $t_{trunc} = 0.1$ & 2 & 9.0 & 3.64 & 0.43 & 0.37 \\ 
446501990 & $4.24_{-0.23}^{+0.17}$ & 0.934 & $12.18_{-0.03}^{+0.36}$ & -27.27 & 0.1 & $t_{trunc} = 1.0$ & 2 & 2.7 & 3.88 & 0.5 & 0.37 \\ 
465281154 & $4.13_{-0.44}^{+0.2}$ & 0.128 & $11.84_{-0.36}^{+0.52}$ & -26.75 & 0.1 & $t_{trunc} = 0.1$ & 2 & 8.61 & 3.4 & 0.39 & 0.25 \\ 
470611726 & $3.88_{-0.5}^{+0.38}$ & 0.538 & $12.04_{-0.06}^{+0.27}$ & -26.93 & 0.1 & $t_{trunc} = 0.1$ & 2 & 7.27 & 3.64 & 0.5 & 0.37 \\ 
470971747 & $3.83_{-0.46}^{+0.52}$ & 0.57 & $12.39_{-0.28}^{+0.0}$ & -27.22 & 1.02 & $t_{trunc} = 1.0$ & 1 & 2.36 & 2.54 & 0.28 & 0.12 \\ 
471106730 & $4.11_{-0.82}^{+0.27}$ & 0.184 & $12.09_{-0.1}^{+0.02}$ & -26.62 & 0.29 & $e^{-t/0.3~\textrm{Gyr}}$ & 2 & 0.54 & 3.38 & 0.42 & 0.25 \\ 
471394809 & $3.88_{-0.46}^{+0.42}$ & 0.932 & $12.22_{-0.25}^{+0.0}$ & -26.89 & 0.32 & $e^{-t/0.3~\textrm{Gyr}}$ & 2 & 4.72 & 3.81 & 0.4 & 0.25 \\ 
471566339 & $3.71_{-0.27}^{+0.3}$ & 0.576 & $11.84_{-0.04}^{+0.38}$ & -26.75 & 0.1 & $t_{trunc} = 0.1$ & 2 & 5.77 & 2.46 & 0.29 & 0.25 \\ 
471600124 & $4.14_{-0.51}^{+0.26}$ & 0.213 & $11.77_{-0.28}^{+0.3}$ & -26.36 & 0.14 & CONSTANT & 2 & 2.32 & 3.25 & 0.49 & 0.25 \\ 
471612288 & $3.92_{-0.45}^{+0.33}$ & 0.448 & $11.63_{-0.31}^{+0.39}$ & -26.21 & 0.1 & $t_{trunc} = 0.1$ & 2 & 0.92 & 3.29 & 0.39 & 0.25 \\ 
471703164 & $3.83_{-0.46}^{+0.43}$ & 0.273 & $11.54_{-0.19}^{+0.26}$ & -26.0 & 0.1 & CONSTANT & 2 & 0.51 & 3.46 & 0.36 & 0.25 \\ 
473133985 & $4.01_{-0.28}^{+0.18}$ & 0.903 & $11.99_{-0.29}^{+0.16}$ & -27.11 & 0.1 & $t_{trunc} = 0.1$ & 2 & 7.56 & 3.58 & 0.38 & 0.25 \\ 
473140970 & $3.96_{-0.09}^{+0.19}$ & 0.896 & $12.11_{-0.28}^{+0.29}$ & -27.43 & 0.1 & $t_{trunc} = 0.1$ & 2 & 9.56 & 3.43 & 0.42 & 0.25 \\ 
473404298 & $3.85_{-0.45}^{+0.53}$ & 0.14 & $11.98_{-0.15}^{+0.42}$ & -26.76 & 0.1 & $e^{-t/1.0~\textrm{Gyr}}$ & 2 & 0.43 & 3.69 & 0.47 & 0.37 \\ 
473408311 & $4.08_{-0.13}^{+0.14}$ & 0.674 & $12.15_{-0.28}^{+0.26}$ & -27.51 & 0.1 & $t_{trunc} = 0.1$ & 2 & 8.66 & 3.65 & 0.44 & 0.25 \\ 
473411673 & $4.1_{-0.33}^{+0.15}$ & 0.533 & $11.91_{-0.35}^{+0.26}$ & -26.91 & 0.1 & $t_{trunc} = 0.1$ & 2 & 5.69 & 3.61 & 0.45 & 0.25 \\ 
473496203 & $4.16_{-0.55}^{+0.22}$ & 0.383 & $11.41_{-0.16}^{+0.51}$ & -26.07 & 0.1 & $t_{trunc} = 0.1$ & 1 & 0.37 & 3.05 & 0.38 & 0.12 \\ 
473498930 & $4.05_{-0.08}^{+0.13}$ & 1.032 & $12.31_{-0.04}^{+0.03}$ & -27.91 & 0.1 & $t_{trunc} = 0.1$ & 2 & 20.72 & 3.53 & 0.41 & 0.25 \\ 
473503196 & $4.13_{-0.25}^{+0.23}$ & 0.547 & $12.15_{-0.01}^{+0.43}$ & -27.19 & 0.1 & $t_{trunc} = 1.0$ & 2 & 3.37 & 3.31 & 0.59 & 0.37 \\ 
473511031 & $3.86_{-0.19}^{+0.34}$ & 1.458 & $12.21_{-0.04}^{+0.36}$ & -27.36 & 0.1 & CONSTANT & 2 & 1.92 & 2.44 & 0.41 & 0.37 \\ 
473512115 & $4.11_{-0.29}^{+0.18}$ & 1.324 & $12.01_{-0.37}^{+0.51}$ & -27.16 & 0.1 & $t_{trunc} = 0.1$ & 2 & 3.44 & 3.62 & 0.43 & 0.25 \\ 
473514761 & $4.2_{-0.47}^{+0.14}$ & 1.842 & $12.25_{-0.45}^{+0.23}$ & -27.03 & 0.72 & $e^{-t/1.0~\textrm{Gyr}}$ & 2 & 3.02 & 3.15 & 0.46 & 0.12 \\ 
473515047 & $3.81_{-0.37}^{+0.41}$ & 0.625 & $12.35_{-0.15}^{+0.03}$ & -27.31 & 0.26 & $e^{-t/0.3~\textrm{Gyr}}$ & 2 & 3.13 & 3.4 & 0.37 & 0.25 \\ 
473515263 & $4.08_{-0.16}^{+0.17}$ & 0.998 & $11.8_{-0.01}^{+0.16}$ & -26.65 & 0.1 & CONSTANT & 2 & 2.27 & 3.34 & 0.45 & 0.25 \\ 
473519025 & $4.01_{-0.31}^{+0.29}$ & 0.714 & $12.18_{-0.02}^{+0.15}$ & -27.27 & 0.1 & $t_{trunc} = 1.0$ & 2 & 0.81 & 3.46 & 0.5 & 0.37 \\ 
473520285 & $3.8_{-0.18}^{+0.37}$ & 1.077 & $12.24_{-0.0}^{+0.34}$ & -27.44 & 0.1 & CONSTANT & 2 & 3.35 & 3.4 & 0.38 & 0.37 \\ 
473520601 & $4.14_{-0.38}^{+0.19}$ & 0.503 & $12.23_{-0.1}^{+0.22}$ & -27.41 & 0.1 & $t_{trunc} = 1.0$ & 2 & 2.05 & 3.55 & 0.56 & 0.37 \\ 
473521671 & $4.13_{-0.18}^{+0.14}$ & 0.866 & $12.3_{-0.03}^{+0.02}$ & -27.57 & 0.1 & $t_{trunc} = 0.1$ & 2 & 2.5 & 3.45 & 0.58 & 0.37 \\ 
473528868 & $4.05_{-0.57}^{+0.31}$ & 0.208 & $11.64_{-0.31}^{+0.37}$ & -26.24 & 0.1 & $t_{trunc} = 0.1$ & 2 & 1.9 & 3.63 & 0.5 & 0.25 \\ 
473530252 & $4.11_{-0.26}^{+0.22}$ & 0.371 & $12.02_{-0.3}^{+0.82}$ & -27.2 & 0.1 & $t_{trunc} = 0.1$ & 2 & 0.76 & 3.75 & 0.46 & 0.25 \\ 
473532585 & $4.04_{-0.31}^{+0.22}$ & 0.523 & $11.91_{-0.35}^{+0.43}$ & -26.91 & 0.1 & $t_{trunc} = 0.1$ & 2 & 0.78 & 3.33 & 0.43 & 0.25 \\ 
476998818 & $3.76_{-0.37}^{+0.38}$ & 1.496 & $12.39_{-0.27}^{+0.07}$ & -27.81 & 0.1 & $t_{trunc} = 1.0$ & 2 & 8.94 & 3.46 & 0.33 & 0.37 \\ 
477008049 & $3.8_{-0.58}^{+0.52}$ & 0.339 & $12.28_{-0.13}^{+0.2}$ & -27.18 & 0.2 & CONSTANT & 1 & 4.79 & 3.45 & 0.51 & 0.37 \\ 
\hline 
\end{tabular}} 
\label{tab:cont:app-Calzetti} 
\end{table*} 
\renewcommand{\arraystretch}{1}
\renewcommand{\arraystretch}{2}
\begin{table*}
\contcaption{} 
\centering
\resizebox{\linewidth}{!}{%
\begin{tabular}{ ccccccccccccc }
\hline 
\textbf{ID} & \textbf{\textit{$z_{phot}$}} & \textbf{$\chi_r^2$} & \textbf{$\log_{10}(M^{*}/M_{\odot}$)} & \begin{tabular}{@{}c@{}}\textbf{Abs.} \\ \textbf{Mag. $(i)$}\end{tabular} & \begin{tabular}{@{}c@{}}\textbf{Age} \\ \textbf{(Gyr)}\end{tabular} & \textbf{SFH} & \begin{tabular}{@{}c@{}}\textbf{[Z/H]} \\ \textbf{(Z$_{\odot}$)}\end{tabular} & \textbf{$\sigma_{AGN}$} & \textbf{$z_{DESonly}$} & \textbf{\textit{$z_{BPZ}$}} & \textbf{E (B-V)} \\
\hline 
477008438 & $4.2_{-0.14}^{+0.08}$ & 2.353 & $11.92_{-0.02}^{+0.03}$ & -26.95 & 0.1 & CONSTANT & 2 & 4.99 & 3.29 & 0.45 & 0.25 \\ 
479472291 & $3.83_{-0.52}^{+0.49}$ & 0.957 & $11.85_{-0.05}^{+0.57}$ & -26.44 & 0.1 & $t_{trunc} = 0.1$ & 2 & 0.38 & 2.4 & 0.56 & 0.37 \\ 
479999051 & $3.46_{-0.06}^{+0.81}$ & 0.255 & $12.35_{-0.12}^{+0.43}$ & -27.57 & 0.11 & $t_{trunc} = 0.1$ & 2 & 3.53 & 3.4 & 0.36 & 0.25 \\ 
480008436 & $3.48_{-0.2}^{+0.82}$ & 0.365 & $12.47_{-0.15}^{+0.67}$ & -27.89 & 0.11 & $t_{trunc} = 0.1$ & 2 & 10.76 & 3.29 & 0.41 & 0.25 \\ 
480339250 & $4.13_{-0.53}^{+0.19}$ & 0.613 & $12.3_{-0.07}^{+0.53}$ & -27.57 & 0.1 & $t_{trunc} = 0.1$ & 2 & 6.05 & 3.34 & 0.55 & 0.37 \\ 
481065880 & $3.87_{-0.29}^{+0.43}$ & 1.102 & $12.27_{-0.01}^{+0.31}$ & -27.51 & 0.1 & $t_{trunc} = 1.0$ & 2 & 4.82 & 2.64 & 0.5 & 0.37 \\ 
481350973 & $3.97_{-0.35}^{+0.33}$ & 0.275 & $11.94_{-0.34}^{+0.39}$ & -26.98 & 0.1 & $t_{trunc} = 0.1$ & 2 & 5.37 & 3.64 & 0.4 & 0.25 \\ 
481989803 & $3.76_{-0.47}^{+0.62}$ & 0.543 & $12.0_{-0.03}^{+0.59}$ & -26.71 & 0.11 & $t_{trunc} = 0.1$ & 2 & 3.48 & 3.29 & 0.43 & 0.25 \\ 
481994767 & $4.2_{-0.63}^{+0.25}$ & 0.245 & $11.5_{-0.25}^{+0.54}$ & -25.89 & 0.1 & $t_{trunc} = 0.1$ & 2 & 0.98 & 3.49 & 0.51 & 0.25 \\ 
482001634 & $3.88_{-0.56}^{+0.38}$ & 1.609 & $12.04_{-0.29}^{+0.01}$ & -26.44 & 0.32 & $e^{-t/0.3~\textrm{Gyr}}$ & 2 & 0.04 & 3.34 & 0.44 & 0.25 \\ 
483918716 & $4.11_{-0.83}^{+0.25}$ & 0.701 & $12.45_{-0.38}^{+0.06}$ & -27.28 & 0.64 & $e^{-t/1.0~\textrm{Gyr}}$ & 2 & 0.5 & 3.37 & 0.43 & 0.25 \\ 
489254835 & $4.11_{-0.39}^{+0.25}$ & 0.866 & $12.21_{-0.11}^{+0.42}$ & -27.34 & 0.1 & $t_{trunc} = 1.0$ & 2 & 4.53 & 3.75 & 0.48 & 0.37 \\ 
490689649 & $4.1_{-0.33}^{+0.11}$ & 1.02 & $12.41_{-0.1}^{+0.07}$ & -27.88 & 0.11 & $e^{-t/0.1~\textrm{Gyr}}$ & 2 & 23.72 & 3.25 & 0.42 & 0.25 \\ 
490704656 & $4.17_{-0.71}^{+0.21}$ & 0.896 & $12.45_{-0.38}^{+0.13}$ & -27.31 & 0.45 & $e^{-t/0.3~\textrm{Gyr}}$ & 2 & 0.75 & 3.29 & 0.51 & 0.25 \\ 
492431224 & $4.17_{-0.8}^{+0.19}$ & 0.868 & $12.22_{-0.05}^{+0.74}$ & -27.36 & 0.1 & $t_{trunc} = 0.1$ & 2 & 7.22 & 3.29 & 0.54 & 0.37 \\ 
492605523 & $3.77_{-0.12}^{+0.33}$ & 0.595 & $11.8_{-0.27}^{+0.36}$ & -26.64 & 0.1 & $t_{trunc} = 0.1$ & 2 & 5.57 & 3.75 & 0.41 & 0.25 \\ 
493212188 & $4.08_{-0.37}^{+0.18}$ & 0.799 & $11.83_{-0.34}^{+0.32}$ & -26.72 & 0.1 & $t_{trunc} = 0.1$ & 2 & 3.81 & 3.5 & 0.4 & 0.25 \\ 
493739755 & $4.21_{-0.45}^{+0.21}$ & 0.94 & $12.1_{-0.24}^{+0.43}$ & -27.07 & 0.1 & $t_{trunc} = 0.3$ & 2 & 4.09 & 3.15 & 0.5 & 0.37 \\ 
493882026 & $3.69_{-0.46}^{+0.4}$ & 0.421 & $11.96_{-0.14}^{+0.19}$ & -26.71 & 0.1 & $e^{-t/1.0~\textrm{Gyr}}$ & 2 & 1.0 & 2.46 & 0.44 & 0.37 \\ 
494789087 & $3.96_{-0.38}^{+0.48}$ & 1.445 & $11.21_{-0.1}^{+0.55}$ & -25.38 & 0.11 & $t_{trunc} = 0.1$ & 2 & 0.73 & 2.63 & 0.4 & 0.0 \\ 
494790027 & $4.04_{-0.38}^{+0.35}$ & 1.06 & $11.84_{-0.05}^{+0.36}$ & -26.62 & 0.11 & $t_{trunc} = 0.1$ & 2 & 3.18 & 2.58 & 0.52 & 0.12 \\ 
494790169 & $4.08_{-0.39}^{+0.32}$ & 0.243 & $11.79_{-0.06}^{+0.55}$ & -26.66 & 0.1 & CONSTANT & 1/5 & 1.53 & 3.96 & 0.42 & 0.25 \\ 
494790792 & $3.95_{-0.65}^{+0.44}$ & 0.243 & $12.24_{-0.05}^{+0.38}$ & -27.49 & 0.1 & $t_{trunc} = 0.1$ & 1 & 4.72 & 3.26 & 0.5 & 0.37 \\ 
494791393 & $3.75_{-0.4}^{+0.61}$ & 0.724 & $11.65_{-0.21}^{+0.14}$ & -25.82 & 0.13 & SSP & 2 & 0.52 & 3.4 & 0.59 & 0.0 \\ 
494792459 & $4.25_{-0.42}^{+0.2}$ & 0.761 & $11.16_{-0.63}^{+1.67}$ & -25.36 & 0.1 & $t_{trunc} = 0.1$ & 2 & 0.36 & 3.15 & 0.44 & 0.12 \\ 
494793098 & $4.02_{-0.32}^{+0.4}$ & 0.179 & $11.85_{-0.1}^{+0.34}$ & -26.65 & 0.11 & $t_{trunc} = 0.1$ & 2 & 0.33 & 3.81 & 0.48 & 0.12 \\ 
494793167 & $4.09_{-0.47}^{+0.24}$ & 0.229 & $11.8_{-0.35}^{+0.38}$ & -26.64 & 0.1 & $t_{trunc} = 0.1$ & 2 & 2.53 & 3.75 & 0.41 & 0.25 \\ 
494800805 & $4.25_{-1.22}^{+0.1}$ & 0.196 & $12.29_{-0.5}^{+0.26}$ & -27.19 & 0.81 & $e^{-t/0.3~\textrm{Gyr}}$ & 2 & 0.64 & 3.15 & 0.39 & 0.0 \\ 
494801634 & $3.34_{-0.26}^{+0.53}$ & 0.373 & $12.04_{-0.06}^{+0.61}$ & -26.95 & 0.1 & SSP & 1/2 & 0.95 & 3.44 & 0.35 & 0.12 \\ 
495323159 & $3.92_{-0.48}^{+0.41}$ & 1.814 & $11.5_{-0.07}^{+0.22}$ & -25.9 & 0.11 & $t_{trunc} = 0.1$ & 1/5 & 0.54 & 2.65 & 0.41 & 0.12 \\ 
495325646 & $3.74_{-0.45}^{+0.58}$ & 0.378 & $11.85_{-0.16}^{+0.25}$ & -26.45 & 0.1 & $t_{trunc} = 1.0$ & 2 & 0.18 & 3.29 & 0.48 & 0.37 \\ 
495342175 & $3.75_{-0.25}^{+0.29}$ & 1.587 & $12.2_{-0.05}^{+0.38}$ & -27.32 & 0.1 & $t_{trunc} = 0.1$ & 2 & 0.81 & 2.46 & 0.4 & 0.37 \\ 
495508558 & $4.25_{-0.18}^{+0.11}$ & 1.819 & $11.86_{-0.06}^{+0.4}$ & -26.79 & 0.1 & $t_{trunc} = 0.3$ & 2 & 0.02 & 3.29 & 0.39 & 0.25 \\ 
495566911 & $3.12_{-0.15}^{+0.21}$ & 0.438 & $11.81_{-0.11}^{+0.03}$ & -26.12 & 0.16 & SSP & 2 & 0.56 & 4.29 & 0.54 & 0.0 \\ 
496787409 & $4.14_{-0.49}^{+0.23}$ & 0.246 & $11.77_{-0.34}^{+0.53}$ & -26.58 & 0.1 & $t_{trunc} = 0.1$ & 2 & 4.22 & 3.19 & 0.45 & 0.25 \\ 
497171956 & $4.16_{-0.08}^{+0.1}$ & 0.851 & $12.24_{-0.29}^{+0.04}$ & -27.75 & 0.1 & $t_{trunc} = 0.1$ & 2 & 6.9 & 3.75 & 0.44 & 0.25 \\ 
\hline 
\end{tabular}} 
\label{tab:cont:app-Calzetti} 
\end{table*} 
\renewcommand{\arraystretch}{1}
\renewcommand{\arraystretch}{2}
\begin{table*}
\contcaption{} 
\centering
\resizebox{\linewidth}{!}{%
\begin{tabular}{ ccccccccccccc }
\hline 
\textbf{ID} & \textbf{\textit{$z_{phot}$}} & \textbf{$\chi_r^2$} & \textbf{$\log_{10}(M^{*}/M_{\odot}$)} & \begin{tabular}{@{}c@{}}\textbf{Abs.} \\ \textbf{Mag. $(i)$}\end{tabular} & \begin{tabular}{@{}c@{}}\textbf{Age} \\ \textbf{(Gyr)}\end{tabular} & \textbf{SFH} & \begin{tabular}{@{}c@{}}\textbf{[Z/H]} \\ \textbf{(Z$_{\odot}$)}\end{tabular} & \textbf{$\sigma_{AGN}$} & \textbf{$z_{DESonly}$} & \textbf{\textit{$z_{BPZ}$}} & \textbf{E (B-V)} \\
\hline 
497174314 & $3.94_{-0.89}^{+0.47}$ & 0.17 & $11.38_{-0.15}^{+0.92}$ & -25.58 & 0.11 & $t_{trunc} = 0.1$ & 1 & 1.76 & 2.56 & 0.59 & 0.12 \\ 
498898550 & $3.67_{-0.34}^{+0.56}$ & 0.39 & $11.84_{-0.02}^{+0.25}$ & -26.42 & 0.1 & $t_{trunc} = 1.0$ & 2 & 2.18 & 2.44 & 0.41 & 0.37 \\ 
499908069 & $3.37_{-0.21}^{+0.85}$ & 0.565 & $11.88_{-0.14}^{+0.25}$ & -26.41 & 0.11 & $t_{trunc} = 0.1$ & 2 & 0.89 & 3.42 & 0.41 & 0.25 \\ 
499909599 & $4.14_{-0.47}^{+0.26}$ & 0.595 & $11.66_{-0.35}^{+0.36}$ & -26.29 & 0.1 & $t_{trunc} = 0.1$ & 2 & 2.04 & 3.21 & 0.43 & 0.25 \\ 
500048125 & $3.75_{-0.45}^{+0.64}$ & 0.585 & $11.87_{-0.18}^{+0.83}$ & -26.37 & 0.13 & SSP & 2 & 2.17 & 3.25 & 0.49 & 0.0 \\ 
500110571 & $3.99_{-0.55}^{+0.26}$ & 0.655 & $11.82_{-0.32}^{+0.36}$ & -26.7 & 0.1 & $t_{trunc} = 0.1$ & 2 & 3.83 & 3.71 & 0.41 & 0.25 \\ 
500571685 & $3.38_{-0.09}^{+0.88}$ & 0.985 & $12.18_{-0.28}^{+0.08}$ & -27.17 & 0.11 & $t_{trunc} = 0.1$ & 2 & 14.94 & 3.26 & 0.4 & 0.25 \\ 
500910602 & $4.22_{-0.55}^{+0.15}$ & 0.558 & $11.63_{-0.09}^{+0.6}$ & -26.23 & 0.1 & $t_{trunc} = 0.1$ & 2 & 1.08 & 3.25 & 0.51 & 0.25 \\ 
501218097 & $4.11_{-0.14}^{+0.22}$ & 0.666 & $11.89_{-0.02}^{+0.43}$ & -26.89 & 0.1 & $t_{trunc} = 1.0$ & 2 & 5.96 & 3.6 & 0.38 & 0.25 \\ 
501511673 & $4.2_{-0.37}^{+0.13}$ & 0.806 & $11.73_{-0.4}^{+0.04}$ & -26.46 & 0.1 & $t_{trunc} = 0.1$ & 2 & 0.23 & 3.25 & 0.49 & 0.25 \\ 
501524910 & $4.13_{-0.32}^{+0.2}$ & 0.38 & $12.01_{-0.36}^{+0.42}$ & -27.17 & 0.1 & $t_{trunc} = 0.1$ & 2 & 4.02 & 3.79 & 0.4 & 0.25 \\ 
501665859 & $3.91_{-0.48}^{+0.37}$ & 1.302 & $11.68_{-0.0}^{+0.42}$ & -26.4 & 0.1 & $t_{trunc} = 0.1$ & 1/5 & 2.39 & 2.65 & 0.29 & 0.25 \\ 
502431214 & $3.67_{-0.29}^{+0.28}$ & 0.805 & $12.4_{-0.01}^{+0.24}$ & -27.8 & 0.1 & $e^{-t/1.0~\textrm{Gyr}}$ & 2 & 7.24 & 3.67 & 0.42 & 0.37 \\ 
502433292 & $3.92_{-0.39}^{+0.33}$ & 1.378 & $12.21_{-0.04}^{+0.38}$ & -27.35 & 0.1 & $t_{trunc} = 1.0$ & 2 & 4.44 & 3.4 & 0.45 & 0.37 \\ 
502449004 & $3.71_{-0.18}^{+0.11}$ & 2.397 & $12.35_{-0.0}^{+0.05}$ & -27.69 & 0.1 & CONSTANT & 2 & 5.02 & 2.46 & 0.45 & 0.37 \\ 
503482151 & $3.68_{-0.32}^{+0.59}$ & 0.318 & $11.95_{-0.0}^{+0.08}$ & -26.71 & 0.1 & CONSTANT & 2 & 5.9 & 3.7 & 0.46 & 0.37 \\ 
503811408 & $3.88_{-0.13}^{+0.28}$ & 0.809 & $12.47_{-0.06}^{+0.22}$ & -28.0 & 0.1 & $t_{trunc} = 0.1$ & 2 & 11.55 & 3.34 & 0.47 & 0.37 \\ 
503973856 & $4.03_{-0.61}^{+0.29}$ & 1.435 & $12.45_{-0.01}^{+0.77}$ & -27.93 & 0.1 & $t_{trunc} = 0.1$ & 2 & 3.56 & 3.35 & 0.61 & 0.37 \\ 
503973990 & $3.73_{-0.37}^{+0.38}$ & 0.86 & $11.66_{-0.14}^{+0.41}$ & -26.3 & 0.1 & $t_{trunc} = 0.1$ & 2 & 2.25 & 2.4 & 0.56 & 0.25 \\ 
503984762 & $3.81_{-0.46}^{+0.3}$ & 0.301 & $11.73_{-0.26}^{+0.05}$ & -26.37 & 0.11 & $t_{trunc} = 0.1$ & 2 & 0.27 & 2.54 & 0.38 & 0.12 \\ 
503985134 & $3.85_{-0.39}^{+0.4}$ & 0.257 & $12.01_{-0.23}^{+0.12}$ & -26.84 & 0.1 & $t_{trunc} = 0.1$ & 2 & 0.82 & 3.48 & 0.48 & 0.37 \\ 
504051667 & $4.0_{-0.64}^{+0.24}$ & 1.718 & $12.18_{-0.25}^{+0.54}$ & -27.06 & 0.26 & CONSTANT & 2 & 1.42 & 3.25 & 0.39 & 0.25 \\ 
504056183 & $3.92_{-0.25}^{+0.33}$ & 0.944 & $12.26_{-0.01}^{+0.18}$ & -27.48 & 0.1 & $t_{trunc} = 1.0$ & 2 & 5.42 & 3.4 & 0.43 & 0.37 \\ 
504194446 & $4.21_{-0.26}^{+0.09}$ & 1.142 & $12.18_{-0.38}^{+0.05}$ & -27.59 & 0.1 & $t_{trunc} = 0.1$ & 2 & 9.77 & 3.48 & 0.48 & 0.25 \\ 
504330828 & $3.86_{-0.1}^{+0.14}$ & 1.841 & $11.86_{-0.37}^{+0.02}$ & -26.8 & 0.1 & $t_{trunc} = 0.1$ & 2 & 4.25 & 2.45 & 0.35 & 0.25 \\ 
504394690 & $3.56_{-0.29}^{+0.59}$ & 1.649 & $12.24_{-0.1}^{+0.61}$ & -27.31 & 0.11 & $t_{trunc} = 0.1$ & 2 & 1.57 & 3.34 & 0.37 & 0.25 \\ 
504825888 & $4.04_{-0.38}^{+0.28}$ & 0.31 & $12.03_{-0.19}^{+0.22}$ & -26.89 & 0.18 & $t_{trunc} = 0.3$ & 2 & 1.46 & 3.3 & 0.37 & 0.25 \\ 
505013250 & $3.86_{-0.16}^{+0.3}$ & 1.22 & $12.5_{-0.0}^{+0.0}$ & -28.07 & 0.1 & $t_{trunc} = 1.0$ & 2 & 12.17 & 3.35 & 0.46 & 0.37 \\ 
505018776 & $3.31_{-0.29}^{+0.65}$ & 0.523 & $12.17_{-0.13}^{+0.75}$ & -26.68 & 0.23 & $t_{trunc} = 0.1$ & 1/5 & 0.29 & 2.48 & 0.42 & 0.25 \\ 
505028285 & $4.13_{-0.32}^{+0.11}$ & 0.528 & $12.22_{-0.35}^{+0.36}$ & -27.68 & 0.1 & $t_{trunc} = 0.1$ & 2 & 17.04 & 3.48 & 0.47 & 0.25 \\ 
506153545 & $3.96_{-0.59}^{+0.32}$ & 1.414 & $11.66_{-0.29}^{+0.38}$ & -26.29 & 0.1 & $t_{trunc} = 0.1$ & 2 & 2.61 & 3.29 & 0.38 & 0.25 \\ 
506329583 & $3.77_{-0.45}^{+0.6}$ & 0.399 & $11.94_{-0.25}^{+0.87}$ & -26.72 & 0.11 & $t_{trunc} = 0.1$ & 1/2 & 1.72 & 2.56 & 0.4 & 0.25 \\ 
506345182 & $4.02_{-0.53}^{+0.29}$ & 0.485 & $12.07_{-0.13}^{+0.39}$ & -27.01 & 0.1 & $t_{trunc} = 1.0$ & 2 & 1.09 & 3.4 & 0.53 & 0.37 \\ 
506383847 & $4.17_{-0.55}^{+0.3}$ & 0.3 & $11.45_{-0.18}^{+0.76}$ & -25.95 & 0.1 & $t_{trunc} = 0.1$ & 1/2 & 4.16 & 3.12 & 0.5 & 0.25 \\ 
506534457 & $3.86_{-0.15}^{+0.1}$ & 1.624 & $12.49_{-0.01}^{+0.05}$ & -28.06 & 0.1 & CONSTANT & 2 & 8.97 & 2.49 & 0.4 & 0.37 \\ 
506537406 & $3.98_{-0.25}^{+0.35}$ & 1.501 & $12.22_{-0.01}^{+0.05}$ & -27.37 & 0.1 & CONSTANT & 2 & 4.12 & 2.59 & 0.5 & 0.37 \\ 
\hline 
\end{tabular}} 
\label{tab:cont:app-Calzetti} 
\end{table*} 
\renewcommand{\arraystretch}{1}
\renewcommand{\arraystretch}{2}
\begin{table*}
\contcaption{} 
\centering
\resizebox{\linewidth}{!}{%
\begin{tabular}{ ccccccccccccc }
\hline 
\textbf{ID} & \textbf{\textit{$z_{phot}$}} & \textbf{$\chi_r^2$} & \textbf{$\log_{10}(M^{*}/M_{\odot}$)} & \begin{tabular}{@{}c@{}}\textbf{Abs.} \\ \textbf{Mag. $(i)$}\end{tabular} & \begin{tabular}{@{}c@{}}\textbf{Age} \\ \textbf{(Gyr)}\end{tabular} & \textbf{SFH} & \begin{tabular}{@{}c@{}}\textbf{[Z/H]} \\ \textbf{(Z$_{\odot}$)}\end{tabular} & \textbf{$\sigma_{AGN}$} & \textbf{$z_{DESonly}$} & \textbf{\textit{$z_{BPZ}$}} & \textbf{E (B-V)} \\
\hline 
506572275 & $3.81_{-0.35}^{+0.44}$ & 0.905 & $12.24_{-0.1}^{+0.59}$ & -27.42 & 0.1 & $t_{trunc} = 0.1$ & 2 & 2.15 & 3.75 & 0.49 & 0.37 \\ 
506589633 & $3.73_{-0.29}^{+0.39}$ & 0.252 & $12.29_{-0.08}^{+0.27}$ & -27.53 & 0.1 & $t_{trunc} = 0.1$ & 2 & 0.77 & 2.55 & 0.43 & 0.37 \\ 
506646930 & $3.62_{-0.17}^{+0.37}$ & 0.67 & $11.78_{-0.01}^{+0.36}$ & -26.62 & 0.1 & $t_{trunc} = 1.0$ & 2 & 13.77 & 3.83 & 3.5 & 0.25 \\ 
506674710 & $3.92_{-0.6}^{+0.35}$ & 0.793 & $11.54_{-0.05}^{+0.4}$ & -26.0 & 0.1 & $t_{trunc} = 0.1$ & 2 & 4.77 & 3.29 & 0.45 & 0.25 \\ 
506674855 & $3.75_{-0.33}^{+0.46}$ & 0.231 & $11.91_{-0.24}^{+0.12}$ & -26.43 & 0.14 & $e^{-t/0.1~\textrm{Gyr}}$ & 2 & 4.12 & 2.48 & 0.35 & 0.25 \\ 
506674909 & $3.53_{-0.58}^{+0.22}$ & 2.948 & $11.96_{-0.62}^{+0.04}$ & -26.35 & 0.81 & $e^{-t/0.3~\textrm{Gyr}}$ & 2 & 4.29 & 3.31 & 0.41 & 0.0 \\ 
506675198 & $3.95_{-0.1}^{+0.19}$ & 1.455 & $12.07_{-0.05}^{+0.17}$ & -27.31 & 0.1 & $t_{trunc} = 0.1$ & 2 & 13.62 & 3.3 & 0.39 & 0.25 \\ 
507681715 & $3.81_{-0.37}^{+0.57}$ & 1.116 & $11.96_{-0.04}^{+0.59}$ & -26.72 & 0.13 & $t_{trunc} = 0.1$ & 2 & 5.25 & 2.56 & 0.41 & 0.12 \\ 
507691551 & $3.87_{-0.43}^{+0.36}$ & 0.75 & $11.66_{-0.05}^{+0.16}$ & -26.3 & 0.1 & $e^{-t/1.0~\textrm{Gyr}}$ & 2 & 3.45 & 3.34 & 0.38 & 0.25 \\ 
507780409 & $3.84_{-0.4}^{+0.16}$ & 0.874 & $12.13_{-0.07}^{+0.15}$ & -27.07 & 0.13 & $e^{-t/0.1~\textrm{Gyr}}$ & 2 & 1.67 & 3.65 & 0.36 & 0.25 \\ 
507785363 & $3.88_{-0.44}^{+0.58}$ & 0.37 & $12.1_{-0.29}^{+0.57}$ & -26.95 & 0.11 & $t_{trunc} = 0.1$ & 2 & 0.67 & 2.64 & 0.48 & 0.25 \\ 
507791066 & $3.75_{-0.21}^{+0.41}$ & 0.553 & $12.42_{-0.02}^{+0.22}$ & -27.88 & 0.1 & CONSTANT & 2 & 10.98 & 3.7 & 0.47 & 0.37 \\ 
507791530 & $3.77_{-0.31}^{+0.53}$ & 0.702 & $12.45_{-0.02}^{+0.08}$ & -27.62 & 0.1 & $t_{trunc} = 0.3$ & 2 & 2.4 & 2.6 & 0.54 & 0.49 \\ 
507803985 & $3.83_{-0.38}^{+0.35}$ & 0.62 & $12.15_{-0.03}^{+0.19}$ & -27.21 & 0.1 & CONSTANT & 2 & 1.63 & 3.5 & 0.4 & 0.37 \\ 
507810919 & $3.77_{-0.21}^{+0.21}$ & 0.563 & $12.37_{-0.04}^{+0.09}$ & -27.73 & 0.1 & $t_{trunc} = 0.1$ & 2 & 9.6 & 2.55 & 0.41 & 0.37 \\ 
508217521 & $3.89_{-0.73}^{+0.37}$ & 0.38 & $11.31_{-0.18}^{+0.4}$ & -25.41 & 0.1 & $t_{trunc} = 0.1$ & 2 & 1.41 & 3.4 & 0.45 & 0.25 \\ 
618652137 & $3.85_{-0.18}^{+0.3}$ & 0.711 & $12.47_{-0.0}^{+0.19}$ & -28.01 & 0.1 & CONSTANT & 2 & 4.1 & 3.34 & 0.45 & 0.37 \\ 
618654757 & $3.94_{-0.27}^{+0.35}$ & 0.554 & $11.65_{-0.02}^{+0.16}$ & -26.28 & 0.1 & $t_{trunc} = 1.0$ & 2 & 0.64 & 3.27 & 0.43 & 0.25 \\ 
618660654 & $3.77_{-0.19}^{+0.38}$ & 1.199 & $12.12_{-0.0}^{+0.19}$ & -27.45 & 0.1 & $t_{trunc} = 1.0$ & 2 & 27.05 & 3.37 & 0.27 & 0.25 \\ 
618664093 & $4.08_{-0.42}^{+0.14}$ & 0.811 & $12.33_{-0.02}^{+0.11}$ & -27.45 & 0.26 & CONSTANT & 2 & 1.29 & 3.34 & 0.47 & 0.25 \\ 
618664306 & $4.09_{-0.24}^{+0.25}$ & 0.964 & $12.06_{-0.02}^{+0.25}$ & -27.38 & 0.1 & $t_{trunc} = 0.1$ & 1 & 1.62 & 3.21 & 0.41 & 0.25 \\ 
618667069 & $3.75_{-0.45}^{+0.2}$ & 1.662 & $12.46_{-0.24}^{+0.1}$ & -27.48 & 0.14 & $e^{-t/0.1~\textrm{Gyr}}$ & 2 & 0.76 & 2.42 & 0.49 & 0.37 \\ 
618667272 & $3.32_{-0.05}^{+0.94}$ & 2.294 & $12.2_{-0.03}^{+0.67}$ & -27.14 & 0.14 & $t_{trunc} = 0.1$ & 2 & 0.4 & 3.29 & 0.4 & 0.12 \\       
		\hline
	\end{tabular}}
	\label{tab:cont:app-Calzetti}
\end{table*}
\renewcommand{\arraystretch}{1}

\section{Photometry of All Candidates}
\label{app:AllPhotometry}

The full DES+VHS photometry, and RA and Dec coordinates (J2000) are provided for all galaxies matching the best candidate criteria. 

\clearpage 
\renewcommand{\arraystretch}{2}
\begin{table*}
	\caption{Photometry for all galaxies matching the best candidate criteria (as in Section~\ref{sec:SelBestCandidates}).}
	\centering
	\resizebox{\linewidth}{!}{%
\begin{tabular}{ cccccccccccc }
		\hline
		\textbf{ID} & \textbf{RA} & \textbf{Dec} & \textbf{$g$} & \textbf{$r$} & \textbf{$i$} & \textbf{$z$} & \textbf{$Y$} & \textbf{$J$} & \textbf{$H$} & \textbf{$K_{s}$} \\
		\hline
		100600870 & 342.30182 & -45.078395 & $22.6555 \pm 0.0389$ & $20.8248 \pm 0.0107$ & $20.4152 \pm 0.0091$ & $20.0676 \pm 0.017$ & $20.0109 \pm 0.0434$ & $19.5742 \pm 0.1168$ & /// & $19.1676 \pm 0.1954$ \\ 
100669215 & 342.03479 & -44.585222 & $23.4434 \pm 0.0705$ & $21.6989 \pm 0.0216$ & $21.3218 \pm 0.0203$ & $21.1502 \pm 0.0362$ & $20.8496 \pm 0.1093$ & $20.3986 \pm 0.2289$ & /// & $20.0326 \pm 0.2853$ \\ 
102002089 & 342.83733 & -44.098439 & $24.601 \pm 0.1949$ & $22.4569 \pm 0.0323$ & $21.8981 \pm 0.0313$ & $21.4884 \pm 0.0422$ & $21.5888 \pm 0.1419$ & $22.069 \pm 0.6841$ & /// & $20.7842 \pm 0.5095$ \\ 
102009403 & 342.92415 & -44.216238 & $23.0138 \pm 0.0534$ & $21.1502 \pm 0.0133$ & $20.7563 \pm 0.0117$ & $20.3417 \pm 0.0217$ & $20.2165 \pm 0.0552$ & $20.2064 \pm 0.2059$ & /// & $19.8197 \pm 0.4318$ \\ 
102009835 & 342.78469 & -44.222644 & $23.2401 \pm 0.0574$ & $21.2488 \pm 0.0128$ & $20.7316 \pm 0.0108$ & $20.3393 \pm 0.0172$ & $20.0885 \pm 0.0449$ & $19.9653 \pm 0.1708$ & /// & $19.5771 \pm 0.301$ \\ 
102009849 & 343.07164 & -44.222642 & $22.9057 \pm 0.0415$ & $20.9317 \pm 0.0096$ & $20.4301 \pm 0.0082$ & $20.0647 \pm 0.0132$ & $19.9559 \pm 0.0384$ & $19.5375 \pm 0.1272$ & /// & $18.8814 \pm 0.1611$ \\ 
102031864 & 342.67377 & -44.586054 & $23.7029 \pm 0.0824$ & $21.6185 \pm 0.0158$ & $21.166 \pm 0.0131$ & $20.6985 \pm 0.0243$ & $20.559 \pm 0.069$ & $20.3562 \pm 0.1998$ & /// & $20.0655 \pm 0.3812$ \\
\hline 
\end{tabular}} 
\label{tab:app:MagGoldenSample} 
\end{table*} 
\renewcommand{\arraystretch}{1}
\renewcommand{\arraystretch}{2}
\begin{table*}
\contcaption{} 
\centering
\resizebox{\linewidth}{!}{%
\begin{tabular}{ cccccccccccc }
\hline 
\textbf{ID} & \textbf{RA} & \textbf{Dec} & \textbf{$g$} & \textbf{$r$} & \textbf{$i$} & \textbf{$z$} & \textbf{$Y$} & \textbf{$J$} & \textbf{$H$} & \textbf{$K_{s}$} \\
\hline 
397764328 & 74.890797 & -52.347374 & $23.5908 \pm 0.0918$ & $21.6711 \pm 0.0211$ & $21.2131 \pm 0.0219$ & $20.6847 \pm 0.0255$ & $20.5936 \pm 0.0656$ & $20.8524 \pm 0.2944$ & $19.809 \pm 0.257$ & $20.0628 \pm 0.3292$ \\ 
397885462 & 73.562242 & -59.48389 & $24.9502 \pm 0.2252$ & $22.5626 \pm 0.0249$ & $21.9313 \pm 0.0258$ & $21.5111 \pm 0.0346$ & $21.5167 \pm 0.1147$ & $21.4336 \pm 0.469$ & /// & $20.1094 \pm 0.3602$ \\ 
398107560 & 74.313676 & -51.699346 & $23.5493 \pm 0.0617$ & $21.6321 \pm 0.014$ & $21.1804 \pm 0.0144$ & $20.7587 \pm 0.0196$ & $20.6536 \pm 0.0679$ & $20.6748 \pm 0.27$ & $20.1912 \pm 0.3498$ & $19.5987 \pm 0.2567$ \\ 
399804681 & 74.814022 & -51.10434 & $24.2284 \pm 0.1626$ & $22.166 \pm 0.0331$ & $21.604 \pm 0.0277$ & $21.032 \pm 0.0389$ & $20.9641 \pm 0.1101$ & /// & $21.3539 \pm 0.8698$ & $19.7706 \pm 0.38$ \\ 
399842053 & 74.771454 & -51.68691 & $23.36 \pm 0.0647$ & $21.5027 \pm 0.0156$ & $21.0551 \pm 0.0165$ & $20.6205 \pm 0.0208$ & $20.5097 \pm 0.0549$ & $20.2867 \pm 0.2367$ & $20.8851 \pm 0.7733$ & $19.7016 \pm 0.3453$ \\ 
399842613 & 75.053674 & -51.696081 & $24.9501 \pm 0.2206$ & $22.4861 \pm 0.029$ & $21.746 \pm 0.0226$ & $21.2557 \pm 0.0297$ & $21.2596 \pm 0.1008$ & /// & $20.7158 \pm 0.4189$ & $20.3271 \pm 0.3888$ \\ 
400998781 & 75.100601 & -45.683346 & $23.7645 \pm 0.0918$ & $21.839 \pm 0.0178$ & $21.3687 \pm 0.0167$ & $21.0268 \pm 0.0233$ & $20.7628 \pm 0.0704$ & $20.4275 \pm 0.2373$ & $20.5278 \pm 0.3386$ & $20.3169 \pm 0.5816$ \\
\hline 
\end{tabular}} 
\label{tab:app:cont:MagGoldenSample} 
\end{table*} 
\renewcommand{\arraystretch}{1}
\renewcommand{\arraystretch}{2}
\begin{table*}
\contcaption{} 
\centering
\resizebox{\linewidth}{!}{%
\begin{tabular}{ cccccccccccc }
\hline 
\textbf{ID} & \textbf{RA} & \textbf{Dec} & \textbf{$g$} & \textbf{$r$} & \textbf{$i$} & \textbf{$z$} & \textbf{$Y$} & \textbf{$J$} & \textbf{$H$} & \textbf{$K_{s}$} \\
\hline 
431827017 & 81.354041 & -58.057944 & $23.2026 \pm 0.0476$ & $21.2471 \pm 0.011$ & $20.7517 \pm 0.0115$ & $20.3163 \pm 0.0167$ & $20.152 \pm 0.0392$ & $19.8805 \pm 0.1737$ & /// & $19.1435 \pm 0.2499$ \\ 
434401854 & 82.755855 & -59.110482 & $22.1099 \pm 0.0321$ & $20.2853 \pm 0.008$ & $19.8901 \pm 0.0083$ & $19.4465 \pm 0.0112$ & $19.3971 \pm 0.029$ & $19.025 \pm 0.0832$ & /// & $18.2387 \pm 0.1132$ \\ 
444147103 & 85.376884 & -58.940568 & $23.5463 \pm 0.0735$ & $21.6129 \pm 0.0154$ & $21.192 \pm 0.0154$ & $20.818 \pm 0.0267$ & $20.775 \pm 0.0862$ & $20.1796 \pm 0.2493$ & /// & $19.7918 \pm 0.3938$ \\ 
444182193 & 85.110945 & -59.440915 & $23.7537 \pm 0.0926$ & $21.8281 \pm 0.0205$ & $21.3662 \pm 0.0227$ & $20.8868 \pm 0.0312$ & $20.7948 \pm 0.0928$ & $20.6415 \pm 0.2817$ & /// & $19.5198 \pm 0.3014$ \\ 
446501990 & 85.426451 & -59.886883 & $24.6979 \pm 0.188$ & $22.3271 \pm 0.0229$ & $21.6196 \pm 0.0219$ & $21.2602 \pm 0.0285$ & $21.088 \pm 0.0768$ & $20.4715 \pm 0.2061$ & /// & $20.2923 \pm 0.4367$ \\ 
465281154 & 88.694639 & -60.880289 & $23.8226 \pm 0.1104$ & $21.8313 \pm 0.0213$ & $21.328 \pm 0.0163$ & $20.9986 \pm 0.0192$ & $20.9853 \pm 0.0918$ & $20.7884 \pm 0.3129$ & /// & $20.0238 \pm 0.4269$ \\ 
470611726 & 68.015452 & -57.726539 & $24.4208 \pm 0.1593$ & $22.3445 \pm 0.0275$ & $21.8636 \pm 0.0288$ & $21.4757 \pm 0.0426$ & $21.2412 \pm 0.1197$ & $21.3266 \pm 0.3366$ & /// & $20.1595 \pm 0.4922$ \\
\hline 
\end{tabular}} 
\label{tab:app:cont:MagGoldenSample} 
\end{table*} 
\renewcommand{\arraystretch}{1}
\renewcommand{\arraystretch}{2}
\begin{table*}
\contcaption{} 
\centering
\resizebox{\linewidth}{!}{%
\begin{tabular}{ cccccccccccc }
\hline 
\textbf{ID} & \textbf{RA} & \textbf{Dec} & \textbf{$g$} & \textbf{$r$} & \textbf{$i$} & \textbf{$z$} & \textbf{$Y$} & \textbf{$J$} & \textbf{$H$} & \textbf{$K_{s}$} \\
\hline 
481350973 & 72.813908 & -47.600626 & $23.1958 \pm 0.0612$ & $21.3706 \pm 0.0158$ & $20.9589 \pm 0.0182$ & $20.6744 \pm 0.0213$ & $20.5618 \pm 0.0759$ & $20.5647 \pm 0.3233$ & /// & $19.9526 \pm 0.5323$ \\ 
481989803 & 72.120899 & -45.61534 & $24.4427 \pm 0.1475$ & $22.4001 \pm 0.0324$ & $21.9192 \pm 0.0285$ & $21.5902 \pm 0.0425$ & $21.6684 \pm 0.1651$ & $20.8491 \pm 0.2714$ & $20.8164 \pm 0.3249$ & /// \\ 
481994767 & 71.896784 & -45.687727 & $24.8643 \pm 0.2282$ & $22.7831 \pm 0.0452$ & $22.2127 \pm 0.0404$ & $21.9062 \pm 0.0618$ & $21.8439 \pm 0.2125$ & $21.8744 \pm 0.5399$ & $22.007 \pm 0.7345$ & /// \\ 
482001634 & 72.307954 & -45.794599 & $24.2255 \pm 0.1234$ & $22.5135 \pm 0.0314$ & $22.1931 \pm 0.0315$ & $21.7569 \pm 0.0548$ & $22.0476 \pm 0.2355$ & $21.3845 \pm 0.3144$ & $21.9101 \pm 0.6719$ & /// \\ 
482208365 & 73.320644 & -51.189418 & $22.8554 \pm 0.0359$ & $20.9531 \pm 0.0099$ & $20.5427 \pm 0.0095$ & $20.1552 \pm 0.014$ & $20.0031 \pm 0.0415$ & $19.931 \pm 0.2528$ & $19.1732 \pm 0.209$ & $19.2358 \pm 0.2245$ \\ 
483918716 & 72.085462 & -48.777383 & $24.1798 \pm 0.1091$ & $22.2827 \pm 0.0232$ & $21.8386 \pm 0.0258$ & $21.435 \pm 0.0344$ & $21.4798 \pm 0.1145$ & $21.4635 \pm 0.3782$ & /// & $19.9161 \pm 0.3324$ \\ 
489254835 & 69.13225 & -55.548991 & $24.316 \pm 0.1564$ & $22.0218 \pm 0.0194$ & $21.4483 \pm 0.0165$ & $21.0965 \pm 0.0303$ & $20.9006 \pm 0.0864$ & $20.7104 \pm 0.2035$ & /// & $20.0555 \pm 0.3584$ \\
\hline 
\end{tabular}} 
\label{tab:app:cont:MagGoldenSample} 
\end{table*} 
\renewcommand{\arraystretch}{1}
\renewcommand{\arraystretch}{2}
\begin{table*}
\contcaption{} 
\centering
\resizebox{\linewidth}{!}{%
\begin{tabular}{ cccccccccccc }
\hline 
\textbf{ID} & \textbf{RA} & \textbf{Dec} & \textbf{$g$} & \textbf{$r$} & \textbf{$i$} & \textbf{$z$} & \textbf{$Y$} & \textbf{$J$} & \textbf{$H$} & \textbf{$K_{s}$} \\
\hline 
502431214 & 65.942299 & -46.553076 & $22.7071 \pm 0.0379$ & $20.9606 \pm 0.0099$ & $20.6372 \pm 0.0112$ & $20.2516 \pm 0.0176$ & $20.1892 \pm 0.0442$ & $19.9818 \pm 0.1994$ & $19.2778 \pm 0.1497$ & $19.0157 \pm 0.1745$ \\ 
502433292 & 66.409402 & -46.568465 & $23.6993 \pm 0.0929$ & $21.727 \pm 0.0198$ & $21.2826 \pm 0.0206$ & $20.8858 \pm 0.0303$ & $20.9898 \pm 0.115$ & $20.1812 \pm 0.2583$ & $19.8842 \pm 0.263$ & $19.974 \pm 0.3567$ \\ 
502449004 & 65.844648 & -46.707587 & $22.8891 \pm 0.0412$ & $21.0685 \pm 0.0131$ & $20.7778 \pm 0.0151$ & $20.474 \pm 0.0207$ & $20.146 \pm 0.0553$ & $19.8829 \pm 0.12$ & /// & $19.5314 \pm 0.3161$ \\ 
503482151 & 66.649327 & -58.546895 & $23.8134 \pm 0.1008$ & $22.062 \pm 0.0256$ & $21.6794 \pm 0.0291$ & $21.3887 \pm 0.0404$ & $21.189 \pm 0.1058$ & $21.1799 \pm 0.2995$ & /// & $20.2835 \pm 0.3403$ \\ 
503811408 & 70.787127 & -52.978216 & $23.2053 \pm 0.0594$ & $21.28 \pm 0.0135$ & $20.8055 \pm 0.0143$ & $20.3545 \pm 0.0178$ & $20.2962 \pm 0.055$ & $20.1231 \pm 0.2981$ & $19.3309 \pm 0.218$ & $19.1173 \pm 0.2316$ \\ 
503973856 & 71.227329 & -44.778327 & $23.6823 \pm 0.1294$ & $21.5272 \pm 0.0331$ & $20.9504 \pm 0.0298$ & $20.6236 \pm 0.0319$ & $20.6429 \pm 0.1169$ & $19.6565 \pm 0.1644$ & /// & $18.9792 \pm 0.2099$ \\ 
503973990 & 71.217655 & -44.780499 & $23.3519 \pm 0.0959$ & $21.6747 \pm 0.0375$ & $21.5158 \pm 0.0491$ & $21.1546 \pm 0.0515$ & $21.0994 \pm 0.1777$ & $21.2706 \pm 0.4969$ & /// & $20.2714 \pm 0.3984$ \\
\hline 
\end{tabular}} 
\label{tab:app:cont:MagGoldenSample} 
\end{table*} 
\renewcommand{\arraystretch}{1}
\renewcommand{\arraystretch}{2}
\begin{table*}
\contcaption{} 
\centering
\resizebox{\linewidth}{!}{%
\begin{tabular}{ cccccccccccc }
\hline 
\textbf{ID} & \textbf{RA} & \textbf{Dec} & \textbf{$g$} & \textbf{$r$} & \textbf{$i$} & \textbf{$z$} & \textbf{$Y$} & \textbf{$J$} & \textbf{$H$} & \textbf{$K_{s}$} \\
\hline 
618652137 & 358.88296 & -54.30947 & $22.8707 \pm 0.0532$ & $21.0181 \pm 0.0112$ & $20.5872 \pm 0.0106$ & $20.1606 \pm 0.0143$ & $20.085 \pm 0.0485$ & $19.9487 \pm 0.2035$ & $19.8445 \pm 0.3877$ & $18.9431 \pm 0.1907$ \\ 
618654757 & 358.67695 & -54.356158 & $23.549 \pm 0.137$ & $21.8112 \pm 0.0272$ & $21.4459 \pm 0.0274$ & $21.1584 \pm 0.0431$ & $21.2514 \pm 0.1468$ & $21.0447 \pm 0.3029$ & $20.8919 \pm 0.477$ & $21.2233 \pm 1.0699$ \\ 
618660654 & 359.14404 & -54.461467 & $22.0072 \pm 0.0258$ & $20.3997 \pm 0.0077$ & $20.1682 \pm 0.009$ & $19.8417 \pm 0.0127$ & $19.8875 \pm 0.0355$ & $19.6534 \pm 0.1604$ & $20.4625 \pm 0.7439$ & $19.1208 \pm 0.2385$ \\ 
618663972 & 358.73193 & -54.517818 & $22.8714 \pm 0.045$ & $20.9306 \pm 0.0113$ & $20.5016 \pm 0.0106$ & $20.0363 \pm 0.0137$ & $20.0142 \pm 0.037$ & $19.4945 \pm 0.1769$ & $19.2994 \pm 0.1598$ & $18.8724 \pm 0.155$ \\ 
618664093 & 359.04338 & -54.519852 & $23.1158 \pm 0.0456$ & $21.2542 \pm 0.0124$ & $20.818 \pm 0.0123$ & $20.4231 \pm 0.0171$ & $20.4425 \pm 0.0485$ & $19.9913 \pm 0.1884$ & $19.9082 \pm 0.4418$ & $19.3846 \pm 0.2939$ \\ 
618664306 & 359.09017 & -54.523087 & $22.9282 \pm 0.0433$ & $21.0268 \pm 0.0135$ & $20.5623 \pm 0.0124$ & $20.2491 \pm 0.0187$ & $20.3538 \pm 0.0456$ & $19.8083 \pm 0.1913$ & $20.1177 \pm 0.5618$ & $19.2618 \pm 0.2835$ \\ 
618667069 & 358.8217 & -54.567628 & $23.8333 \pm 0.1179$ & $21.8771 \pm 0.0233$ & $21.5924 \pm 0.0206$ & $21.0456 \pm 0.0259$ & $20.9855 \pm 0.105$ & $20.6079 \pm 0.323$ & $20.1235 \pm 0.4174$ & $19.4716 \pm 0.2758$ \\
\hline
	\end{tabular}}
	\label{tab:app:cont:MagGoldenSample}
\end{table*}
\renewcommand{\arraystretch}{1}

\bsp	
\label{lastpage}
\end{document}